\documentclass[twocolumn]{aastex7}

\usepackage{verbatim}
\usepackage{xcolor}
\usepackage{float}
\usepackage{tabularx}
\usepackage{makecell}
\usepackage{multirow}
\usepackage{enumitem}
\usepackage{graphicx}
\usepackage[T1]{fontenc}
\usepackage{textcomp}
\usepackage{hyperref}
\usepackage{longtable}
\usepackage{orcidlink}

\begin{document}

\shortauthors{Ratajczak et al.}
\shorttitle{DESI-COSMOS and DESI-XMMLSS Catalogs}

\title{The Compilation and Validation of the Spectroscopic Redshift Catalogs for the DESI-COSMOS and DESI-XMMLSS Fields}


\author[0009-0000-5242-7549]{J.~Ratajczak}
\email{joshua.ratajczak@utah.edu}
\affiliation{Department of Physics and Astronomy, The University of Utah, 115 South 1400 East, Salt Lake City, UT 84112, USA}
\author[0000-0002-0553-3805]{K.~S.~Dawson}
\email{kdawson@astro.utah.edu}
\affiliation{Department of Physics and Astronomy, The University of Utah, 115 South 1400 East, Salt Lake City, UT 84112, USA}
\author[0000-0001-9382-5199]{N.~Weaverdyck}
\email{nweaverdyck@lbl.gov}
\affiliation{Lawrence Berkeley National Laboratory, 1 Cyclotron Road, Berkeley, CA 94720, USA}
\author{J.~Aguilar}
\email{jaguilar@lbl.gov}
\affiliation{Lawrence Berkeley National Laboratory, 1 Cyclotron Road, Berkeley, CA 94720, USA}
\author[0000-0001-6098-7247]{S.~Ahlen}
\email{ahlen@bu.edu}
\affiliation{Department of Physics, Boston University, 590 Commonwealth Avenue, Boston, MA 02215 USA}
\author[0000-0001-7600-5148]{E.~Armengaud}
\email{eric.armengaud@cea.fr}
\affiliation{IRFU, CEA, Universit\'{e} Paris-Saclay, F-91191 Gif-sur-Yvette, France}
\author[0000-0003-4162-6619]{S.~Bailey}
\email{stephenbailey@lbl.gov}
\affiliation{Lawrence Berkeley National Laboratory, 1 Cyclotron Road, Berkeley, CA 94720, USA}
\author[0000-0001-9712-0006]{D.~Bianchi}
\email{davide.bianchi1@unimi.it}
\affiliation{Dipartimento di Fisica ``Aldo Pontremoli'', Universit\`a degli Studi di Milano, Via Celoria 16, I-20133 Milano, Italy}
\affiliation{INAF-Osservatorio Astronomico di Brera, Via Brera 28, 20122 Milano, Italy}
\author{D.~Blanco}
\email{blanco@ucsc.edu}
\affiliation{Department of Astronomy and Astrophysics, UCO/Lick Observatory, University of California, 1156 High Street, Santa Cruz, CA 95064, USA}
\author[0000-0002-8934-0954]{A.~Brodzeller}
\email{AllysonBrodzeller@lbl.gov}
\affiliation{Lawrence Berkeley National Laboratory, 1 Cyclotron Road, Berkeley, CA 94720, USA}
\author{D.~Brooks}
\email{david.brooks@ucl.ac.uk}
\affiliation{Department of Physics \& Astronomy, University College London, Gower Street, London, WC1E 6BT, UK}
\author[0000-0001-7316-4573]{F.~J.~Castander}
\email{fjc@ice.csic.es}
\affiliation{Institut d'Estudis Espacials de Catalunya (IEEC), c/ Esteve Terradas 1, Edifici RDIT, Campus PMT-UPC, 08860 Castelldefels, Spain}
\affiliation{Institute of Space Sciences, ICE-CSIC, Campus UAB, Carrer de Can Magrans s/n, 08913 Bellaterra, Barcelona, Spain}
\author{T.~Claybaugh}
\email{tmclaybaugh@lbl.gov}
\affiliation{Lawrence Berkeley National Laboratory, 1 Cyclotron Road, Berkeley, CA 94720, USA}
\author[0000-0002-2169-0595]{A.~Cuceu}
\email{acuceu@lbl.gov}
\affiliation{Lawrence Berkeley National Laboratory, 1 Cyclotron Road, Berkeley, CA 94720, USA}
\author[0000-0002-1769-1640]{A.~de la Macorra}
\email{macorra@fisica.unam.mx}
\affiliation{Instituto de F\'{\i}sica, Universidad Nacional Aut\'{o}noma de M\'{e}xico,  Circuito de la Investigaci\'{o}n Cient\'{\i}fica, Ciudad Universitaria, Cd. de M\'{e}xico  C.~P.~04510,  M\'{e}xico}
\author[0000-0002-4928-4003]{Arjun~Dey}
\email{arjun.dey@noirlab.edu}
\affiliation{NSF NOIRLab, 950 N. Cherry Ave., Tucson, AZ 85719, USA}
\author[0000-0002-5665-7912]{Biprateep~Dey}
\email{b.dey@utoronto.ca}
\affiliation{Department of Astronomy \& Astrophysics, University of Toronto, Toronto, ON M5S 3H4, Canada}
\affiliation{Department of Physics \& Astronomy and Pittsburgh Particle Physics, Astrophysics, and Cosmology Center (PITT PACC), University of Pittsburgh, 3941 O'Hara Street, Pittsburgh, PA 15260, USA}
\author{P.~Doel}
\email{apd@star.ucl.ac.uk}
\affiliation{Department of Physics \& Astronomy, University College London, Gower Street, London, WC1E 6BT, UK}
\author[0000-0002-3033-7312]{A.~Font-Ribera}
\email{afont@ifae.es}
\affiliation{Institut de F\'{i}sica d’Altes Energies (IFAE), The Barcelona Institute of Science and Technology, Edifici Cn, Campus UAB, 08193, Bellaterra (Barcelona), Spain}
\author[0000-0002-2890-3725]{J.~E.~Forero-Romero}
\email{je.forero@uniandes.edu.co}
\affiliation{Departamento de F\'isica, Universidad de los Andes, Cra. 1 No. 18A-10, Edificio Ip, CP 111711, Bogot\'a, Colombia}
\affiliation{Observatorio Astron\'omico, Universidad de los Andes, Cra. 1 No. 18A-10, Edificio H, CP 111711 Bogot\'a, Colombia}
\author[0000-0001-9632-0815]{E.~Gaztañaga}
\email{gaztanaga@gmail.com}
\affiliation{Institut d'Estudis Espacials de Catalunya (IEEC), c/ Esteve Terradas 1, Edifici RDIT, Campus PMT-UPC, 08860 Castelldefels, Spain}
\affiliation{Institute of Cosmology and Gravitation, University of Portsmouth, Dennis Sciama Building, Portsmouth, PO1 3FX, UK}
\affiliation{Institute of Space Sciences, ICE-CSIC, Campus UAB, Carrer de Can Magrans s/n, 08913 Bellaterra, Barcelona, Spain}
\author[0000-0003-3142-233X]{S.~Gontcho A Gontcho}
\email{satyagontcho@lbl.gov}
\affiliation{Lawrence Berkeley National Laboratory, 1 Cyclotron Road, Berkeley, CA 94720, USA}
\affiliation{University of Virginia, Department of Astronomy, Charlottesville, VA 22904, USA}
\author{G.~Gutierrez}
\email{gaston@fnal.gov}
\affiliation{Fermi National Accelerator Laboratory, PO Box 500, Batavia, IL 60510, USA}
\author[0000-0001-9822-6793]{J.~Guy}
\email{jguy@lbl.gov}
\affiliation{Lawrence Berkeley National Laboratory, 1 Cyclotron Road, Berkeley, CA 94720, USA}
\author[0009-0007-2936-1124]{T.~Hagen}
\email{tyler.hagen@utah.edu}
\affiliation{Department of Physics and Astronomy, The University of Utah, 115 South 1400 East, Salt Lake City, UT 84112, USA}
\author[0000-0002-9136-9609]{H.~K.~Herrera-Alcantar}
\email{hiram.herreraalcantar@cea.fr}
\affiliation{Institut d'Astrophysique de Paris. 98 bis boulevard Arago. 75014 Paris, France}
\affiliation{IRFU, CEA, Universit\'{e} Paris-Saclay, F-91191 Gif-sur-Yvette, France}
\author[0000-0002-6550-2023]{K.~Honscheid}
\email{kh@physics.osu.edu}
\affiliation{Center for Cosmology and AstroParticle Physics, The Ohio State University, 191 West Woodruff Avenue, Columbus, OH 43210, USA}
\affiliation{Department of Physics, The Ohio State University, 191 West Woodruff Avenue, Columbus, OH 43210, USA}
\affiliation{The Ohio State University, Columbus, 43210 OH, USA}
\author[0000-0001-6558-0112]{D.~Huterer}
\email{huterer@umich.edu}
\affiliation{Department of Physics, University of Michigan, 450 Church Street, Ann Arbor, MI 48109, USA}
\affiliation{University of Michigan, 500 S. State Street, Ann Arbor, MI 48109, USA}
\author[0000-0002-6024-466X]{M.~Ishak}
\email{mishak@utdallas.edu}
\affiliation{Department of Physics, The University of Texas at Dallas, 800 W. Campbell Rd., Richardson, TX 75080, USA}
\author[0000-0001-8528-3473]{J.~Jimenez}
\email{jjimenez@ifae.es}
\affiliation{Institut de F\'{i}sica d’Altes Energies (IFAE), The Barcelona Institute of Science and Technology, Edifici Cn, Campus UAB, 08193, Bellaterra (Barcelona), Spain}
\author[0000-0003-0201-5241]{R.~Joyce}
\email{richard.joyce@noirlab.edu}
\affiliation{NSF NOIRLab, 950 N. Cherry Ave., Tucson, AZ 85719, USA}
\author[0000-0002-0000-2394]{S.~Juneau}
\email{stephanie.juneau@noirlab.edu}
\affiliation{NSF NOIRLab, 950 N. Cherry Ave., Tucson, AZ 85719, USA}
\author{R.~Kehoe}
\email{kehoe@physics.smu.edu}
\affiliation{Department of Physics, Southern Methodist University, 3215 Daniel Avenue, Dallas, TX 75275, USA}
\author[0000-0002-8828-5463]{D.~Kirkby}
\email{dkirkby@uci.edu}
\affiliation{Department of Physics and Astronomy, University of California, Irvine, 92697, USA}
\author[0000-0003-3510-7134]{T.~Kisner}
\email{tskisner@lbl.gov}
\affiliation{Lawrence Berkeley National Laboratory, 1 Cyclotron Road, Berkeley, CA 94720, USA}
\author[0000-0003-2644-135X]{S.~E.~Koposov}
\email{skoposov@ed.ac.uk}
\affiliation{Institute for Astronomy, University of Edinburgh, Royal Observatory, Blackford Hill, Edinburgh EH9 3HJ, UK}
\affiliation{Institute of Astronomy, University of Cambridge, Madingley Road, Cambridge CB3 0HA, UK}
\author[0000-0001-6356-7424]{A.~Kremin}
\email{akremin@lbl.gov}
\affiliation{Lawrence Berkeley National Laboratory, 1 Cyclotron Road, Berkeley, CA 94720, USA}
\author{O.~Lahav}
\email{o.lahav@ucl.ac.uk}
\affiliation{Department of Physics \& Astronomy, University College London, Gower Street, London, WC1E 6BT, UK}
\author{A.~Lambert}
\email{arlambert@lbl.gov}
\affiliation{Lawrence Berkeley National Laboratory, 1 Cyclotron Road, Berkeley, CA 94720, USA}
\author[0000-0002-6731-9329]{C.~Lamman}
\email{clamman@g.harvard.edu}
\affiliation{Center for Astrophysics $|$ Harvard \& Smithsonian, 60 Garden Street, Cambridge, MA 02138, USA}
\author[0000-0003-1838-8528]{M.~Landriau}
\email{mlandriau@lbl.gov}
\affiliation{Lawrence Berkeley National Laboratory, 1 Cyclotron Road, Berkeley, CA 94720, USA}
\author[0000-0001-7178-8868]{L.~Le~Guillou}
\email{llg@lpnhe.in2p3.fr}
\affiliation{Sorbonne Universit\'{e}, CNRS/IN2P3, Laboratoire de Physique Nucl\'{e}aire et de Hautes Energies (LPNHE), FR-75005 Paris, France}
\author[0000-0002-3677-3617]{A.~Leauthaud}
\email{alexie@ucsc.edu}
\affiliation{Department of Astronomy and Astrophysics, UCO/Lick Observatory, University of California, 1156 High Street, Santa Cruz, CA 95064, USA}
\affiliation{Department of Astronomy and Astrophysics, University of California, Santa Cruz, 1156 High Street, Santa Cruz, CA 95065, USA}
\author{J.~Lee}
\email{joey.s.lee@utah.edu}
\affiliation{Department of Physics and Astronomy, The University of Utah, 115 South 1400 East, Salt Lake City, UT 84112, USA}
\author[0000-0003-1887-1018]{M.~E.~Levi}
\email{melevi@lbl.gov}
\affiliation{Lawrence Berkeley National Laboratory, 1 Cyclotron Road, Berkeley, CA 94720, USA}
\author[0000-0003-3616-6486]{Q.~Li}
\email{qinxun.li@utah.edu}
\affiliation{Department of Physics and Astronomy, The University of Utah, 115 South 1400 East, Salt Lake City, UT 84112, USA}
\author{I.~Longhurst}
\email{u1047163@umail.utah.edu}
\affiliation{Department of Physics and Astronomy, The University of Utah, 115 South 1400 East, Salt Lake City, UT 84112, USA}
\author[0000-0001-7729-6629]{Y.~Luo}
\email{yifeiluo@lbl.gov}
\affiliation{Lawrence Berkeley National Laboratory, 1 Cyclotron Road, Berkeley, CA 94720, USA}
\author[0000-0003-4962-8934]{M.~Manera}
\email{mmanera@ifae.es}
\affiliation{Departament de F\'{i}sica, Serra H\'{u}nter, Universitat Aut\`{o}noma de Barcelona, 08193 Bellaterra (Barcelona), Spain}
\affiliation{Institut de F\'{i}sica d’Altes Energies (IFAE), The Barcelona Institute of Science and Technology, Edifici Cn, Campus UAB, 08193, Bellaterra (Barcelona), Spain}
\author[0000-0002-4279-4182]{P.~Martini}
\email{martini.10@osu.edu}
\affiliation{Center for Cosmology and AstroParticle Physics, The Ohio State University, 191 West Woodruff Avenue, Columbus, OH 43210, USA}
\affiliation{Department of Astronomy, The Ohio State University, 4055 McPherson Laboratory, 140 W 18th Avenue, Columbus, OH 43210, USA}
\affiliation{The Ohio State University, Columbus, 43210 OH, USA}
\author[0000-0002-4475-3456]{J.~McCullough}
\email{jmccullough@princeton.edu}
\affiliation{SLAC National Accelerator Laboratory, 2575 Sand Hill Road, Menlo Park, CA 94025, USA}
\author[0000-0002-1125-7384]{A.~Meisner}
\email{aaron.meisner@noirlab.edu}
\affiliation{NSF NOIRLab, 950 N. Cherry Ave., Tucson, AZ 85719, USA}
\author{R.~Miquel}
\email{rmiquel@ifae.es}
\affiliation{Instituci\'{o} Catalana de Recerca i Estudis Avan\c{c}ats, Passeig de Llu\'{\i}s Companys, 23, 08010 Barcelona, Spain}
\affiliation{Institut de F\'{i}sica d’Altes Energies (IFAE), The Barcelona Institute of Science and Technology, Edifici Cn, Campus UAB, 08193, Bellaterra (Barcelona), Spain}
\author[0000-0002-2733-4559]{J.~Moustakas}
\email{jmoustakas@siena.edu}
\affiliation{Department of Physics and Astronomy, Siena College, 515 Loudon Road, Loudonville, NY 12211, USA}
\author[0000-0001-9070-3102]{S.~Nadathur}
\email{seshadri.nadathur@port.ac.uk}
\affiliation{Institute of Cosmology and Gravitation, University of Portsmouth, Dennis Sciama Building, Portsmouth, PO1 3FX, UK}
\author[0000-0001-8684-2222]{J.~ A.~Newman}
\email{janewman@pitt.edu}
\affiliation{Department of Physics \& Astronomy and Pittsburgh Particle Physics, Astrophysics, and Cosmology Center (PITT PACC), University of Pittsburgh, 3941 O'Hara Street, Pittsburgh, PA 15260, USA}
\author[0000-0003-3188-784X]{N.~Palanque-Delabrouille}
\email{npalanque-delabrouille@lbl.gov}
\affiliation{IRFU, CEA, Universit\'{e} Paris-Saclay, F-91191 Gif-sur-Yvette, France}
\affiliation{Lawrence Berkeley National Laboratory, 1 Cyclotron Road, Berkeley, CA 94720, USA}
\author[0000-0002-0644-5727]{W.~J.~Percival}
\email{will.percival@uwaterloo.ca}
\affiliation{Department of Physics and Astronomy, University of Waterloo, 200 University Ave W, Waterloo, ON N2L 3G1, Canada}
\affiliation{Perimeter Institute for Theoretical Physics, 31 Caroline St. North, Waterloo, ON N2L 2Y5, Canada}
\affiliation{Waterloo Centre for Astrophysics, University of Waterloo, 200 University Ave W, Waterloo, ON N2L 3G1, Canada}
\author{C.~Poppett}
\email{clpoppett@lbl.gov}
\affiliation{Lawrence Berkeley National Laboratory, 1 Cyclotron Road, Berkeley, CA 94720, USA}
\affiliation{Space Sciences Laboratory, University of California, Berkeley, 7 Gauss Way, Berkeley, CA  94720, USA}
\affiliation{University of California, Berkeley, 110 Sproul Hall \#5800 Berkeley, CA 94720, USA}
\author[0000-0001-7145-8674]{F.~Prada}
\email{fprada@iaa.es}
\affiliation{Instituto de Astrof\'{i}sica de Andaluc\'{i}a (CSIC), Glorieta de la Astronom\'{i}a, s/n, E-18008 Granada, Spain}
\author[0000-0001-6979-0125]{I.~P\'erez-R\`afols}
\email{ignasi.perez.rafols@upc.edu}
\affiliation{Departament de F\'isica, EEBE, Universitat Polit\`ecnica de Catalunya, c/Eduard Maristany 10, 08930 Barcelona, Spain}
\author[0000-0001-5999-7923]{A.~Raichoor}
\email{araichoor@lbl.gov}
\affiliation{Lawrence Berkeley National Laboratory, 1 Cyclotron Road, Berkeley, CA 94720, USA}
\author[0000-0002-3500-6635]{C.~Ravoux}
\email{corentin.ravoux@clermont.in2p3.fr}
\affiliation{Universit\'{e} Clermont-Auvergne, CNRS, LPCA, 63000 Clermont-Ferrand, France}
\author{G.~Rossi}
\email{graziano@sejong.ac.kr}
\affiliation{Department of Physics and Astronomy, Sejong University, 209 Neungdong-ro, Gwangjin-gu, Seoul 05006, Republic of Korea}
\author[0000-0001-9897-576X]{Y.~Salcedo Hernandez }
\email{yos47@pitt.edu}
\affiliation{Department of Physics \& Astronomy and Pittsburgh Particle Physics, Astrophysics, and Cosmology Center (PITT PACC), University of Pittsburgh, 3941 O'Hara Street, Pittsburgh, PA 15260, USA}
\author[0000-0002-9646-8198]{E.~Sanchez}
\email{eusebio.sanchez@ciemat.es}
\affiliation{CIEMAT, Avenida Complutense 40, E-28040 Madrid, Spain}
\author[0000-0002-0408-5633]{C.~Saulder}
\email{csaulder@mpe.mpg.de}
\affiliation{Max Planck Institute for Extraterrestrial Physics, Gie\ss enbachstra\ss e 1, 85748 Garching, Germany}
\author{D.~Schlegel}
\email{djschlegel@lbl.gov}
\affiliation{Lawrence Berkeley National Laboratory, 1 Cyclotron Road, Berkeley, CA 94720, USA}
\author{M.~Schubnell}
\email{schubnel@umich.edu}
\affiliation{Department of Physics, University of Michigan, 450 Church Street, Ann Arbor, MI 48109, USA}
\affiliation{University of Michigan, 500 S. State Street, Ann Arbor, MI 48109, USA}
\author[0000-0002-6588-3508]{H.~Seo}
\email{seoh@ohio.edu}
\affiliation{Department of Physics \& Astronomy, Ohio University, 139 University Terrace, Athens, OH 45701, USA}
\author{D.~Sprayberry}
\email{david.sprayberry@noirlab.edu}
\affiliation{NSF NOIRLab, 950 N. Cherry Ave., Tucson, AZ 85719, USA}
\author[0000-0003-1704-0781]{G.~Tarl\'{e}}
\email{gtarle@umich.edu}
\affiliation{University of Michigan, 500 S. State Street, Ann Arbor, MI 48109, USA}
\author{B.~A.~Weaver}
\email{benjamin.weaver@noirlab.edu}
\affiliation{NSF NOIRLab, 950 N. Cherry Ave., Tucson, AZ 85719, USA}
\author[0000-0003-2229-011X]{R.~H.~Wechsler}
\email{rwechsler@stanford.edu}
\affiliation{Kavli Institute for Particle Astrophysics and Cosmology, Stanford University, Menlo Park, CA 94305, USA}
\affiliation{Physics Department, Stanford University, Stanford, CA 93405, USA}
\affiliation{SLAC National Accelerator Laboratory, 2575 Sand Hill Road, Menlo Park, CA 94025, USA}
\author[0000-0001-9912-5070]{M.~White}
\email{mwhite@berkeley.edu}
\affiliation{Department of Physics, University of California, Berkeley, 366 LeConte Hall MC 7300, Berkeley, CA 94720-7300, USA}
\affiliation{University of California, Berkeley, 110 Sproul Hall \#5800 Berkeley, CA 94720, USA}
\author[0000-0001-5381-4372]{R.~Zhou}
\email{rongpuzhou@lbl.gov}
\affiliation{Lawrence Berkeley National Laboratory, 1 Cyclotron Road, Berkeley, CA 94720, USA}
\author[0000-0002-6684-3997]{H.~Zou}
\email{zouhu@nao.cas.cn}
\affiliation{National Astronomical Observatories, Chinese Academy of Sciences, A20 Datun Road, Chaoyang District, Beijing, 100101, P.~R.~China}

\begin{abstract}
Over several dedicated programs that include targets beyond the main cosmological samples, the Dark Energy Spectroscopic Instrument (DESI) collected spectra for 304,970 unique objects in two fields centered on the COSMOS and XMM-LSS fields. In this work, we develop spectroscopic redshift robustness criteria for those spectra, validate these criteria using visual inspection, and provide two custom Value-Added Catalogs with our redshift characterizations. With these criteria, we reliably classify 212,935 galaxies below $z<1.6$, 9,713 quasars and 35,222  stars. As a critical element in characterizing the selection function, we provide the description of 70 different algorithms that were used to select these targets from imaging data. To facilitate joint imaging/spectroscopic analyses, we provide row-matched photometry from the Dark Energy Camera, Hyper-Suprime Cam, and public COSMOS2020 photometric catalogs. Finally, we demonstrate example applications of these large catalogs to photometric redshift estimation, cluster finding, and completeness studies.
\end{abstract}
\section{Introduction}
\label{intro}
The Dark Energy Spectroscopic Instrument (DESI) is a Stage-IV dark energy spectroscopic survey currently being used to explore the cosmological model through measurements of large scale structure \citep{desi16,desi_yr1,desi_yr3}. The instrument is designed to measure the baryonic acoustic oscillation (BAO) feature at sub-percent precision over each of three redshift intervals from $0 < z < 2.5$.  Making the measurement at such high precision requires measuring redshifts of tens of millions of galaxies and quasars \citep{sv_overview}. With a high fiber count and wide field of view to meet this requirement, DESI is also an ideal instrument for building customized spectroscopic samples at very high surface density over a few square degrees. 

The COSMOS and XMM-LSS fields are two equatorial regions, with a legacy of multiband-photometry and spectroscopy. The COSMOS Survey \citep{COSMOS_Survey}, covering 2 deg$^2$, pioneered deep photometry over many bandpasses to facilitate a broad range of astronomical studies. The COSMOS2020 data release \citep{Weaver22} includes imaging programs that span radio to X-ray wavelengths from the following facilities:  Very Large Array \citep{VLA_cosmos},  Max-Planck Millimeter Bolometer Array \citep{COSBO_cosmos, Aztec_COSMOS}, Spitzer \citep{SPITZER_COSMOS},  Visible and Infrared Survey Telescope for Astronomy \citep{VISTA_Cosmos}, unWISE \citep{unWise_cosmos}, Subaru, Hubble Space Telescope, Mayall, Blanco, CFHT,  \citep{SUBURU_COSMOS,HST_COSMOS, CFHT}, GALEX \citep{UVGALEX_COSMOS}, XMM-Newton \citep{XMM-Newton_COSMOS} and Chandra \citep{CHANDRA_cosmos}. The public COSMOS catalogs also includes a large collection of photometric redshift, stellar mass, and star formation rate estimates. 
The latest COSMOS2025 data release \citep{COSMOS2025} includes imaging from JWST NIRcam over the central 0.54 deg$^{2}$ of the COSMOS field. The XMM-LSS field also benefits from an overlap in multi-band photometry and spectroscopic data. The multi-band photometry spans from radio to X-ray from the following facilities: GMRT \citep{GMRT_XMM}, LoFAR \citep{LOFAR_XMM}, HerMES  \citep{Hermes_XMM}, SWIRE \citep{SWIRE_XMM}, Dark Energy Survey\citep{DES_XMM}, Hyper Suprime-Cam \citep{Aihara_2022},  Galex \citep{UVGALEX_COSMOS}, XMM-SERVS and XXL \citep{XMM_servs,XXL_XMM}. 

The COSMOS and XMM-LSS deep, multiband-photometry has been complemented by various spectroscopic programs including SDSS, BOSS, and eBOSS \citep{sdss2000,Eisenstein_2011,SDSS_2013,eboss_2016, blanton_sdssiv}, VLT/VIMOS \citep{VLT_COSMOS}, OzDES \citep{OzDES_XMM}, DEVILS \citep{DEVILS_XMM}, MMT/Binospec \citep{fabricant2019}, 
C3R2 DR3 \citep{stanford2021}, VIPERS \citep{scodeggio2018}, DEIMOS \citep{DEIMOS_COSMOS}, LEGA-C DR2 \citep{straatman2018}, MOSDEF \citep{kriek2015}, VVDS \citep{lefevre2015}, and DEEP2 \citep{newman2013}. Recently, the COSMOS spectroscopic redshift compilation\footnote{\url{https://github.com/cosmosastro/speczcompilation}}  was publicly released to summarize all published redshifts within the 2 deg$^2$ COSMOS field \citep{COSMOS_spec}.

The wealth of existing photometric and spectroscopic data has supported a large set of astrophysical studies such as the calibration of photometric redshifts \citep[e.g.,][]{DES_photz} for cosmic shear and weak lensing measurements \citep[e.g.,][]{DES_WL}. In addition, these data sets have allowed estimates of the stellar mass function \citep[e.g.,][]{SMF_cosmos}, studies of the integrated stellar, gas, dust, and dark matter properties for high redshift galaxies \citep[e.g.,][]{COSMOS_highz_gal}, and constraints on the UV-luminosity function and cosmic star formation rate density \citep[e.g.,][]{COSMOS_UV}. Other examples include the confirmation of proto-clusters and cluster members \citep[e.g.,][]{Clusters_cosmos} and spectroscopic confirmation of Ly-$\alpha$ emitting galaxies at $z>7$ to probe the high redshift universe \citep[e.g.,][]{cosmos_z7}.

While published data have supported a range of science, the existing spectroscopic samples come from multiple instruments with varying wavelength coverage and robustness in redshift classification. In addition, the algorithms for selecting targets from imaging data are rarely defined, leaving uncertainty in the selection function.  DESI, with well-described target selection algorithms and consistent data processing methodology, provides an opportunity for a well-calibrated spectroscopic sample to complement the existing photometry.

\begin{figure*}
    \centering
    \includegraphics[width=\textwidth]{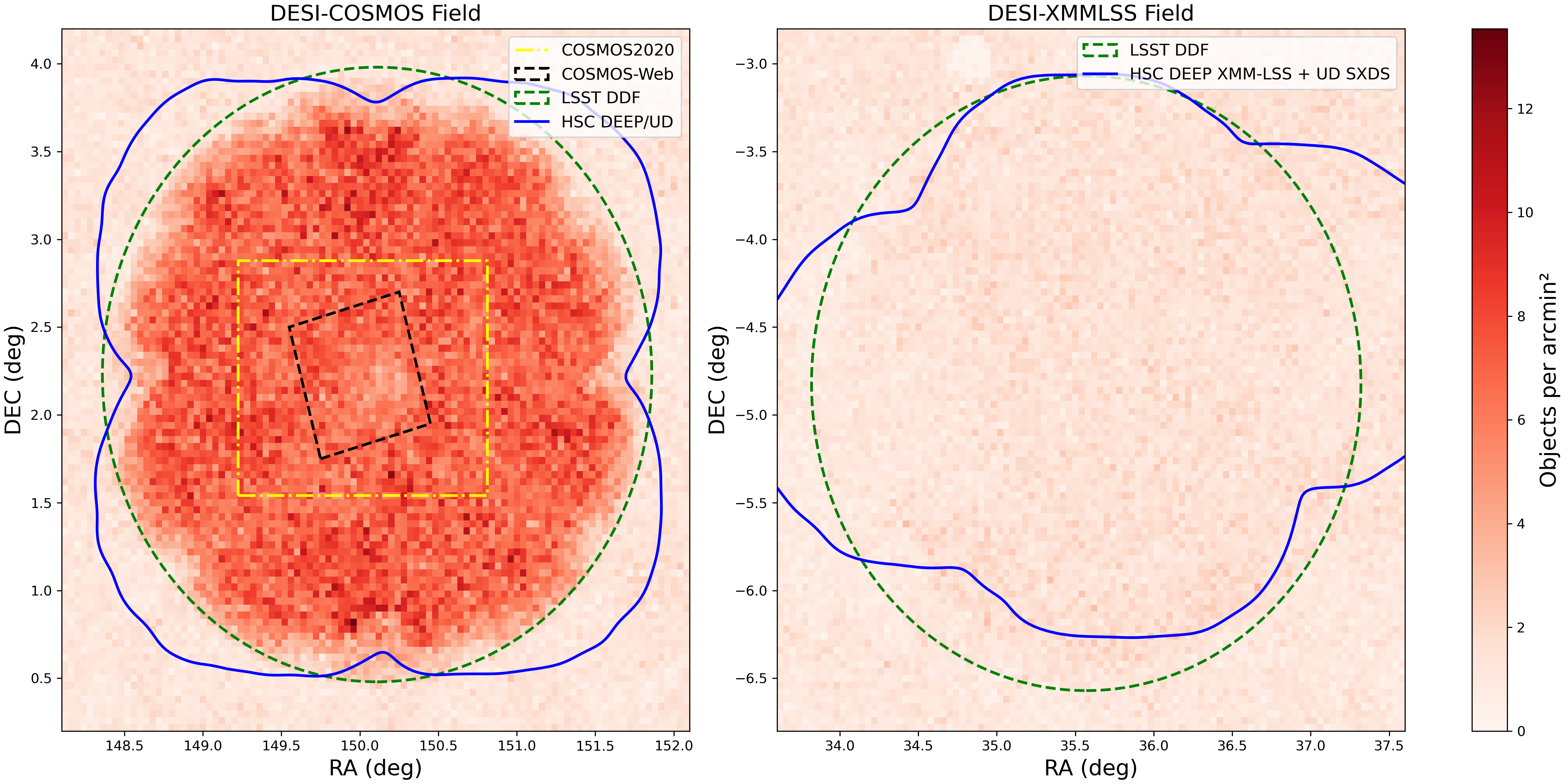}
    \caption{The spectroscopic coverage of the DESI Value-Added catalogs. The boundaries of the plot enclose the 16 deg$^2$ DESI-COSMOS and DESI-XMMLSS fields.  The colorbar depicts the number of objects per arcmin$^2$. \textbf{Left:} Over-plotted on the DESI-COSMOS field are overlapping deep drilling fields for COSMOS2020, COSMOS2025, LSST and HSC COSMOS-Deep/Ultradeep fields. \textbf{Right:} Over-plotted on the DESI-XMMLSS field is the LSST deep drilling field and combined HSC Deep XMM-LSS and Ultradeep SXDS fields.}
    \label{fig:footprint}
\end{figure*}

DESI dedicated 50 hours to spectroscopy in a 8 deg$^2$ region centered on the COSMOS field and 17 hours in a 8 deg$^2$ region centered on the XMM-LSS field. With its 5000-fiber multiplex, when also accounting for its main survey, DESI produced 233,247 and 74,990 unique spectra over 16 deg$^2$ centered on the COSMOS and XMM-LSS fields, respectively. We define the DESI-COSMOS ($\alpha = 150.1^{\circ}$, $\delta = 2.2^{\circ}$) and DESI-XMMLSS ($\alpha = 35.6^{\circ}$, $\delta = -4.8^{\circ}$)fields in Figure~\ref{fig:footprint} to cover 16 deg$^2$, ensuring complete overlap with existing and future deep imaging areas. Because the DESI-COSMOS field has more dedicated observation time, that field has a higher density sample over the central 8 deg$^2$ region. 

In these two fields, DESI's five primary samples were supplemented with a variety of customized target samples. The five primary targeting programs include Luminous Red Galaxies \citep[LRG;][]{zhou_2023}, the Bright Galaxy Survey \citep[BGS;][]{Hahn_2023}, the Milky Way Survey \citep[MWS;][]{Cooper_2023}, Emission Line Galaxies \citep[ELG;][]{Raichoor_2023} and Quasars \citep{Chaussidon_2023}. The supplementary programs included samples at a low surface density, allocated spare fibers during the primary program, and samples at a higher surface density that required dedicated observations with customized fiber assignments. 

In this work, we present a custom evaluation of all spectra within the COSMOS and XMM-LSS fields. We modify the redshift quality cuts beyond those that were informed by visual inspection for DESI galaxies \citep{lan23} and quasars \citep{alexander23}. The modified quality cuts increase the redshift purity for the main samples while minimally reducing the completeness in these fields. We use these new criteria to develop an algorithm for determining reliable redshifts for the wealth of supplemental programs within these two fields. In total, over 32 deg$^2$, we find robust spectroscopic classification of 212,935 galaxies below $z<1.6$, 9,713 quasars and 35,222  stars. In addition, several dedicated programs acquired spectra for potential Lyman-break and Lyman-alpha emission galaxy targets \citep{lbg_vanina, lae_white} with full redshift performance to be reported in the future. We exclude these targets from our study, as the automated redshift classification is still in development.

Complementing this study, we are publicly releasing a Value-Added Catalog that provides redshift and quality estimates for galaxies and quasars and radial velocities and derived parameters for stars. These catalogs also include photometric measurements from DECaLs DR9\footnote{\url{https://www.legacysurvey.org/dr9/description/}}, Hyper Suprime-Cam PDR3 Wide/Ultra Deep \citep{Aihara_2022}, DECam DR10\footnote{\url{https://www.legacysurvey.org/dr10/description/}}, COSMOS2020 \citep{Weaver22} and the Merian medium-band survey from the Victor M. Blanco telescope \citep{Luo_2024}. 

In addition to the legacy value, this catalog has potential to support future work such as photometric observations from the Vera C. Rubin Observatory.  With its 9.6 deg$^2$ field of view, the Vera C. Rubin Observatory Legacy Survey of Space and Time (LSST) will produce a large photometric sample spanning 20,000 deg$^2$ of the sky \citep{LSST}. This survey will include frequent, dedicated observations in the COSMOS and XMM-LSS fields \citep{LSST_DDF}, which are well-matched to the DESI 3-degree-diameter field of view.  Of the many science goals within LSST, weak lensing and cosmic shear serve as clear examples of science drivers that will benefit from having spectroscopic observations in these two dedicated fields. 

Here, we present the samples, classification scheme, and redshift robustness statistics for the Value-Added Catalog. In Section~\ref{target}, we describe the DESI target selection algorithms. In Section~\ref{spectroscopy}, we describe the observations of the COSMOS and XMM-LSS fields, the automated redshift classification pipeline, redshift consistency tests, spectral modeling, and visual inspection procedures. In Section~\ref{performancetesting} we present the quality cuts for each of the main target classes and how those cuts inform the quality assessment for each of the supplemental programs. We also present the results of consistency and robustness checks. In Section~\ref{VAC}, we describe the Value-Added Catalog of the DESI spectroscopic redshifts in the COSMOS and XMM-LSS fields. Finally, in Section~\ref{exampleuses} we provide examples on the utility of these spectroscopic redshifts for other cosmological studies and provide a summary in Section~\ref{summary}. The Value-Added Catalogs contain large heterogeneous samples of extragalactic and stellar sources. Due to the complexity of all the samples, we provide selection functions and redshift robustness criteria for each target class to enable various scientific studies.  All magnitudes in the paper are in AB magnitude system \citep{oke83} and are corrected for
Galactic extinction. The photometry included in the matched catalogs remains in the standard format, without corrections from Galactic extinction.

\section{DESI Target Classes}
\label{target}
 The majority of targets chosen for DESI observation were selected from the DESI Legacy Imaging Surveys \citep{dey19,zou17}.  Additional targets were chosen from Hyper Suprime Cam (HSC) PDR3 Wide/Ultra Deep, the Dark Energy Camera Deep observations, and Merian photometry. The selection procedure for targets is given in \cite{myers23}.  

DESI observations were planned in four different phases. The first phase of the DESI program was Target Validation (December 2020 to March 2021), hereafter referred to as \texttt{SV1}. \texttt{SV1} consisted of deep observations of extended target selections for each of LRG, ELG, quasar, BGS and MWS targets \citep{sv_overview}. \texttt{SV1} data were then used to finalize the \texttt{MAIN} Survey target selections. The next major phase of the Survey Validation program was the One-Percent Survey, hereafter called \texttt{SV3} (April – May 2021). Targets in this survey were selected by algorithms that closely represented those in the \texttt{MAIN} Survey that were assigned fibers for spectroscopic observation at a very high completeness. Planning for the \texttt{MAIN} Survey, included fiber assignments over 14,000 deg$^2$ and is ongoing with the primary goal of constraining the nature of dark energy \citep{rsd, y3_bao}. Finally, customized observations were planned as special programs in dedicated fields. 

The dark time targets (LRGs, ELGs, quasars) for the main survey were assigned fibers over a tiling pattern that yielded an average of roughly five observation opportunities per coordinate. The bright time targets (BGS and MWS) yielded an average of roughly three observation opportunities per coordinate. For \texttt{SV1}, \texttt{SV3} and in the special programs, customized tile patterns provided more observation opportunities per coordinate and thus a higher surface density of targets that were assigned fibers In total 146 and 44 special dedicated tiles were assigned to COSMOS and XMM-LSS, respectively.

\subsection{DESI Target Bitmasks}

DESI targets are identified for spectroscopy via the algorithm outlined in \cite{myers23}, where each object is assigned a bitwise number indicating the target class for which that object was targeted. \texttt{DESINAME} was introduced in 2024 to help identify objects that undergo multiple observations across DESI phases. \texttt{DESINAME} is particularly important as it serves as the main identifier for all targets across the catalog.

For each phase, the selection of targets was recorded via bitmasks. For each of \texttt{SV1}, \texttt{SV3}, and \texttt{MAIN}, there is a primary, secondary, MWS, and BGS bitmask, totaling 12 bitmasks. For targets in special programs, target information is recorded in text files external to the DESI target bitmasks as indicated by the \texttt{DARK\_TOO\_HIP} bit in the secondary mask.

Each target has an associated \texttt{DESI\_TARGET}, \texttt{DESI\_SCND\_TARGET}, \texttt{BGS\_TARGET}, and/or \texttt{MWS\_TARGET} bit that indicates the targeting classes and subclass. In this scheme, the \texttt{DESI\_TARGET} bitmask relays the top-level information including the class of primary targets (LRG, ELG, quasar, BGS and MWS) and where subclass information can be found in additional bitmasks. Subclass information for MWS targets can be found in \texttt{MWS\_TARGET} and BGS subclass information can be found \texttt{BGS\_TARGET}. Secondary subclass information can be found in \texttt{DESI\_SCND\_TARGET}. Prefixes for \texttt{SV1} and \texttt{SV3} are appended to the bitmask if an object was observed prior to the \texttt{MAIN} survey. 

This bitmask structure leads to a selection algorithm that is easily recorded for each of the targets. The overall selection function is therefore well understood in both the COSMOS and XMM-LSS fields. In what follows, we describe the target selection philosophy for the primary, secondary and special program targets. We provide a full description of each target selection in Appendix~\ref{appendix}. 

\subsection{Primary Target Selection}

The primary target classes of LRG, ELG, quasar, BGS and MWS targets comprise the majority of spectra produced by DESI. The philosophy for the primary DESI target selection is outlined below.  

\subsubsection{Luminous Red Galaxies (LRG)}
The LRG sample is selected from imaging data based on a strong continuum break at 4000 \AA. The sample is selected using $g,r,z$ optical bands with infrared photometry from the WISE $W1$ band with an emphasis on the $r - W1$ color. This  $r - W1$ color serves as good proxy for redshift from $0.4 < z < 1.1$ for a passively evolving population with a strong 4000 \AA~ break \citep{zhou_2023}. This selection is intended to cover an approximately constant number density over as wide of a redshift range as possible.

The DESI LRG sample has a significantly higher target density and redshift range when compared to previous surveys such as eBOSS and SDSS \citep{zhou_2023}. In addition to the main selection of LRG targets, special programs were designed to increase the number density by almost an order of magnitude in the COSMOS field.

For all these samples, these objects are dominated by older stellar populations, yielding redder spectra with relatively strong metal absorption lines allowing us to measure robust redshifts.

\subsubsection{Emission Line Galaxies (ELG)}

The ELG sample is selected based on relatively blue colors in the $g-r$ vs. $r-z$ plane. In the approximate redshift range from $0.6 < z < 1.6$, these colors signify the presence of a young stellar population. By focusing on this higher redshift range, this selection identifies a population of galaxies in a redshift regime where LRGs are too faint to be efficiently classified through DESI spectroscopy. 

The DESI ELG sample has a target density about 10 times more dense than the ELG sample of eBOSS and stretches one magnitude fainter \citep{Raichoor_2023}. Several special programs were designed to explore emission line galaxies in the COSMOS field, further increasing the number density.
 
These galaxies become very faint in continuum flux at higher redshift but commonly feature strong [OII] emission lines. The [OII] doublet $\lambda\lambda 3726,3279$ \r{A} resulting from star formation \citep{madau2014} can be used to uniquely determine the spectroscopic redshift of the target.

\subsubsection{Quasars}
Quasars are extremely luminous active galactic nuclei powered by accretion onto supermassive black holes. Their intrinsic brightness makes them efficient tracers of large scale structure at redshifts $ z > 0.9$. Quasars are selected through a combination of optical (\textit{g, r, z}) and IR colors using the $W1$ and $W2$ measurements from WISE. Due to the dusty torus, quasars are brighter in the mid-infrared at all redshifts compared to stars of similar optical magnitude and color. The DESI selection uses this unique feature as a powerful tool to reject stars from the quasar selection \citep{Chaussidon_2023}. 

The bulk of the quasar sample comes from the main target selection, with small secondary programs supplementing the complete quasar sample in the COSMOS and XMM-LSS fields.

Quasar spectra are characterized by multiple broad emission lines that extend from the  UV to visible wavelengths in the rest frame. These broad emission lines provide a useful tool for robust spectroscopic classification.

\subsubsection{Bright Galaxy Survey (BGS)}
The BGS sample will be a collection of over 10 million galaxies spanning $0 < z < 0.4$. The goal of the BGS sample is to probe the epoch where dark energy is dominant \citep{Hahn_2023}.

The BGS sample is divided into two primary samples, \texttt{BGS BRIGHT} and \texttt{BGS FAINT}.  \texttt{BGS BRIGHT} constitutes an $r < 19.5$ magnitude-limited sample that is ten times larger than that of the Sloan Digital Sky Survey I and II \citep{sdss2000, strauss}. The \texttt{BGS FAINT} sample is color-selected to prioritize strong H$\alpha$ emission to fainter magnitudes, $19.5 < r < 20.175$ \citep{Hahn_2023}. Subsamples of bright targets supplement the remainder of the Bright Galaxy Survey. 

Similar to the LRG sample, the relatively low redshift and intrinsic brightness of these targets lead to a strong continuum that enables the detection of a plethora of absorption lines, prime for redshift classification.

\subsubsection{Milky Way Survey (MWS)}

Making use of DESI bright time, the primary Milky Way Survey program will include approximately ten million stars to a limiting magnitude of $r < 19$ with the goal of providing radial velocities, metallicities, and other stellar parameters \citep{Cooper_2023}. The main target selection algorithm covers three target categories, \texttt{MAIN-BLUE}, \texttt{MAIN-RED} and \texttt{MAIN-BROAD}, which in combination cover the full color–magnitude space within $16 < r < 19$ \citep{Cooper_2023}. In addition, there are 23 stellar target classes spanning the observational programs in DESI and stars often appear as interlopers in other selections as well. 

\subsection{Secondary Targets}
Secondary programs, designed by members of the DESI collaboration, extend beyond DESI's primary cosmology science drivers. Across both the COSMOS and XMM-LSS fields, there are targets that span 32 secondary programs. Examples of these programs include point sources with extreme colors, low redshift dwarf galaxies, high proper-motion stars, supernova host galaxies, and a variety of other extragalactic and stellar sources. Examples of results from these programs include calibration of the Fundamental Plane relation \citep{fp_said}, relationships between radio emission and dust extinction in quasars \citep{secondary_red_qso}, and much more. We provide a full description of the secondary target selections in Appendix~\ref{appendix}.

\subsection{Special Programs}
DESI conducted dedicated observations of high-value, high-density targets to explore options for future surveys and enrich the science from the COSMOS and XMM-LSS fields.  For each of these programs, a set of dedicated tiles, each with a customized target list, was designed for observation. The observations of these tiles were prioritized to be completed in a limited time frame.  

The extended selections for LRGs and ELGs mentioned above were conducted in these special programs. Other samples from these programs include samples for photometric redshift calibration of LSST weak lensing sources, LAEs and LBGs. Publications from these programs include LAE clustering studies \citep{lae_white}, LBG target selection techniques \citep{lbg_vanina} and extracting the Lyman-$\alpha$ forest from LBG spectra \citep{herrera2025}.  There were six special programs in the COSMOS field and five in the XMM-LSS field completed in the first three years of the DESI survey. We provide a full description of the special target selections in Appendix~\ref{appendix}.

\section{Spectroscopy}
\label{spectroscopy}
DESI is a multi-object spectroscopic instrument installed on the Mayall 4-m telescope at Kitt Peak National Observatory in Arizona. The instrument makes use of an 8 deg$^2$ field of view and robotic fiber positioners to simultaneously capture spectra across a focal plane of 5000 fibers \citep{abareshi22, poppett2024, miller2024}. Ten petals of 500 fibers each connect to one of ten high-efficiency spectrographs with a  spectroscopic wavelength coverage from 3600-9800 \AA. The high throughput, efficient fiber positioning, and high fiber count allow DESI to capture thousands of spectra over a short period of time. 

A standard observation in DESI consists of a 1000s effective exposure time \citep{schlafly_desi}.
In addition to standard DESI observations, several hours of effective spectroscopic exposure time were dedicated to special tiles in the COSMOS field  and the XMM-LSS field \citep{dr1}. To facilitate studies in these two regions and in neighboring areas, we produce two catalogs covering 16 deg$^2$ each. The footprint of this study includes coverage from the standard tiles in addition to special tiles centered on the COSMOS and XMM-LSS fields. The fields covered in these catalogs hereafter are referred to as the DESI-COSMOS and DESI-XMMLSS fields.


These observations are processed through a data reduction pipeline which converts the raw CCD images to one-dimensional spectra \citep{guy_2023} followed by redshift and spectroscopic classification. Using these redshift estimates, we then apply a spectral fitting algorithm that simultaneously models the spectrum's continuum and emission lines. To further quantify automated redshift efficiency, we compare the pipeline redshift estimates across multiple observations of the object in different phases of DESI, when possible. We visually inspect all discrepancies in redshift to estimate the catastrophic failure rate and inform quality cuts customized to each target class.

\subsection{The DESI Data Reduction Pipeline and Redshift Estimation}
\label{redrock+afterburner}
We summarize the data reduction pipeline which is described in full detail in \cite{guy_2023}. The raw images are corrected for dark current, flat fields and bad pixels. Each science exposure undergoes wavelength calibration, and the spectroscopic Point Spread Function (PSF) is measured and used in a spectral extraction. Sky spectrum models are subtracted from spectra using fibers placed on empty areas of the sky (called ``sky fibers'') and flux calibration is performed using appropriate stars. For each survey (\texttt{SV1}, \texttt{SV3}, \texttt{MAIN}, \texttt{SPECIAL}), a unique coadded spectrum is created for each object. That is, multiple coadds exist for the same object if it was observed in different surveys, allowing us to test the consistency of the redshift pipeline. With these co-added spectra, spectral classification and redshift estimates are performed.

DESI utilizes a Principal Component Analysis (PCA) fitting algorithm, called \textit{Redrock}\footnote{https://github.com/desihub/\textit{Redrock}}, to determine spectral class and redshift estimates for all spectra \citep{redrock2024, anand2024}. The algorithm uses a set of templates that represent broad spectral classes of galaxies, stars and quasars. The combination of these templates provides composite solutions constructed to best model each spectrum. \textit{Redrock} fits these composite solutions across a complete range of redshift-parameter space, choosing the best fitting solution based on the lowest $\chi^2$ value following the philosophy in \cite{bolton_2012}.

\textit{Redrock} produces many fit parameters useful for assessing the quality of the template fit and redshift estimation. The $\chi^{2}$ can be used to determine the quality of the best template fit. Another fit statistic, $\Delta \chi^{2}$ is the difference between the $\chi^{2}$ of two neighboring fits (for \textit{Redrock}, the best and next best fit). A high $\Delta \chi^{2}$ indicates that the best fit is highly preferred compared to the next best fit and indicates a higher confidence in the redshift estimate. A small $\Delta \chi^{2}$ indicates that there is little distinction between the first and second-best fit, leading to a lower confidence in the redshift estimate. All fits are given a \texttt{ZWARN} flag, indicating any issues in the observation or data reduction pipeline for that spectrum. \textit{Redrock} also classifies the spectrum, attributing the spectral class, or \texttt{SPECTYPE}, to the template of the best fit.

For quasars specifically, we utilize two afterburner pipelines, Mg\textrm{II} and \textit{QuasarNET} \citep{qsonet}, to ensure proper spectral classification of these targets. The Mg\textrm{II} afterburner tests for a broad Mg\textrm{II} $\lambda2800$ \AA ~emission line, characteristic of qusar spectra. The method consists of fitting a Gaussian in a 250 \AA ~window centered at the position of Mg\textrm{II} line given by \textit{Redrock}. We consider the Mg\textrm{II} as a broad line if the improvement of $\chi^{2}$ is better than 16, the width of the Gaussian greater than 10 \AA ~and the significance of the amplitude of the Gaussian greater than 3. The algorithm possibly changes the source classification but never modifies the redshift given by \textit{Redrock} \citep{Chaussidon_2023}. 

The \textit{QuasarNET} algorithm is a deep neural-network that looks for the following six emission lines: Ly$\alpha$, C\textrm{IV}, C\textrm{III}], Mg\textrm{II}, H$\alpha$ and  H$\beta$. An object is classified as a quasar (QSO) if the confidence probabilities of at least one of the six aforementioned lines is greater than 0.5.

\subsection{\textit{FastSpecFit}} 
Redshift estimation for objects with a low signal-to-noise ratio per pixel rely on prominent emission lines to accurately estimate a spectrum's redshift. The strength of these emission lines can be an indicator of the success of the redshift estimation and used as a metric for redshift quality. To get a more accurate measurement of the strength of emission lines in these spectra, we use \textit{FastSpecFit}\footnote{https://github.com/desihub/fastspecfit}\citep{fastspecfit}, a physically-motivated algorithm that measures emission line strength and physical properties of a galaxy from its spectrum and photometry. \textit{FastSpecFit} produces measurements for fluxes and uncertainties of emission lines such as, the [OII] $\lambda\lambda3726, 3729$ \AA ~and [OIII] $\lambda\lambda4959,5007$ \AA ~lines. These measurements from \textit{FastSpecFit} will be used to examine redshift quality for target classes with strong emission line properties.

\subsection{Assessing Redshift Estimations}
To assess the accuracy of the data reduction and \textit{Redrock} pipeline, we compare objects across multiple observations. If the estimated redshifts for the same object are consistent, we assume that the redshift estimation is correct. In cases where we have discrepant redshift estimations, we assume one of the redshift estimations is incorrect. 

We match objects that have the same \texttt{DESINAME} across the all programs, \texttt{MAIN}, \texttt{SV3}, and \texttt{SV1}. Before matching, we require \texttt{ZWARN} = 0 or 4  with a $\Delta\chi^{2} \geq 40$ threshold, informed by prior survey validation of the BGS sample \citep{lan23}, to remove lower quality spectra. \texttt{ZWARN} = 0 indicates there was no issue with the redshift fit, and \texttt{ZWARN} = 4 indicates that the $\chi^{2}$ of best fit is too close to that of second best ($\Delta\chi^{2} < 9$. A \texttt{ZWARN} = 4 is non-problematic as we apply our own $\Delta\chi^{2}$ cuts, customized to each sample. We then compare the redshifts via Eq.~\ref{3} and Eq.~\ref{4}. 

Following the science requirements for DESI, a threshold for catastrophic failures is set at 1000 km/s \citep{DESI_science_req}. If the difference between reported redshifts of the same target in different programs is within 1000 km/s for galaxies and stars, the target is considered to have a robust and consistent redshift:
    \begin{equation}
        \Delta v = \frac{|z_{i} - z_{j}|*c}{(1+z_{i})} \leq 1000 ~\rm{km\,s^{-1}}
        \label{3}
    \end{equation}
If the difference between reported redshifts is within 3000 km/s for quasars, the target is considered to have a robust and consistent redshift:
    \begin{equation}
        \Delta v = \frac{|z_{i} - z_{j}|*c}{(1+z_{i})} \leq 3000 ~\rm{km\,s^{-1}}
        \label{4}
    \end{equation}
We find a high consistency in redshift estimation of better than $\sim$99.3\% across programs for both the DESI-COSMOS and DESI-XMMLSS field as seen in Table~\ref{tab:cross_match}. Examining the discrepant pairs allows us to identify failure modes and better understand catastrophic failures.

\begin{table}[h]
\centering
\begin{tabular}{l r r}
\hline
Survey & \# of Matches & Catastrophic Failure Rate \\ 
\hline\hline
\multicolumn{3}{l}{\textbf{COSMOS}} \\
\hline
\texttt{MAIN-SV3} & 70250 & 0.35\% \\
\texttt{MAIN-SV1} & 14094 & 0.72\% \\
\texttt{SV3-SV1} & 20719 & 0.41\% \\
\hline
\multicolumn{3}{l}{\textbf{XMM-LSS}} \\
\hline
\texttt{MAIN-SV1} & 10871 & 0.37\% \\
\hline
\end{tabular}
\caption{Cross matched objects between DESI programs. The catastrophic failure rate is the implied rate of failure for all objects with \texttt{ZWARN} = 0 or 4 and $\Delta\chi^{2} \geq 40$.}
\label{tab:cross_match}
\end{table}

\subsection{Visual Inspection}
In cases where measurements disagree, we visually inspect the individual spectra to investigate which are catastrophic failures. The goals of visual inspection of are to: (1) assess the DESI redshift estimation pipeline; and (2) investigate catastrophic failure modes of spectra to inform quality cuts. The visual inspection (VI) procedure is conducted through  the \textit{Prospect}\footnote{\url{https://github.com/desihub/prospect}} spectral visualization tool.

The standardized process for inspecting DESI spectra is outlined in \cite{lan23} and is summarized here:
\begin{enumerate}
    \item Each spectrum has 2 visual inspectors.
    \item Each inspector reports four key features:
    \begin{itemize}
        \item Redshift
        \item A quality flag where spectra are assigned 1 of 5 discrete values ranging from 0-4. 4 denotes the highest quality a spectrum can receive and 0 denotes the lowest score.
        \item Type of source: galaxy, star or quasar
        \item Any anomalies, systematic features, or artifacts within the spectrum
        \item Additional comments regarding the spectra (e.g., two objects in the spectrum)
    \end{itemize}
    
    \item The outputs of the two inspectors are analyzed and compared. For each spectrum, the VI report must satisfy the following conditions:
        \begin{itemize}
            \item the \texttt{SPECTYPE} from both inspectors is the same;
            \item the difference between VI quality is within 1, i.e., $|$VI quality$_i$ - VI quality$_j|$ $\leq$ 1;
            \item the difference between VI redshift is within 1000 km s$^{-1}$.
        \end{itemize}
    \item If any of these conditions are not met, a third inspector inspects the spectrum and determines the final VI results. This third inspector might be one of the original inspectors.
\end{enumerate}

The criteria for the quality of each spectra, denoted by a discrete value 0-4 is as follows \citep{lan23}:

\begin{itemize}
    \item \textbf{Quality 4}: confident classification based on clear two or more spectral features (e.g., spectra with multiple absorption lines or emission lines);
    \item \textbf{Quality 3}: probable classification with at least one secure spectral feature and continuum or many weak spectral features;
    \item \textbf{Quality 2}: possible classification with one strong spectral feature;
    \item \textbf{Quality 1}: unlikely classification with some signal but no distinct feature;
    \item \textbf{Quality 0}: no meaningful signal;
\end{itemize}

With the above procedure and criteria, a final VI redshift with an average VI quality $\ge$ 2.5 in considered a secure VI redshift. 

In addition to the visual inspection on objects with discrepant redshifts, we spot check objects that have multiple redshift estimations across DESI without any quality cuts. Due to the size of this sample, systematically visually inspecting over 3,000 spectra is not tractable and spot checks were simply used to inform quality cuts.


\section{Redshift Quality Assessment}
\label{performancetesting}
Using the trends identified through visual inspection of objects with failed redshift estimations, we develop quality cuts to ensure robust redshift estimates for each target class.
For all objects, we only include spectra with $\texttt{ZWARN} = 0$ or $\texttt{ZWARN} = 4$.

We find for the BGS and LRG samples, failures in redshift estimation stem from spectra with low SNR per pixel. ELG spectra without strong [OII] emission lead to poor redshift estimates as well. Line confusion for broad-line emission in quasars leads to incorrect classification and redshift failures.  The primary target classes then serve as models for developing customized quality cuts for all of the secondary and special program targets. We use the LRG sample as the archetype for strong-continuum, high SNR per pixel spectra and the ELG sample for strong emission line spectra. All targets spectroscopically confirmed as quasars or targeted as a quasar are run through the quasar afterburners. For all spectra classified as stars, we simply report derived stellar parameter measurements from the \textit{RVSpecfit} pipeline \citep{koposov2025,rvs_paper}.

\subsection{Bright Galaxy Survey}
The BGS selection was developed to prioritize objects with a strong continuum and therefore a high SNR across the spectrum. We find objects with a weak continuum have poor redshift estimates while objects with a high SNR continuum yield accurate redshift estimates.  

Before any quality cuts are assigned, we assign all BGS spectra with a spectroscopic classification of `STAR' or `QSO' to be evaluated with the stellar and quasar quality assessments, respectively. We also impose a hard redshift cut of $z < 1.5$ as there are known failure modes for classifications past this redshift.

\begin{figure*}
    \centering
    \includegraphics[width=0.45\textwidth]{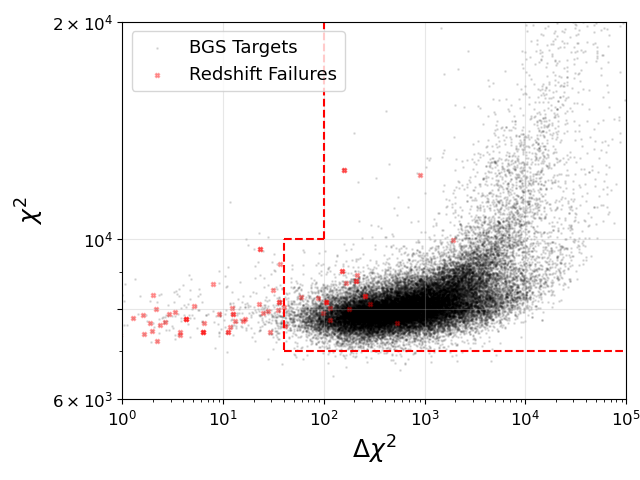}
    \hspace{0.02\textwidth}
    \raisebox{4mm}{\includegraphics[width=0.45\textwidth]{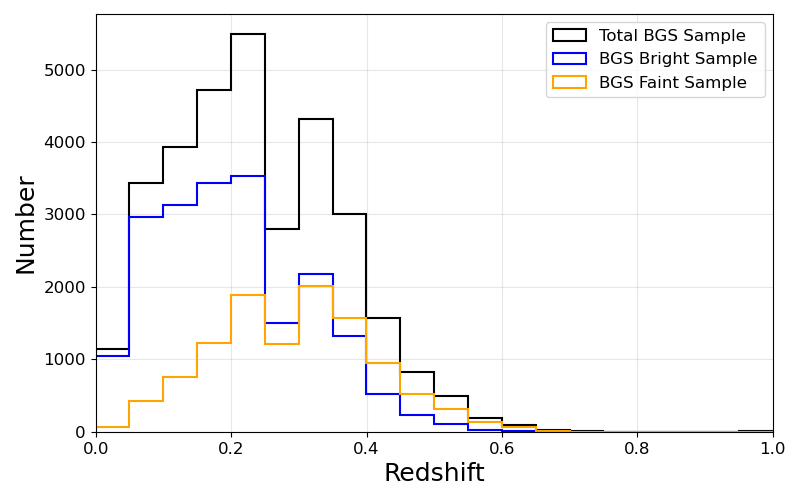}}
    \caption{Statistics of the BGS sample in the COSMOS field. \textbf{Left}: Distribution of the BGS sample in $\chi^{2}$ as a function of $\Delta\chi^{2}$. A large fraction of objects cluster in the locus around high $\Delta\chi^{2}$ and $\chi^{2} \sim 8000$. The red data points indicate visually inspected BGS spectra with incorrect redshifts. The red dashed line displays the $\chi^{2}$ and $\Delta\chi^{2}$ quality cuts. Spectra in the upper right are assumed to have robust redshifts. \textbf{Right}: The number of BGS spectra with robust redshifts in bins of $\Delta z = 0.05$. The total BGS sample (black) is split into a BGS Bright sample ($r < 19.5$, blue) and a BGS Faint sample (orange), which extends the selection to fainter galaxies.}
    \label{fig:bgs_deltacut}
\end{figure*}

The wide wavelength coverage of DESI spectra has the potential to provide information on many different spectral features for high-SNR spectra. We characterize a sample to have high SNR per pixel if the sample has an median SNR per pixel greater than two (SNR$_{\rm{b/r/z}} >2$). This leads to greater differentiability between fits, making $\Delta\chi^{2}$ a strong indicator of a quality redshift estimate.
The characteristics of BGS targets in $\chi^{2}$ vs $\Delta\chi^{2}$  parameter space are seen in Figure ~\ref{fig:bgs_deltacut}. The BGS sample populates this plane in a locus around high $\Delta\chi^{2}$ with $\chi^{2}$ values around 8000. Through visual inspection of a subsample of spectra found in this locus, a very large fraction of these spectra yielded accurate redshift estimations. An additional trend can be seen along this locus at high $\Delta\chi^{2}$, where objects tend to have an increasing $\chi^{2}$. A visual inspection of a random subsample of these spectra verifies that the redshift estimates are still accurate. In these cases, the large $\chi^{2}$ is a result of the high SNR of the spectra where the models provide the correct redshift but lack the flexibility to fit all features. 

Turning our attention to catastrophic redshift failures identified from visual inspection, we find poor redshift estimates at low $\Delta\chi^{2}$.  This motivates a requirement that spectra have $\Delta\chi^{2} > 40$ in order to be considered a robust redshift. 

Spectra that have $\chi^{2} < 7000$ tend to have a large fraction of their pixels flagged in the data reduction. We find that all of the pixels from one or more cameras were flagged, indicating at least one camera malfunctioned or was corrupted during observation. We therefore require spectra to have a $\chi^{2} > 7000$ in order to be considered a robust redshift.

Spectra with high $\chi^{2}$ yield a significant fraction of catastrophic failures in the regime where $\Delta\chi^{2} < 100$. A combination of low $\Delta\chi^{2}$ and high $\chi^{2}$ indicates a poor overall fit of the model to the data with little differentiability between models.  For this reason, we require $7000 < \chi^{2} < 10000$ (effectively $\chi^2_{dof} \sim 1$) for objects with $\Delta\chi^{2} < 100$. This requires a good model fit for objects with lower differentiability to be determined as a robust redshift.

\begin{figure}
    \includegraphics[width=\linewidth]{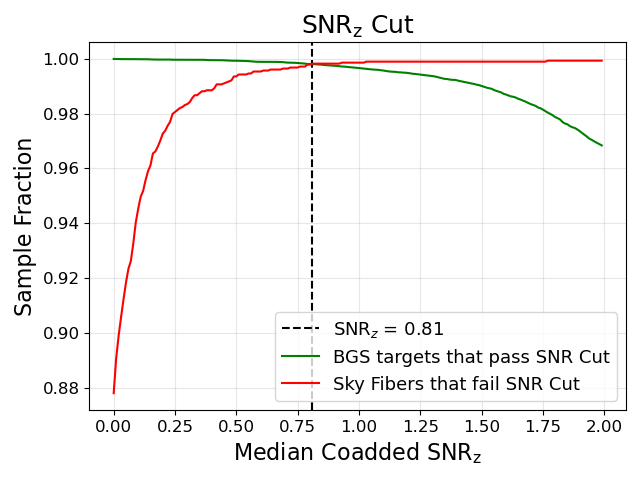}
    \caption{The cumulative fraction of spectra exceeding a median SNR$_{z}$ per pixel. The green line denotes the fraction of BGS spectra that are included at the given SNR$_{\rm{z}}$ cut. The red line denotes the fraction of sky spectra that are removed at a given SNR$_{z}$ cut. The vertical line at SNR$_{\rm{z}}=0.81$ represents the optimal SNR$_{z}$ cut for which we exclude the most sky spectra while minimally impacting the quality BGS sample. }
    \label{bgs_sky}
\end{figure}

Spectra from sky fibers (Section~\ref{spectroscopy}) are all run through the redshift fitting pipeline, and because there is no expected source, they act as a robustness test in the regime of no signal. We use sky fibers that appear to produce robust redshift estimates as indicators of catastrophic failures to probe the low SNR regime of BGS targets. We find a non-negligible fraction ($2.7\%$) of sky fibers pass our $\chi^{2}$ vs $\Delta\chi^{2}$ quality cuts.  We mitigate the contamination in the low SNR range of the BGS sample by imposing a SNR cut in the z-band (SNR$_{\rm{z}}$) region of the DESI spectrum.
In Figure~\ref{bgs_sky}, we demonstrate that a simple SNR cut at SNR$_{\rm{z}} = 0.81$ significantly reduces the number of sky fibers that pass a $\chi^{2}$ vs $\Delta\chi^{2}$ cut while having minimal impact on the overall completeness BGS sample. The SNR cut reduces the number of sky spectra that pass from 2767 to 5 while reducing our sample of BGS redshifts by only 0.19\%. Spectra from sky fibers are not included in the Value-Added Catalog and are simply used as test of our robustness criteria. 

We consider all BGS spectra to yield a robust redshift if they meet these conditions:
\begin{enumerate}
    \item \texttt{SPECTYPE} = `GALAXY'
    \item $z < 1.5$
    \item $\mathrm{SNR}_z > 0.81$
    \item $\chi^2 \geq 7000$
    \item Either 5a \textbf{or} 5b must be satisfied:
    \begin{enumerate}
        \item[(5a)] $\Delta\chi^2 \geq 100$
        \item[(5b)] $40 \leq \Delta\chi^2 < 100$ \textbf{and} $\chi^2 < 10000$
    \end{enumerate}
\end{enumerate}
 
\begin{figure*}
    \centering
    \includegraphics[width=0.95\textwidth]{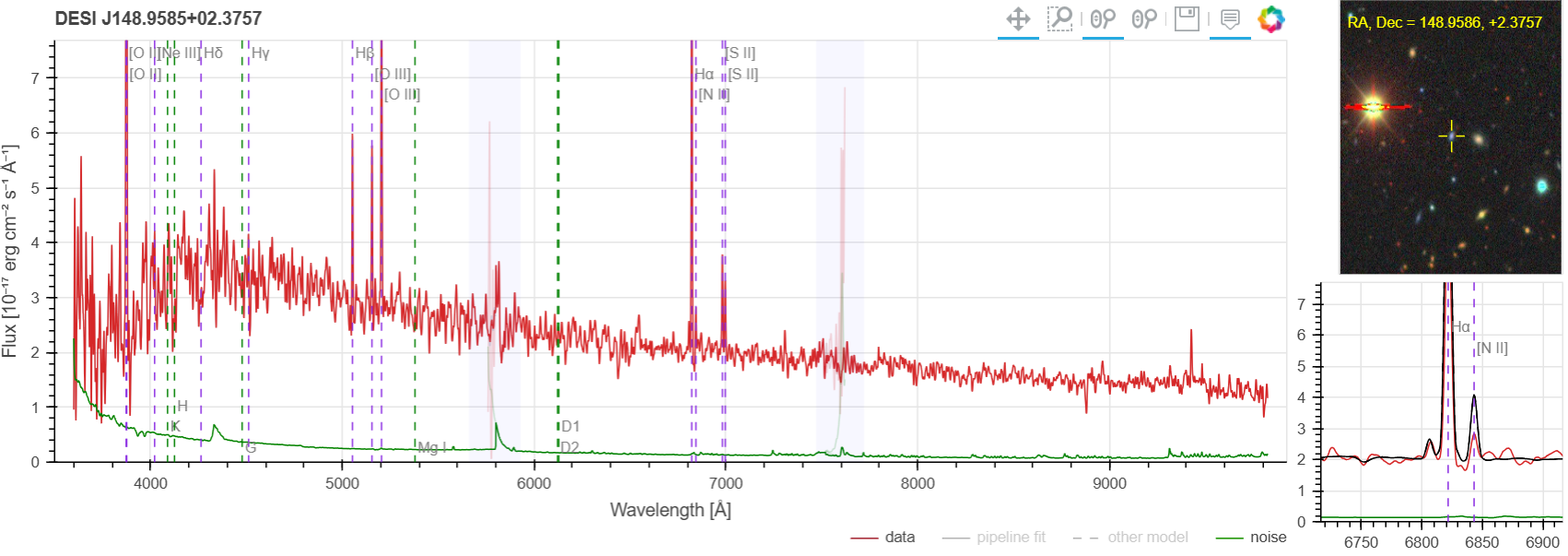}
    \vspace{0.5cm}
    \includegraphics[width=0.95\textwidth]{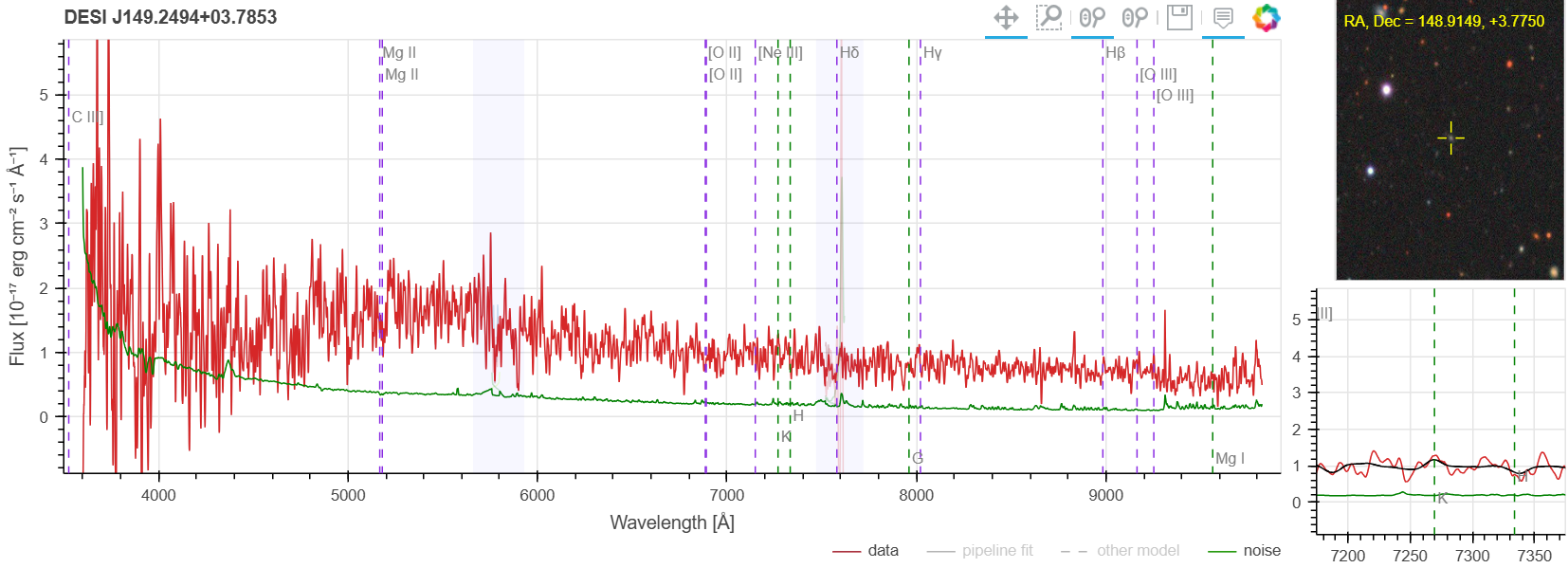}
    \caption{Examples of two BGS spectra that pass (top) and fail (bottom) our quality cuts plotted using the \textit{Prospect} visual inspection tool. 
    The red line is the spectrum with $\sim$8000 flux density measurements across the entire wavelength range. The green line shows the uncertainty in the flux density measurement at each wavelength. Common emission lines (dashed purple) and absorption lines (dashed green) are shown. The top postage stamp is the color composite image of the object from Legacy Survey DR9 imaging. The lower panel is a magnified region around significant absorption/emission lines. For quality spectra (top) the absorption lines are clearly distinguishable. For spectra with poor redshift estimations the emission lines or absorption lines appear nonphysical indicating a poor model fit. The \texttt{DESINAME} of each object is found in the top left corner of each figure and both spectra were obtained in SV1 as part of the `bright’ program. }
    \label{fig:bgs_spectra}
\end{figure*}

 In Figure~\ref{fig:bgs_spectra}, we show an example of a BGS spectrum that passes the cut and one that does not. The BGS spectrum in the top panel meets the quality redshift criteria and indeed has very high SNR, a strong continuum, along with good $\chi^{2}$ and $\Delta\chi^{2}$ statistics. For the spectrum shown in the bottom panel that fails the cut, we can see a clear continuum with a model that does not match the absorption features.

A majority of BGS spectra (98.3\%) pass the cuts defined in this section. Based on these classifications, there are 21,684 unique BGS targets spectroscopically confirmed as galaxies in the DESI-COSMOS field and 18,021 unique BGS targets spectroscopically confirmed as galaxies in the DESI-XMMLSS field. The redshift distribution of BGS galaxies with reliable redshift estimations in the DESI-COSMOS field is shown in the right panel of Figure~\ref{fig:bgs_deltacut}. In comparing our robustness criteria to the `fiducial' criteria in \cite{Hahn_2023}, we estimate the catastrophic failure rate by comparing the redshift estimates for objects with multiple observations. Using the fiducial criteria of \texttt{ZWARN} = 0 and $\Delta\chi^2 > 40$, the main BGS sample retains 98.9\% of its objects and has an implied catastrophic failure rate of 1.4\% (144/10064). Our robustness criteria 
retains 98.3\% of objects, with an implied catastrophic failure rate of 0.72\% (73/9974).

\subsection{Luminous Red Galaxies}
Spectra from the LRG sample typically have a strong continuum at redder wavelengths characterized by a distinct 4000 \AA~break. The high SNR (median SNR per pixel $\sim$2), especially at redder wavelengths, enables clear detection of prominent absorption lines like Ca\textrm{II} H and K. Due to their similar spectroscopic properties, the redshift quality cuts for the LRG spectra are similar to those of the BGS. We impose a SNR cut of SNR$_{z} = 0.53$ that only removes 0.2\% of our LRG sample while reducing the catastrophic failure rate from sky spectra to 0.009\%.

\begin{figure*}
    \centering
    \includegraphics[width=0.45\textwidth]{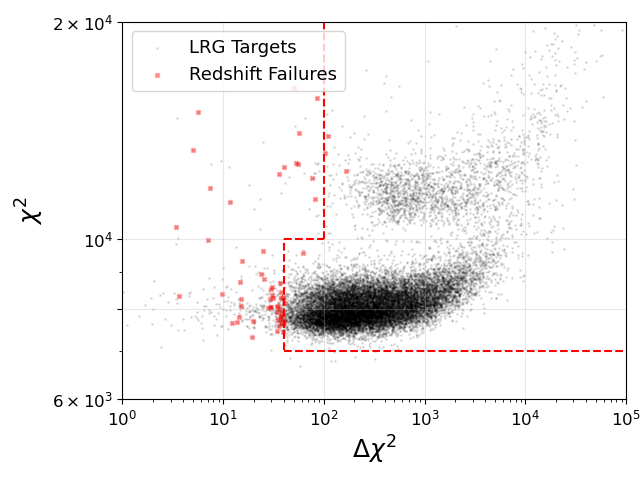}
    \hfill
    \includegraphics[width=0.45\textwidth]{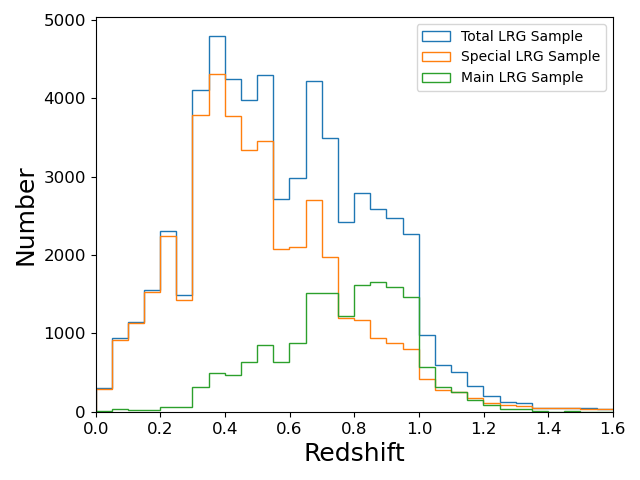}
    \caption{Statistics of the LRG sample in the COSMOS field. \textbf{Left}: Distribution of the LRG sample in $\chi^{2}$ vs $\Delta\chi^{2}$. The red dashed line displays the $\chi^{2}$ and $\Delta\chi^{2}$ quality cuts. Spectra in the upper right are quality redshifts.  \textbf{Right}: The redshift distribution of LRG spectra with robust redshifts. Objects are placed into bins of $\Delta$z = 0.05. The total LRG sample (blue) is split into two sub samples. The primary LRG sample (green) is taken from the main DESI target selection. The special LRG sample (orange) is the extended selection as part of a special program to probe low redshift LRGs.}
    \label{fig:lrg_deltacut}
\end{figure*}

In Figure~\ref{fig:lrg_deltacut}, the LRG spectra largely populate the same accepted region as the BGS program. The upper sequence in the left panel of Figure~\ref{fig:lrg_deltacut} is populated by LRGs that were observed in \texttt{SV1} with longer exposure times than the LRGs in \texttt{SV3} and \texttt{MAIN}. This leads to LRGs with higher SNR and therefore higher $\chi^2$ values. LRG spectra with robust redshifts meet the following condition: 

\begin{enumerate}
    \item \texttt{SPECTYPE} = `GALAXY'
    \item $z < 1.6$
    \item SNR$_{\rm{z}} > 1.28$
    \item $\chi^{2} \geq 7000$
    \item Either 5a \textbf{or} 5b must be satisfied:
    \begin{enumerate}
        \item[(5a)] $\Delta\chi^2 \geq 100$
        \item[(5b)] $40 \leq \Delta\chi^2 < 100$ \textbf{and} $\chi^2 < 10000$
    \end{enumerate}
\end{enumerate}

\begin{figure*}
    \centering
    \includegraphics[width=0.95\textwidth]{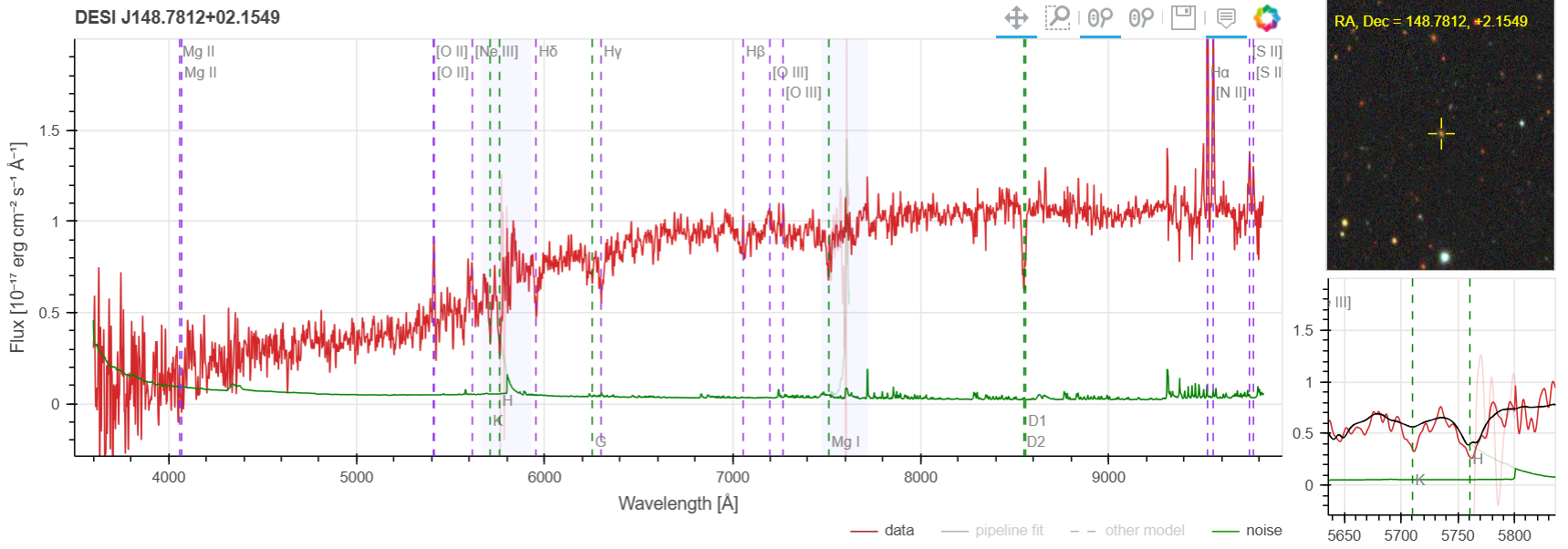}
    \vspace{0.5cm}
    \includegraphics[width=0.95\textwidth]{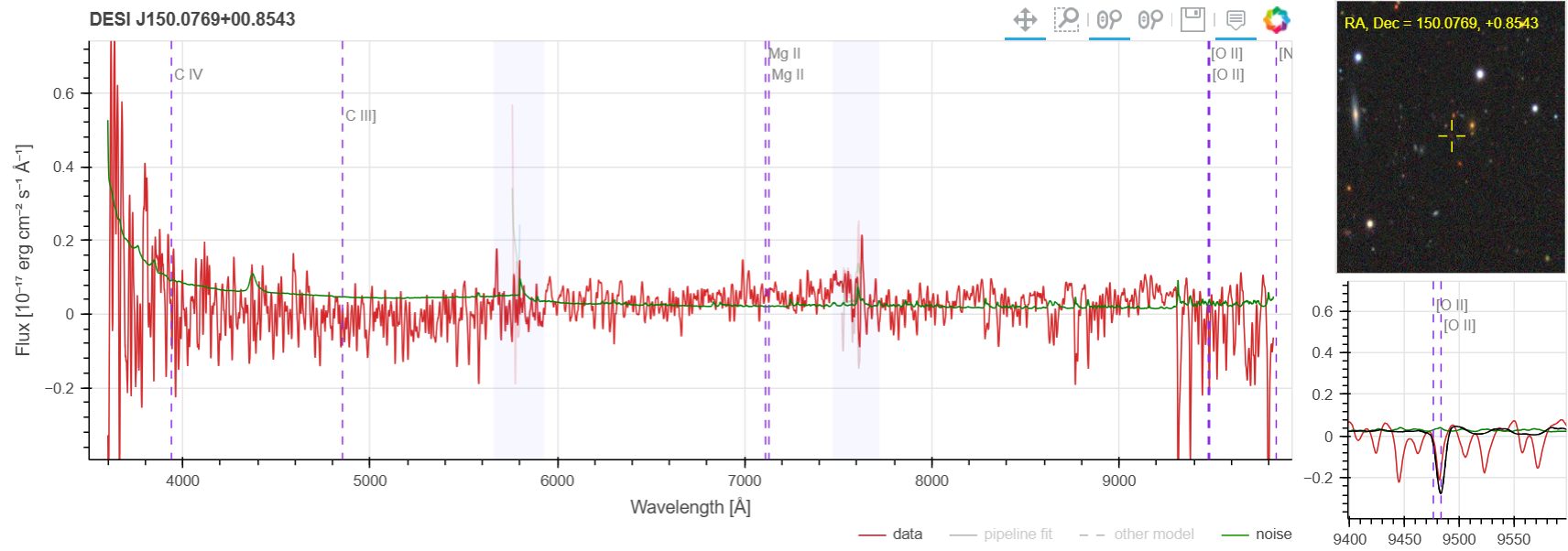}
    \caption{Examples of two LRG spectra that pass (top) and fail (bottom) our quality cuts. 
    The structure and color scheme is identical to that in Figure~\ref{fig:bgs_spectra}. For quality spectra (top) the absorption lines, line Ca II H and K, are clearly distinguishable. For spectra with poor redshift estimations the emission lines appear nonphysical indicating a poor model fit. The \texttt{DESINAME} of each object is found in the top left corner of each figure and both spectra were obtained in SV1 as part of the `dark’ program.}
    \label{fig:lrg_spectra}
\end{figure*}

In Figure~\ref{fig:lrg_spectra} we show an example of an LRG spectrum that passes the cut and one that does not. The LRG spectrum that passes the cut indeed has very high SNR, a strong continuum break at 4000 \AA~ and clearly identified Ca II H and K features. The spectrum that fails the cut, was rejected due to a low $\Delta \chi^{2}$ indicating the fitted features are difficult to differentiate from similar features from fits at another redshift. 

There are 11,565 LRG targets spectroscopically confirmed as galaxies in the DESI-COSMOS field and 9,736 LRG targets spectroscopically confirmed as galaxies in the DESI-XMMLSS field with robust redshifts. The redshift distribution of the LRG sample is shown in the right panel of Figure~\ref{fig:lrg_deltacut}.

We can compare our robustness criteria to the `fiducial' criteria in \cite{lan23} and estimate the catastrophic failure rate by comparing the redshift estimates for objects with multiple observations. Using the fiducial criteria of \texttt{ZWARN} = 0, $z<1.5$ and $\Delta\chi^2 > 15$, the main LRG sample retains 99.5\% of its objects and has an implied catastrophic failure rate of 0.56\% (30/5304). Our robustness criteria 
retains 97.9\% of objects with an implied catastrophic failure rate of 0.29\% (15/5179).

\subsection{Emission Line Galaxies}

ELG spectra typically exhibit low SNR in the continuum (median SNR per pixel $<1$) with strong emission lines, primarily the [OII] doublet. These [OII] $\lambda\lambda 3726,3279$ \r{A} lines are sufficient for estimating a redshift even without informative signal in the rest of the spectrum as is often the case at high redshift ($z>1$).  In other cases, particularly at lower redshift, the continuum and associated features can offer sufficient signal for robust classification, even when [OII] lines are weak. 
The combination of the SNR measurements of the [OII] doublet along with $\Delta\chi^{2}$ has been proven sufficient to reliably classify ELG spectra with quality redshifts \citep{Raichoor_2023}, according to the following relation:

\begin{equation}
    \rm{log}_{10}(SNR_{\rm{[OII]}}) > 0.9 - 0.2 \rm{log}_{10}(\Delta\chi^{2})
    \label{eq:oii}
\end{equation}

Instead of using a custom routine within the main pipeline, we use the [OII] flux measurements provided by \textit{FastSpecFit} \citep{fastspecfit}. As with the BGS and LRG targets, we only include ELGs with a \texttt{SPECTYPE} = `GALAXY' classification. We exclude all ELGs with $z>1.6$ as the [OII] doublet along with other distinguishable features are redshifted to the far red or out of the DESI wavelength coverage.

\begin{figure}
    \centering
    \includegraphics[width=0.45\textwidth]{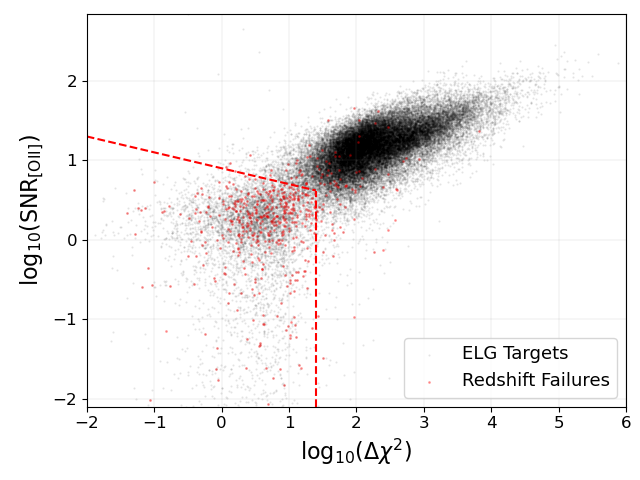}
    \hfill
    \includegraphics[width=0.45\textwidth]{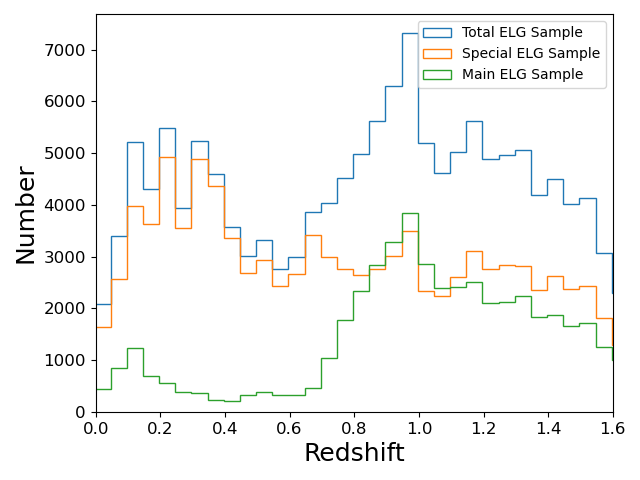}
    \caption{Statistics of the ELG sample in the COSMOS field. \textbf{Top}:  The distribution of the ELG sample in [OII] SNR vs $\Delta\chi^{2}$. The dashed red lines indicate the cuts in [OII] SNR and $\Delta\chi^{2}$ where excluded spectra are located in the bottom left section.  A large fraction (80.9\%) of ELG spectra fall into the accepted region. \textbf{Bottom}: The redshift distribution of ELG spectra with robust redshifts. Objects are placed into bins of $\Delta$z = 0.05. The total ELG sample (blue) is split into two sub samples. The primary ELG sample (green) is the main DESI target selection. The special ELG sample (orange) is the extended selection as part of a special program to probe a broader selection of ELG spectra.}
    \label{fig:elg_pop}
\end{figure}

 We demonstrate how the ELG sample populates the [OII] SNR vs $\Delta\chi^{2}$ parameter space in Figure~\ref{fig:elg_pop}, where the ELG sample populates a locus that increases in [OII] SNR with increasing $\Delta\chi^{2}$.  The sliding [OII] SNR vs $\Delta\chi^{2}$ cut allows for robust redshift estimations for spectra with low $\Delta\chi^{2}$ as long as the the [OII] SNR is sufficiently high. A $\Delta\chi^{2}\geq 25$ threshold is conservatively set to ensure spectra with confident fits are not cut regardless of the [OII] SNR. 

\begin{figure}
    \includegraphics[width=\linewidth]{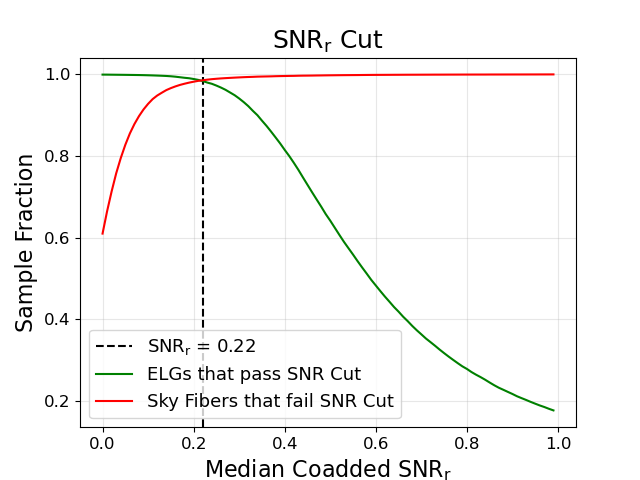}
    \caption{The cumulative fraction of ELG spectra exceeding a median SNR$_{\rm{r}}$ per pixel. The green line denotes the fraction of ELG spectra that are included at the given SNR$_{\rm{r}}$ cut. The red line denotes the fraction of sky spectra that are removed at a given SNR$_{\rm{r}}$ cut. The vertical line at SNR$_{\rm{r}}=0.22$ represents the optimal SNR$_{\rm{r}}$ cut for which we exclude the most sky spectra while minimally impacting the completeness of the ELG sample.}
    \label{fig:elgs_snr}
\end{figure}

Reducing sky fiber contamination is important for the ELG sample as sky subtraction of inherently low SNR spectra can cause catastrophic failures. Using a representative sample of sky fibers, we examine what fraction of sky spectra pass a [OII] SNR vs $\Delta\chi^{2}$ cut in Figure~\ref{fig:elgs_snr}. A SNR cut in the r-band ($\textrm{SNR}_{\rm{r}}$) was determined to balance a reduction in the sky fiber contamination with a desire to maintain high completeness in the ELG sample. We find that an additional SNR cut ($\textrm{SNR}_{\rm{r}} <0.22$), excludes 98.9\% of sky fibers that pass the [OII] SNR vs $\Delta\chi^{2}$ criteria. In total 99.97\% of sky fibers are excluded with all ELG quality cuts applied implying a 0.026\% catastrophic failure rate for low SNR ELG spectra. 

Spectra in the ELG sample that satisfy the following conditions yield a robust redshift:

\begin{enumerate}
    \item \texttt{SPECTYPE} = `GALAXY'
    \item $z<1.6$
    \item Either 3a \textbf{or} 3b must be satisfied:
    \begin{enumerate}
        \item[(3a)] $\Delta\chi^{2} \geq 25$
        \item[(3b)] $\log_{10}(\rm{SNR}_{\rm{[OII]}}) > 0
.9 - 0.2 \log_{10}(\Delta\chi^{2})$
    \end{enumerate}
    \item $\rm{SNR}_r > 0.22$
\end{enumerate}

\begin{figure*}
    \centering
    \includegraphics[width=0.95\textwidth]{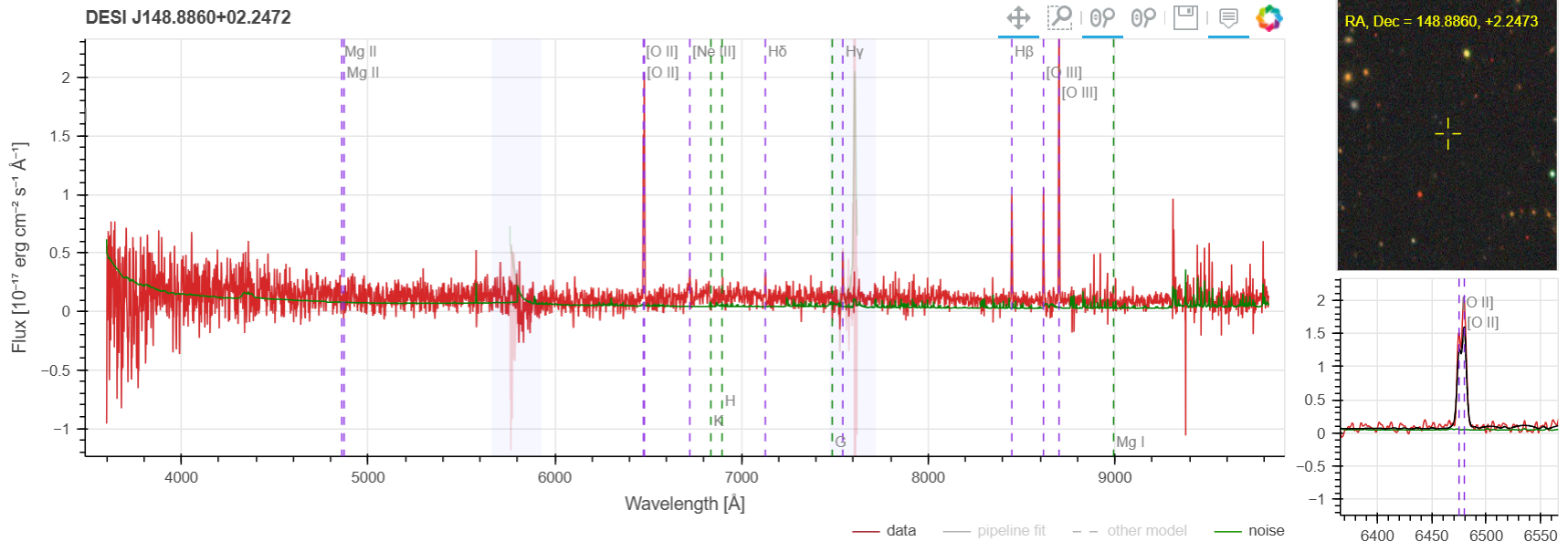}
    \vspace{0.5cm}
    \includegraphics[width=0.95\textwidth]{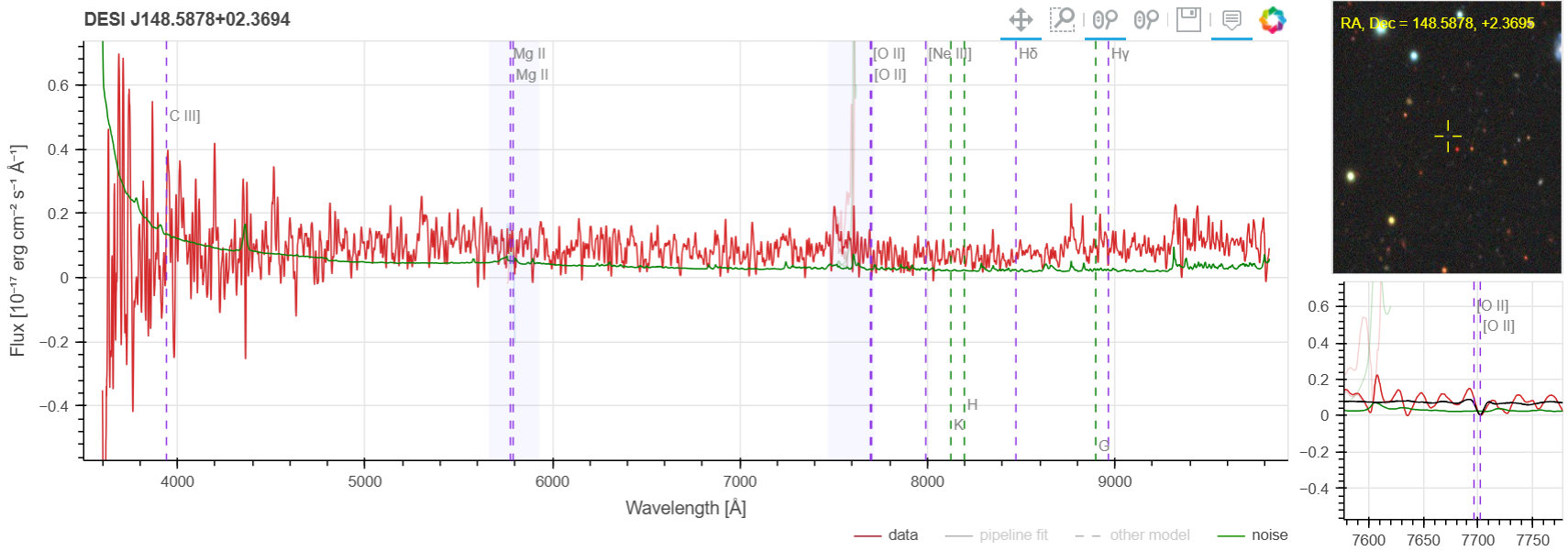}
    \caption{Examples of two ELG spectra that pass (top) and fail (bottom) our quality cuts. 
    The structure and color scheme is identical to that in Figure~\ref{fig:bgs_spectra}. For quality spectra (top) the emission lines like the [OII] doublet are clearly distinguishable. For spectra with poor redshift estimations the emission lines appear nonphysical indicating a poor model fit. The \texttt{DESINAME} of each object is found in the top left corner of the figure and both were observed in SV1 as part of the `dark' program.}
    \label{fig:elg_spectra}
\end{figure*}

An example of an ELG spectrum that passes the quality cuts is shown in the top panel of Figure~\ref{fig:elg_spectra}. It has very strong emission lines and a well resolved [OII] doublet. The spectrum in the bottom panel of Figure~\ref{fig:elg_spectra} fails the quality cuts and has very weak signal with almost no detectable signal in emission lines.

With these cuts, 80.9\% of the primary ELG sample is retained. In total, there are 30,264 ELG galaxies with reliable redshift estimates in the DESI-COSMOS field and 22,427 in the DESI-XMMLSS field. The redshift distribution of the ELG sample is shown in the bottom panel of Figure~\ref{fig:elg_pop}.

We can compare our robustness criteria to the `fiducial' criteria in \cite{Raichoor_2023} and estimate the catastrophic failure rate by comparing the redshift estimates for objects with multiple observations.  The fiducial criteria identifies reliable redshift estimates by their [OII] emission and $\Delta\chi^2$.  This criteria retains 89.0\% of its objects and has an implied catastrophic failure rate of 1.35\% (157/11630). Our robustness criteria, requires the object to be spectroscopically confirmed as a galaxy and limits the redshift to below the sensitivity of DESI. Our criteria retains 80.9\% of objects with an implied catastrophic failure rate of 0.68\% (71/10428). 

\subsection{Quasars}
\begin{figure}
    \includegraphics[width=\linewidth]{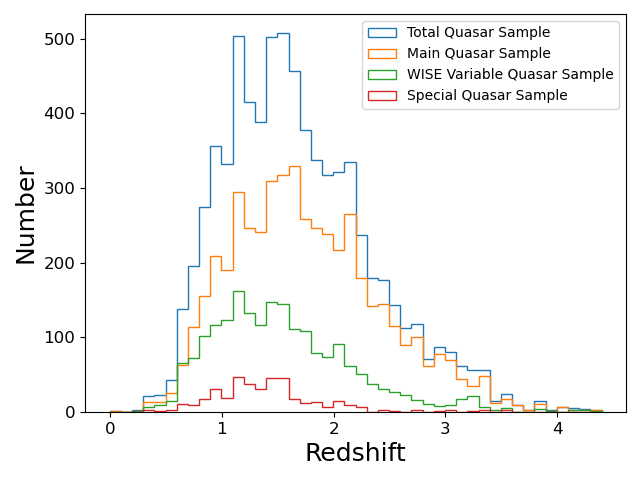}
    \caption{The redshift distribution of quasar spectra with robust redshifts in the DESI-COSMOS field. Objects are placed into bins of $\Delta$z = 0.1. The total quasar sample (blue) is split into three sub samples. The primary quasar sample (orange) results from the main DESI target selection. The quasar subsample (green) is a selection of variable quasars while the other sample (red) represents quasars observed in special programs.}
    \label{fig:qso_nz}
\end{figure}

Quasars span a large redshift range, as shown in Figure~\ref{fig:qso_nz}. This expansive redshift range leads a diverse sample with various broad emission lines. This spectral diversity motivates the multi-step approach of using \textit{Redrock}+afterburners in Section~\ref{redrock+afterburner} to ensure quality spectral classification and redshift estimations for this sample. We apply MgII and \textit{QuasarNET} afterburners to all objects targeted as a quasar or objects from other classes spectroscopically confirmed as a quasar (i.e., \texttt{SPECTYPE} = `QSO'). 

\begin{figure*}
    \includegraphics[width=\linewidth]{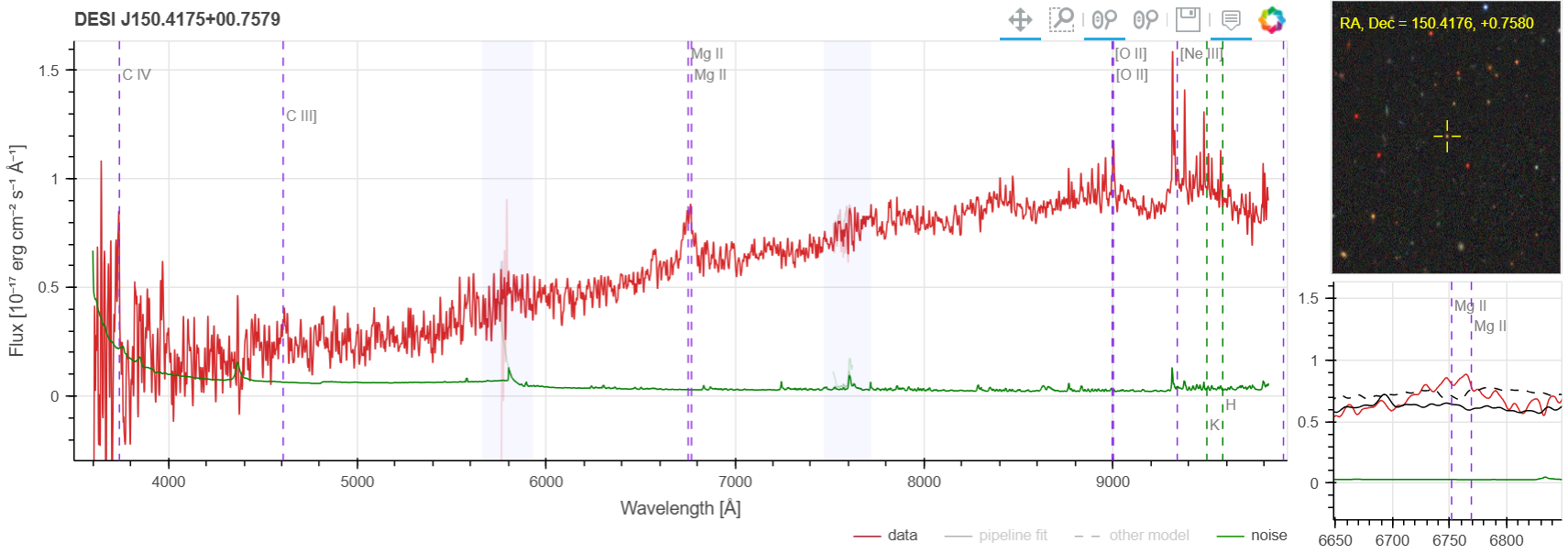}
    \caption{A quasar spectrum with a corrected redshift estimation and spectral classification after being processed through the MgII and \textit{QuasarNET} afterburners. The original spectrum analysis from \textit{Redrock} yielded a redshift estimation of z = 1.68 with \texttt{SPECTYPE} = `GALAXY'. With clear broad line emission, \textit{QuasarNET} detected an incorrect classification of galaxy and the spectral classification was switched to quasar. \textit{Redrock} also provided a more accurate redshift estimation of z = 1.4138 after using the \textit{QuasarNET} classification and redshift as a prior. The \texttt{DESINAME} of this object is found in the top left corner of the figure. This object was observed in SV1 as part of the `dark' program.}
    \label{fig:qso_qn}
\end{figure*}

Figure~\ref{fig:qso_qn} shows an example of a red quasar that was incorrectly redshifted and misclassified as a galaxy by \textit{Redrock} but was reclassified as a quasar with afterburners. A new redshift using \textit{Redrock} was determined using the \textit{QuasarNET} prior.

The primary quasar population for the DESI-COSMOS field consists of 4,471 quasars while the DESI-XMMLSS field consists of 4,711 unique quasars.  Using visual inspection of repeat observations, the combination of \textit{Redrock} with these two afterburners, we report a catastrophic failure rate of $\sim 2.6\%$.

\subsection{Stars}
For the stellar population in our catalogs, we report the radial velocity and physical stellar parameters given by \textit{RVSpecfit} \citep{rvs_paper,rvspecfit,RVS_EDR, koposov2025}. \textit{RVSpecfit} fits stellar templates directly at the pixel level in order to determine the objects radial velocities and other physical parameters. This direct-pixel level fitting allows for more accurate uncertainty estimates and can account for the diversity of stellar objects.

In order for an object to be fit by \textit{RVSpecfit}, it must satisfy the following conditions:

\begin{enumerate}
    \item has MWS\_ANY targeting bit OR
    \item has SCND\_ANY bit set OR
    \item classified as STAR by \textit{Redrock} OR
    \item has radial velocities within -1500 and 1500 km/s
    \item Median SNR per pixel in every camera must be greater than 2 (SNR$_{\rm{b/r/z}} > 2$).
\end{enumerate}
 From the primary selections, we report the RVS measurements for 19,399 stars in the DESI-COSMOS field and 10,998 in the DESI-XMMLSS field.

\subsection{Other Target Classes}
The DESI-COSMOS and DESI-XMMLSS catalogs include smaller samples of other targets from secondary and special programs. The description of all these samples is given in Appendix~\ref{appendix}. With the framework established for LRGs, ELGs, and QSOs, quality cuts are applied to each sample depending on target class. Each sample was analyzed to determine appropriate quality cuts by examining the prominent features, redshift range, SNR per pixel, and spectroscopic classification. Each sample is then sorted into one of four groups for redshift quality assessment. For target classes with high SNR and strong continuum features, we impose quality cuts using $\Delta \chi^{2}$ and $\chi^{2}$ similar to the BGS and LRG samples. For targets that have low SNR but prominent features, an emission line quality cut is imposed as with the ELG sample. Secondary and special program spectra that have been targeted or spectroscopically confirmed as quasars are treated using the same afterburners as the primary quasar sample. Secondary and special program spectra that are stellar targets or spectroscopically confirmed as stars are processed through the RVS pipeline. A full list of all target classes with the number of robust redshifts for both fields can be found in Appendix~\ref{appendix2}.

\subsubsection{Secondary and Special Continuum-Characterized Samples}
The secondary and special program samples described in this section have spectra that exhibit strong continuum and typically high SNR. Since these spectra have properties that are `LRG-like', spectra in these samples have a robust redshift estimations if they meet these criteria:

\begin{enumerate}
    \item \texttt{SPECTYPE} = `GALAXY'
    \item $z < 1.6$
    \item SNR$_{\rm{z}} > 1.28$
    \item $\chi^{2} \geq 7000$
    \item Either 5a \textbf{or} 5b must be satisfied:
    \begin{enumerate}
        \item[(5a)] $\Delta\chi^2 \geq 100$
        \item[(5b)] $40 \leq \Delta\chi^2 < 100$ \textbf{and} $\chi^2 < 10000$
    \end{enumerate}
\end{enumerate}

In order to demonstrate the robustness of applying these criteria to these samples, we show one larger subsample that falls into this continuum-characterized group.

The \texttt{LOWZ} sample consists of low redshift, dwarf galaxies intended to supplement the BGS sample at fainter magnitudes. This sample is intended to constrain the dwarf galaxy luminosity function and identify possible transient hosts.

\begin{figure}
    \includegraphics[width=\linewidth]{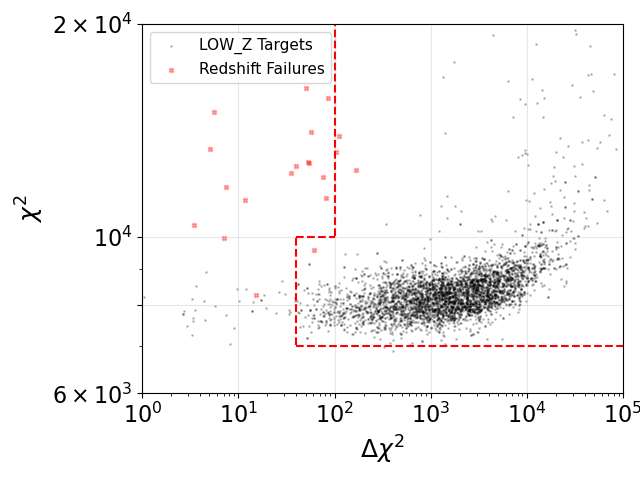}
    \caption{ Distribution of the \texttt{LOW\_Z} sample in the DESI-COSMOS field in $\chi^{2}$ vs $\Delta\chi^{2}$ parameter space. The red dashed line displays the $\chi^{2}$ and $\Delta\chi^{2}$ quality cuts. Spectra in the upper right are assumed to have quality redshifts. }
    \label{lowz_dchi2}
\end{figure}

The \texttt{LOWZ} sample has high SNR per pixel with a strong continuum similar to the LRG population, motivating a $\Delta\chi^{2}$ vs $\chi^{2}$ cut. As seen in Figure~\ref{lowz_dchi2}, a majority of the \texttt{LOWZ} sample falls within the accepted range of  $\Delta\chi^{2}$ vs $\chi^{2}$. With all cuts imposed, 95.6\% of \texttt{LOWZ} objects pass our robustness criteria.

\begin{table}[h]
\centering
\begin{tabular}{l r}
 \hline
 Sample &  Quality Redshifts \\ 
 \hline\hline
 PRINCIPAL (LRG) & 32124 \\
 LOWZ\_TIER3 & 1875 \\
 LOWZ\_FAINT & 1819 \\
 BGS\_BGS\_BRIGHT & 1748 \\
 DESI\_LRG & 1079 \\
 LOW\_Z & 842 \\
 BGS\_BGS\_FAINT & 679 \\
 LOWZ\_TIER2 & 655 \\
 LOWZ\_BRIGHT & 569 \\
 BGS\_BGS\_FAINT\_HIP & 263 \\
 UNWISE\_BLUE & 296 \\
 LOWZ\_TIER1 & 141 \\
 UNWISE\_GREEN & 53 \\
 \hline
\end{tabular}
 \caption{Secondary and Special Continuum Samples. The number indicates the number of robust redshifts in the DESI-COSMOS field. These samples observed in secondary and special programs exhibit spectral properties similar to the primary LRG and BGS sample. Redshift quality was assessed using similar algorithms to the LRG samples.}
 \label{tab:sec_cont}
\end{table}

A summary of all `Continuum-Characterized' target classes is shown in in Table~\ref{tab:sec_cont}. Spectra in this group all are subject to the quality cuts outlined above.

\subsubsection{Secondary and Special Emission Line Characterized Samples}
These secondary and special program samples are characterized by prominent emission lines such as the [OII] doublet and [OIII]. Spectra in these target samples have redshift quality cuts modeled after the primary ELG sample and have quality redshifts if they meet these criteria:

\begin{enumerate}
    \item \texttt{SPECTYPE} = `GALAXY'
    \item $z<1.6$
    \item Either 3a \textbf{or} 3b must be satisfied:
    \begin{enumerate}
        \item[(3a)] $\Delta\chi^{2} \geq 25$
        \item[(3b)] $\log_{10}(\rm{SNR}_{\rm{[OII]}}) > 0
.9 - 0.2 \log_{10}(\Delta\chi^{2})$
    \end{enumerate}
    \item $\rm{SNR}_r > 0.22$
\end{enumerate}

To demonstrate the robustness of applying these criteria to these samples, we examine two of the larger subsamples in the emission line group, SNe Hosts and the [OIII] sample for the DESI-COSMOS field.

\begin{figure*}
    \centering
    
    \includegraphics[width=0.95\textwidth]{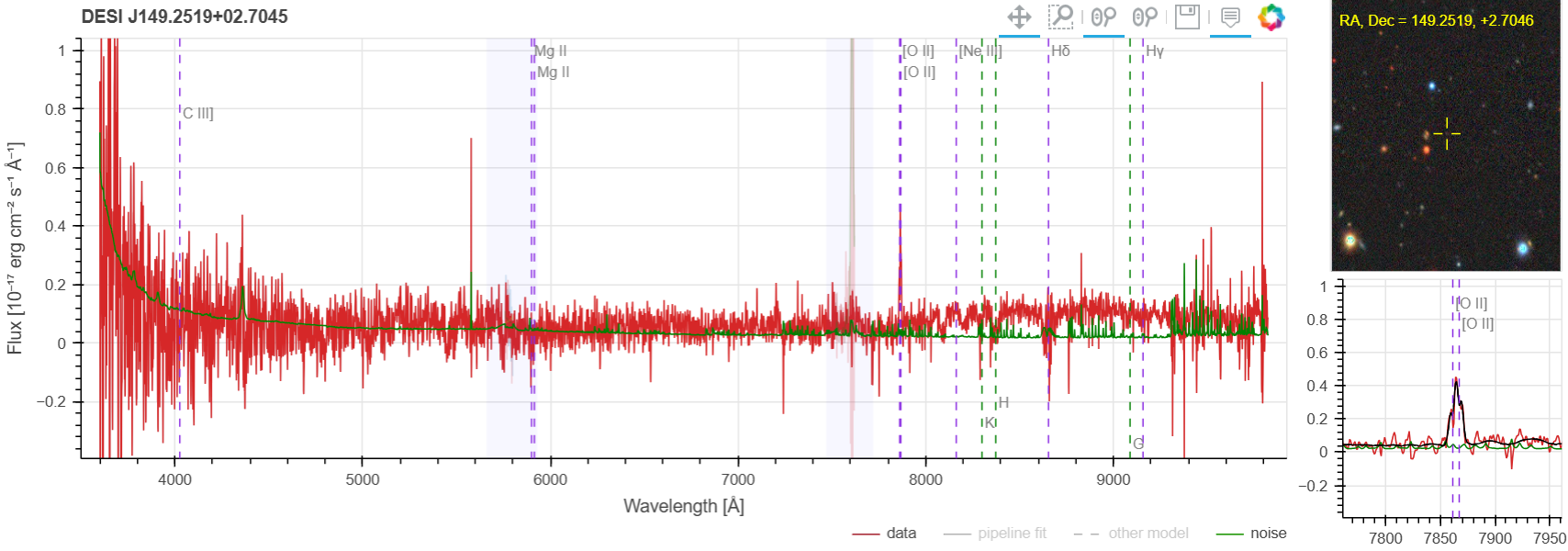}
    \caption{Example of a spectrum in our emission line sample for secondary and special program targets. The structure and color scheme is identical to that in Figure~\ref{fig:bgs_spectra}. A representative spectrum for the secondary and special program targets that use the strength of the [OII] doublet  for determining the redshift quality is shown. The \texttt{DESINAME} of this object is found in the top left corner of the figure. This spectrum was observed in SV1 as part of the `other’ program.}
    \label{fig:sec_emlin}
\end{figure*}

The SNe Host program is designed to target the host galaxies of SNe Ia in order to obtain redshifts for dark energy constraints. Targets in these programs are similar to the ELG samples with prominent [OII] emission. In Figure~\ref{fig:sec_emlin}, an example of a SNe host spectrum is shown with strong [OII] emission lines.

The [OIII] sample is comprised of two samples: low redshift galaxies selected using photometry from the Merian project \citep{Luo_2024} and a sample of galaxies intended for LSST-like weak lensing analysis (See Appendix~\ref{appendix}). The [OIII] emission lines for these samples provide a clean feature for quality redshifts.

\begin{figure*}
    \centering
        \includegraphics[width=0.45\textwidth]{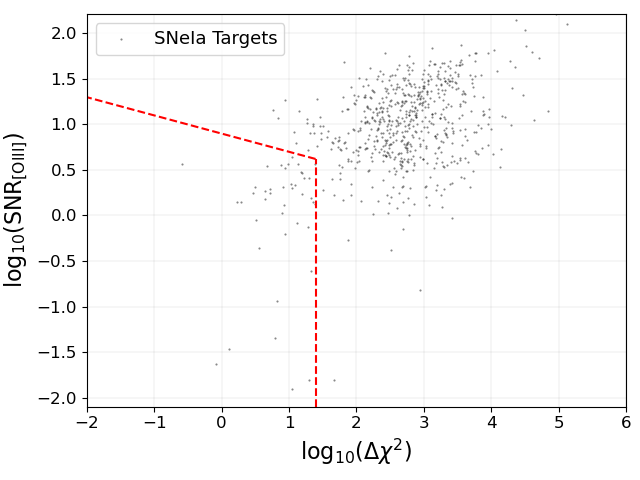}
        \hfill
        \includegraphics[width=0.45\textwidth]{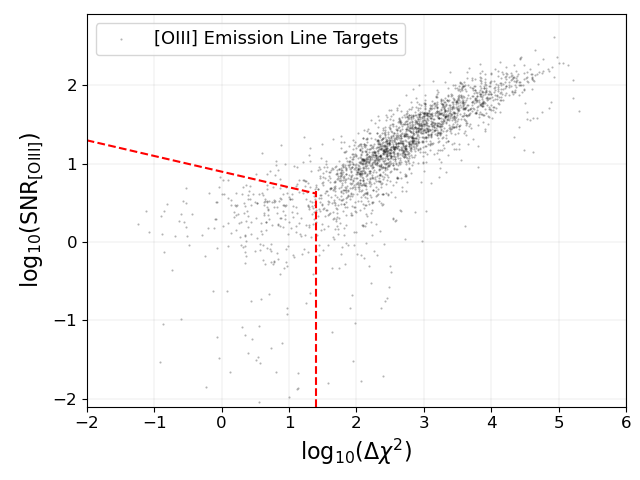}
    
    \caption{Statistics of the secondary and special emission line samples in the COSMOS field (See Table~\ref{tab:second_emline} for samples in the [OII] and [OIII] samples) \textbf{Left}:  The distribution of the secondary SNe Ia Host sample in SNR$_{\rm{[OII]}}$ vs $\Delta\chi^{2}$ parameter space. The dashed red lines indicate the cuts in SNR$_{\rm{[OII]}}$ and $\Delta\chi^{2}$ where excluded spectra are located in the bottom left section.  \textbf{Right}:  The distribution of the secondary and special [OIII] sample in SNR$_{\rm{[OIII]}}$ vs $\Delta\chi^{2}$ parameter space. The dashed red lines indicate the cuts in SNR$_{\rm{[OIII]}}$ and $\Delta\chi^{2}$ where excluded spectra are located in the bottom left section.}
    \label{fig:emline}
\end{figure*}

In Figure~\ref{fig:emline}, spectra populate the accepted regions for both the [OII] and [OIII] samples, similar to the ELG sample. We summarize all the secondary and special programs that are included in each sample and report the number of quality redshifts for each sample in Table~\ref{tab:second_emline}.

\begin{table}[h]
\centering
\small
\begin{tabular}{l r}
 \hline
 Sample & Quality Redshifts \\ 
 \hline\hline
 \multicolumn{2}{l}{\textbf{[OII] Objects}} \\
 \hline
 ELG\_PRINCIPAL & 58057 \\
 ELG (SPECIAL) & 13671 \\
 FILLER\_HIP & 4051 \\
 FILLER\_LOP & 1679 \\
 HSC\_HIZ\_SNE & 625 \\
 DESI\_ELG\_LOP & 547 \\
 DESI\_ELG\_HIP & 420 \\
 MERIAN (z $\geq$ 0.5) & 195 \\
 \shortstack[l]{WISE\_VAR\_QSO \\ (\texttt{SPECTYPE} = `GALAXY'}) & 184 \\
 \shortstack[l]{PSF\_OUT \\ (\texttt{SPECTYPE} = `GALAXY'}) & 153 \\
 DESI\_ELG & 85 \\
 STRONG\_LENS & 17 \\
 \hline
 \multicolumn{2}{l}{\textbf{[OIII] Objects}} \\
 \hline
 MERIAN (z $<$ 0.5) & 1180 \\
 DESI\_DEEP\_HIP & 657 \\
 DESI\_DEEP\_LOP & 484 \\
 DESI\_DEEP\_HIP2 & 132 \\
 \hline
\hline
\end{tabular}
 \caption{Secondary and Special Emission Line Samples. The number indicates the number of robust redshifts in the DESI-COSMOS field. These targets used [OII] or [OIII] emission line SNR to determine spectroscopic redshift quality.}
 \label{tab:second_emline}
\end{table}

\subsection{Secondary and Special Quasar samples}
Target classes in Table~\ref{tab:sec_qso} are objects that were targeted as a quasar or quasar-like object and/or have been spectroscopically confirmed as quasars. All secondary and special program targets are processed through the MgII and \textit{QuasarNET} afterburners to confirm spectral classification. 
\begin{table}[h]
\centering
\begin{tabular}{l r}
 \hline
 Sample & Quality Redshifts \\ 
 \hline\hline
 WISE\_VAR\_QSO & 1494 \\
 PSF\_OUT(\texttt{SPECTYPE}=`QSO') & 867 \\
 DESI\_QSO & 596 \\
 ISM\_CGM\_QGP & 55 \\
 QSO\_RED & 22 \\
 LOW\_MASS\_AGN & 3 \\
 QSO\_Z5 & 2 \\
 
\hline
\end{tabular}
 \caption{Secondary and Special Quasar Line Samples. The number indicates the number of objects with robust redshifts in the DESI-COSMOS field.
Spectroscopic redshift quality was assessed using the same algorithms as the primary quasar sample.}
 \label{tab:sec_qso}
\end{table}

\subsection{Secondary and Special Stellar samples}

Target classes in Table~\ref{tab:sec_star} are objects that were targeted as stars and/or have been spectroscopically confirmed as stars and processed through the \textit{RVSpecfit} pipeline. A summary of all secondary and special program stellar targets is given in Table~\ref{tab:sec_star}.

\begin{table}[h]
\centering
\begin{tabular}{l r}
 \hline
 Sample & Number \\ 
 \hline\hline
 MWS\_MWS\_MAIN\_BLUE & 265 \\
 MWS\_MWS\_BROAD & 188 \\
 MWS\_MWS\_MAIN\_RED & 65 \\
 MWS\_FAINT\_BLUE & 59 \\
 MWS\_CALIB & 24 \\
 BACKUP\_CALIB & 24 \\
 MWS\_FAINT\_RED & 18 \\
 FAINT\_HPM & 18 \\
 MWS\_CLUS\_GAL\_DEEP & 15 \\
 MWS\_MWS\_WD & 15 \\
 MWS\_MWS\_BHB & 7 \\
 MWS\_RR\_LYRAE & 4 \\
 WD\_BINARIES\_BRIGHT & 6 \\
 HPM\_SOUM & 4 \\
 BHB & 3 \\
 WD\_BINARIES\_DARK & 6 \\
 MWS\_MWS\_NEARBY & 1 \\
 
\hline
\end{tabular}
 \caption{Secondary and Special stellar samples. Similar to the primary stellar samples, objects in these samples are processed through the \textit{RVSpecfit} pipeline.}
 \label{tab:sec_star}
\end{table}

\nocite{rsd,yr3}

\section{Value-Added Catalog}
\label{VAC}
We release all information in this study in two public catalogs, DESI-COSMOS and DESI-XMMLSS. These two catalogs can be found at the webpage for all DESI Value-Added Catalogs\footnote{\url{https://data.desi.lbl.gov/public/papers/c3/cosmos-xmmlss/}}. The data model for the catalogs can be found publicly here\footnote{\url{https://data.desi.lbl.gov/public/papers/c3/cosmos-xmmlss/README.md
}} and all columns referenced in this section are described in Appendix~\ref{appendix3}. For both fields, the data product is a FITS table with several extensions: \texttt{SPECZ\_CAT}, \texttt{DECALS\_DR9\_PHOT}, \texttt{DECAM\_PHOT}, \texttt{HSC\_WIDE\_PHOT}, and \texttt{HSC\_UD\_PHOT}. In the DESI-COSMOS catalog, we include two additional extensions, \texttt{MERIAN\_PHOT} and \texttt{COSMOS2020\_PHOT}.

The catalogs have two main components: the main spectroscopic redshift catalog and supplemental photometric measurements. The redshift catalogs consist of 304,970 unique objects which are row matched across the photometric extensions. Here we describe the data provided in the publicly available VACs for both the DESI-COSMOS and DESI-XMMLSS fields.

\subsection{Redshift Catalog}
The main data product of the spectroscopic catalog is found in the \texttt{SPECZ\_CAT} extension. This extension models the DESI EDR spectroscopic redshift catalog data model\footnote{\url{https://desidatamodel.readthedocs.io/en/latest/index.html}} including target location, identification and targeting information. These columns include: \texttt{DESINAME, TARGET\_RA, TARGET\_DEC, PMRA, PMDEC, OBJTYPE, DESI\_TARGET, SCND\_TARGET, MWS\_TARGET, BGS\_TARGET, SV1\_DESI\_TARGET, SV1\_SCND\_TARGET, SV1\_MWS\_TARGET, SV1\_BGS\_TARGET, SV3\_DESI\_TARGET, SV3\_SCND\_TARGET, SV3\_MWS\_TARGET, SV3\_BGS\_TARGET, SURVEY, PROGRAM, CHI2, DELTACHI2, SPECTYPE, SUBTYPE, ZWARN, IS\_QSO\_MGII, IS\_QSO\_QN\_NEW\_RR},\text{\, and\ } \texttt{TSNR2\_BGS/LRG/ELG/QSO/LYA}. Objects in the catalog are unique to a resolution of 0.36 arcsec. Above a threshold of 0.36 arcsec, it is ambiguous if objects at close separations ($< 1$ arcsec) are distinct.
Information indicating if the target has corresponding RVS measurements or photometry from the appended photometric catalogs is found in columns: \texttt{HAS\_RVS, HAS\_DECALS\_PHOT, HAS\_DECAM\_PHOT,  HAS\_HSC\_UD\_PHOT,HAS\_HSC\_ WIDE\_PHOT, HAS\_MERIAN\_PHOT, HAS\_COSMOS2020\_PHOT}.

\subsubsection{Unique Targets}
For all unique targets observed in a single program, we additionally report the parameters that describe the redshift and quality of the redshift. The \texttt{QUALITY\_Z} column indicates whether the spectrum meets the quality criteria for that target class and the reported redshift is robust. The \texttt{BEST\_Z} column is the best estimate of the redshift for each object.  We report emission line flux and emission line variance in the \texttt{LINENAME\_FLUX} and \texttt{LINENAME\_FLUX\_IVAR} columns and include measurements for [OII], [OIII], H$\alpha$ and H$\beta$ from \textit{FastSpecFit}. The stellar fit parameters from the \textit{RVSpecfit} pipeline \citep{koposov_edr,koposov2025} are found in the following columns: \texttt{VRAD, VRAD\_ERR, VRAD\_SKEW, VRAD\_KURT, LOGG, TEFF, ALPHAFE, FEH, LOGG\_ERR, TEFF\_ERR, ALPHAFE\_ERR, FEH\_ERR} and \texttt{VSINI}.

\subsubsection{Targets observed in multiple programs}
For objects observed in multiple epochs, we introduce several new parameters from the \textit{Redrock} and quasar afterburners pipeline, attributed to the best available spectrum.  Barring exceptions defined below, we defined the best available spectrum as the spectrum with the highest $\Delta\chi^{2}$ parameter.
We merged redshift estimations, excluding spectra that do not have a quality redshift. For all redshifts, we report two quantities:

\begin{itemize}
    \item \texttt{BEST\_Z}: The mean of all quality redshifts (\texttt{QUALITY\_Z} = 1). For targets only observed in a single epoch, the \texttt{BEST\_Z} is simply the redshift estimation from the data pipeline.
    \item \texttt{DZ}: The largest difference in quality redshifts for the targets observed in multiple epochs, where  $dz =\frac{z_{max}-z_{min}}{1+z_{min}}$. This parameter is set to 0 if the object was observed only in one epoch.
\end{itemize}

For all parameters associated with \textit{Redrock} and quasar afterburners, we simply report the values associated with the best quality spectrum. The \texttt{SURVEY} and \texttt{PROGRAM} columns indicate the epoch from which the best spectrum was observed. We also include two columns describing the total coadded exposure time of the best spectrum, \texttt{TOTAL\_COADD\_EXPTIME}, and number of observations for the target, \texttt{TOTAL\_NUM\_COADDS}. All targeting information from each epoch will be included in the corresponding target mask columns. The reported emission line flux, emission line uncertainty and stellar fit parameters from the RVS pipeline represent the error weighted mean and error on the weighted mean over all quality spectra. 

Visual inspection was performed on spectra with quality redshifts that disagree ($dz > 0.0033$). For cases where only two observations were made of the same object, visual inspection was performed.  If the visual inspection yielded a confident redshift (\texttt{VI\_quality} $\geq 2.5$), the redshift from visual inspection is given in the \texttt{BEST\_Z} column.  If the visual inspection did not yield a confident redshift (\texttt{VI\_quality} $< 2.5$), the \texttt{QUALITY\_Z} bit is flipped to 
\texttt{False} and the \texttt{BEST\_Z} column reports the redshift from the spectrum with the highest $\Delta\chi^{2}$. In cases with more than two observations of the same object with only one discrepant redshift estimate, we simply took the redshift estimate identified by the majority of observations. In total, 70,799 objects had repeat observations with robust redshift classifications, for a total of 180,381 individual spectra. Of those individual spectra, the merging process revealed that 179,967 were given the correct redshift (99.77\% purity) from the automated classification scheme.

A full list and description of each column can be found in Appendix~\ref{appendix3}.

\subsection{Photometry Extensions}

In order to supplement the well-described target selection algorithms in Appendix~\ref{appendix}, we provide row-matched photometry from various imaging surveys for all targets. We matched all spectroscopic objects within $1^{\prime\prime}$ to the public imaging data sets, taking the closest match when multiple objects were located within $1^{\prime\prime}$ of the spectroscopic object.

The photometry includes DECaLs DR9\footnote{\url{https://www.legacysurvey.org/dr9/description/}}, DECaLS DR10\footnote{\url{https://www.legacysurvey.org/dr10/description/}}, Hyper Suprime-Cam PDR3 Ultra Deep/Wide \citep{Aihara_2022}, photometry from the public COSMSOS2020 photometric catalogs \citep{Weaver22} and finally, photometry from the Merian medium-band survey (only for DESI-COSMOS) from the Victor M. Blanco telescope \citep{Luo_2024}. The appended photometry is provided in extensions to the \texttt{SPECZ\_CAT} for each photometric sample:
\begin{itemize}
    \item \texttt{HDU2:DECaLS DR9} is an extension containing row matched  magnitudes, flux estimates ($g,r,z,W1,W2$), flux uncertainties and selected columns from the public DECaLs DR9 catalog\footnote{\url{https://www.legacysurvey.org/dr9/files/\#sweep-brickmin-brickmax-fits}}.
    \item \texttt{HDU3:DECaLs DR10} is an extension containing row matched  magnitudes, flux ($g,r,i,z,W1,W2$), flux uncertainties and selected columns from the public DECaLs DR10 data\footnote{\url{https://www.legacysurvey.org/dr10/files/\#sweep-brickmin-brickmax-fits}}.
    \item \texttt{HDU4:HSC PDR3 ULTRA DEEP} is an extension containing row matched photometry from HSC Ultra Deep PDR3 and has selected data columns from the public data catalog\footnote{\url{https://hsc-release.mtk.nao.ac.jp/schema/\#pdr3.pdr3\_wide.forced2}}. The selected data columns cover photometry across ($g,r,z,i,y$) and include magnitudes, flux estimates, and flux uncertainties for different models.
     \item \texttt{HDU5:HSC PDR3 WIDE} is an extension containing row matched photometry from HSC Wide PDR3 and has selected data columns from the public data catalog\footnote{\url{https://hsc-release.mtk.nao.ac.jp/schema/\#pdr3.pdr3\_dud\_rev.forced2}}. The selected data columns cover photometry across ($g,r,z,i,y$) and include magnitudes, flux estimates, and flux uncertainties for different models.
    \item \texttt{HDU6:COSMOS2020} (DESI-COSMOS ONLY) is an extension containing row matched  magnitudes, flux estimates, and flux uncertainties, photometric redshifts, galaxy parameter estimates from the COSMOS2020 public catalog including $\sim$30 broadband and medium-band optical and near-infrared filters \citep{Weaver22}. 
    \item \texttt{HDU7:MERIAN} (DESI-COSMOS ONLY) is an extension containing row matched magnitudes, flux estimates, and flux uncertainties from the Merian Survey.  The  Merian filter set includes: N708 ($\lambda_{c} =7080\AA$, $\Delta\lambda = 275\AA$) and N540 ( $\lambda_{c} = 5400\AA$, $\Delta\lambda = 210\AA$).

\end{itemize}

\section{Example Applications}
\label{exampleuses}
In this section, we provide example applications of the DESI-COSMOS and DESI-XMMLSS
VAC catalogs demonstrating completeness with respect to the parent galaxy population, comparison to other spectroscopic compilations, photometric redshift quality, and galaxy cluster member identification.

\subsection{Sample Completeness}
A key benefit of adding high-fidelity spectroscopic information in the XMM-LSS and COSMOS fields is that both fields are observed by many different surveys, and so are used for cross-calibration and comparison across different data sets. For example low-noise photometric observations of objects in the deep drilling fields for DES \citep{OzDES_XMM} and LSST \citep{LSST_DDF} provide an estimate of the true source fluxes. With an estimate of ``truth'', these samples can be used to probe the selection function or can be used to infer the redshift distribution of samples used for cosmological analyses with photometric data \citep{obiwan}.

Crucially, characterizing the selection function works well only for a high quality reference spanning a wide range in color-magnitude space.
We opt to demonstrate the completeness of our DESI-COSMOS spectroscopic catalog using HSC Wide photometry, as this is most directly comparable to what will be available with Rubin. 

We limit the completeness assessment to a radius 1.6 deg$^2$ from the field center, RA, DEC = (150.1, 2.182). The full area of the DESI-COSMOS field includes more spectra, but with lower typical completeness. We apply a series of quality and footprint cuts on the HSC Wide sample photometry to obtain a ``clean'' sample based on pixel, centroiding, and flux flags, removing any objects that evaluate to \texttt{TRUE} for any of the following columns, for each filter in $\texttt{BAND}\in \{g, r, i, z, y\}$:
\begin{itemize}
    \item \texttt{\{BAND\}\_PIXELFLAGS\_EDGE},\\ \texttt{\{BAND\}\_PIXELFLAGS\_BAD}, \\\texttt{\{BAND\}\_PIXELFLAGS\_INTERPOLATEDCENTER}, \\\texttt{\{BAND\}\_PIXELFLAGS\_SATURATEDCENTER}, \\\texttt{\{BAND\}\_PIXELFLAGS\_CRCENTER}
    \item \texttt{\{BAND\}\_SDSSCENTROID\_FLAG}
    \item \texttt{\{BAND\}\_CMODEL\_FLUX\_FLAG}, \\\texttt{\{BAND\}\_PSFFLUX\_FLUX\_FLAG},\\ \texttt{\{BAND\}\_KRONFLUX\_FLUX\_FLAG},\\\texttt{\{BAND\}\_APERTUREFLUX\_15\_FLUX\_FLAG}
    \item \texttt{\{BAND\}\_INPUTCOUNT\_VALUE} $< 3$
\end{itemize}
Note that we do \textit{not} apply a \texttt{BRIGHTOBJECT\_CENTER} flag, as we find this to be too conservative a cut that removes 40\% of the objects that have high-quality DESI redshifts. In total, these cuts remove $\sim15\%$ of the objects in the DESI COSMOS region with $i<24.1$, mostly due to the pixel and centroiding flags. 

We then characterize the fraction of all objects with clean photometry that have quality DESI redshifts in two different ways, emphasizing completeness in either apparent magnitude or apparent colors. Note that here we define completeness in the strict sense of what fraction of all objects in the imaging data have good DESI redshifts, which is distinct from and more stringent than the characterization of what parts of color-magnitude space have been observed with DESI. In reality, the latter is what has to be practically used for extrapolating existing spectroscopic catalogs outside of their domain of support, and a key challenge is identifying if and how such extrapolation is biased \citep{jaime2024, dircal25prep}.

\begin{figure}
    \centering
    \includegraphics[width=1\linewidth]{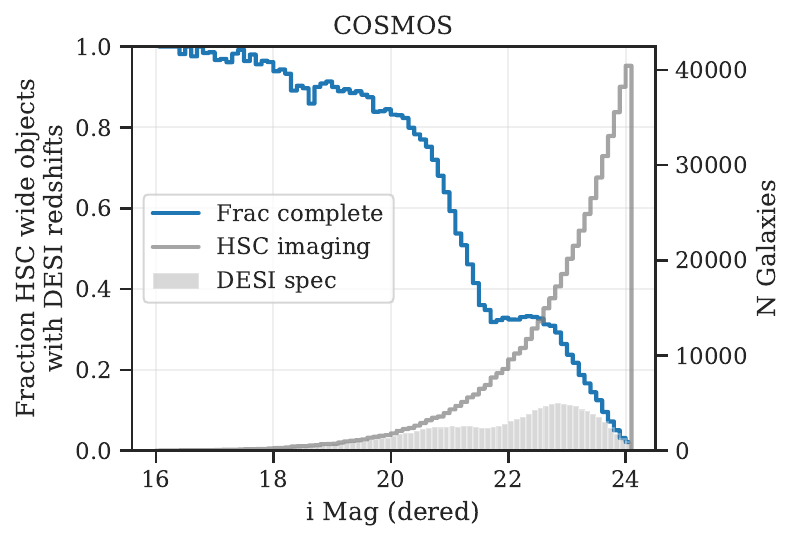}
    \caption{Completeness of extended objects in DESI-COSMOS as a function of dereddened $i$ magnitude as measured from HSC. The gray curve gives the total number of extended HSC objects with clean photometry, the filled gray histogram gives the number of those objects with good DESI redshifts, and the blue curve gives the resultant fraction complete, i.e., the ratio of the two gray histograms.}
    \label{fig:mag_completeness}
\end{figure}

Figure~\ref{fig:mag_completeness} shows the completeness as a function of dereddened $i$-band magnitude\footnote{All HSC magnitudes are \texttt{CMODEL} magnitudes unless otherwise indicated.}, indicating that such samples are all but complete up to $i < 17.5$, and with a majority of extended objects having spectroscopic classification up to $i < 21.3$. The number of DESI spectra peaks at $i \sim 22.9$, and after this the completeness drops precipitously to effectively 0 at $i\sim24.1$, as the number of photometric objects grows exponentially (shaded histogram).

This  increase of galaxies at fainter magnitudes highlights a pernicious challenge for spectroscopic surveys: both the observational time required to get a good redshift per object \textit{and} the number of objects increase geometrically as magnitude increases. The resulting lack of completeness at the faint end means that as photometric samples go deeper, the fractional number of spectra that are representative of such samples goes down rapidly.
As a consequence, it is common to assume that various galaxy population characteristics (such as redshift) are dependent on color, but not on overall magnitude \citep{masters2015,hildebrandt2020}, such that observations from well-characterized bright populations can be applied to the faint objects that dominate photometric samples. However, it is known that such assumptions do not hold in practice \citep{jaime2024}, such that representative  spectroscopic measurements for faint samples are crucial for understanding their true properties.

\begin{figure*}
    \centering
    \includegraphics[width=0.45\textwidth]{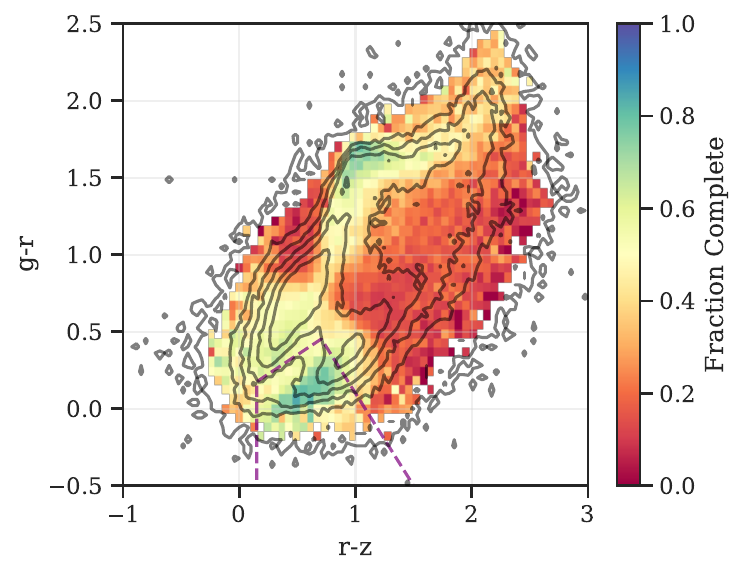}
    \includegraphics[width=0.45\textwidth]{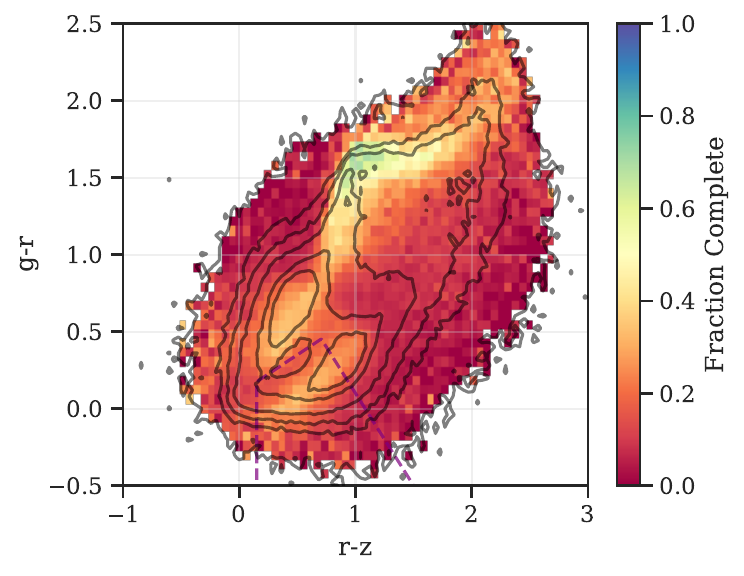}
    \caption{Marginal density and completeness of clean HSC extended objects in the DESI-COSMOS area in the $(g-r, r-z)$ plane. The black curves give the $\{0.25, 0.50, 0.75, 0.9, 0.95, 0.99\}$ object density iso-contours, while colors indicate the fraction of those objects with good DESI redshifts (limited to cells with at least 10 galaxies). The dashed purple line shows the boundaries of the main DESI ELG selection, for reference. \textbf{Left:} Sample selected with a limiting magnitude of $i<23$. \textbf{Right:} Sample selected with a limiting magnitude of $i<24.1$.}
    \label{fig:colorspace_completeness}
\end{figure*}

Figure ~\ref{fig:colorspace_completeness} shows the DESI spectroscopic completeness of clean HSC extended objects projected into a simpler $(g-r, r-z)$  color space, and limited to $i < 23$ magnitude (left) or  $i < 24.1$ magnitude (right). Contours show the density of the population, with colors illustrating completeness. There is clear inhomogeneity in the sampling of the $grz$ color space, with over-representation of some parts of the colorspace (e.g., due to the ELG selection, indicated via the purple dashed line). 

\begin{figure*}
    \centering
    \includegraphics[width=1\linewidth]{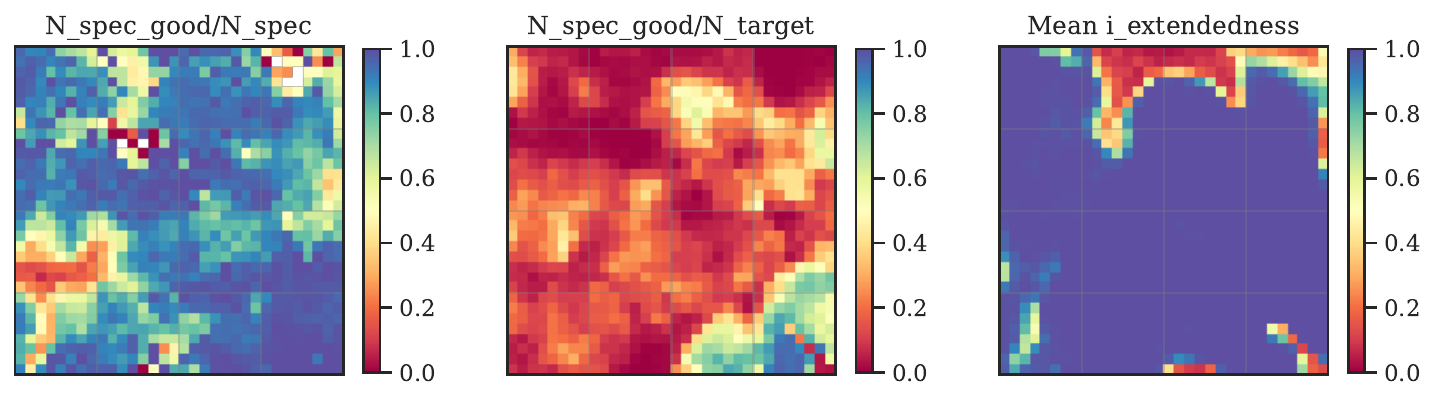}
    \caption{Completeness properties in a 2D color manifold defined via \textit{grizy} + $i_{\rm fiber}$ fluxes for a limiting magnitude of $i<24.1$ . Axes are arbitrary, but nearby SOM cells are also nearby in the SOM-learned colorspace. 
    \textbf{Left:} Fraction of spectroscopically observed objects that pass our quality criteria ($\texttt{QUALITY\_Z}= True$) in each cell, identifying clear regions in color space with high failure rates and high redshift success rates. \textbf{Center:} Fraction of all clean HSC objects with good DESI spectra. 
    \textbf{Right:} Average 
    $\texttt{I\_EXTENDEDNESS\_VALUE}$ of objects in each cell. DESI undersamples point sources, with relatively fewer good DESI spectra in cells with $\texttt{I\_EXTENDEDNESS\_VALUE} \sim 0$.}
    \label{fig:som_completeness}
\end{figure*}

In Figure~\ref{fig:som_completeness} we show completeness of a self-organizing map (SOM). This is a learned 2D manifold of the $grizy+i_{\rm{fiber}}$ HSC wide color space up to a limiting magnitude of $i<24.1$. We use the SOM algorithm described in \cite{Sanchez:2020reg}, which has several improvements to traditional SOMs, including a tunable hyperparameter that dictates how much the SOM cares about differences in the overall magnitude between objects vs. differences in the objects' photometric colors, as well as accounting for measurement uncertainties in the fluxes. 

The hyperparameter $\sigma_F$ defines a Gaussian prior on an overall amplitude rescaling of the flux in all bands, which is marginalized over when determining the ``distance" between any given object and the SOM cells covering the magnitude-color-space. In the extremes, setting $\sigma_F=0$ allows for no rescaling, such that objects with the same color but rather different overall magnitudes will populate different SOM cells, whereas $\sigma_F \rightarrow \infty$ enables free rescaling of the overall flux without penalty, erasing magnitude information and resulting in a SOM that is \textit{only} sensitive to color.

This avoids a common practical challenge when working with SOMs, which is whether to include only colors or also an overall magnitude when defining the relevant distance metric for learning the manifold. 

Here we use $\sigma_F = 0.4$, which corresponds to a $1\sigma$ penalty for an overall magnitude difference of 0.4. This gives a view that is  more sensitive to colors than overall magnitude, and is largely complementary to the magnitude-only completeness characterization of Figure~\ref{fig:mag_completeness}. This metric is more likely to group objects with like photo-z's, which depend much more strongly on color than magnitude \citep[c.f.][for more details on the benefits of this SOM algorithm, particularly App. A]{Sanchez:2020reg}.

Figure~\ref{fig:som_completeness} illustrates the selection effects present in the DESI-COSMOS sample in the self-organizing map. The left plot shows clear regions with higher spectroscopic failure rates, requiring further spectroscopic follow-up. The center plot shows how quality DESI redshifts are highly concentrated in certain regions of color-space. The right plot shows how the SOM identifies and separates point sources naturally through the inclusion of both \textit{i} and $i_{\rm{fiber}}$ fluxes. Cells characterizing point-like objects tend to be in regions with very low completeness, illustrating how DESI preferentially targets extended objects.

The central plot can be used to construct completeness weights \citep{lange2025}. These would be computed as the inverse completeness fraction of each galaxy's SOM cell, and enable one to reweight information coming from the DESI sample to better represent the respective imaging samples, accounting for selection along the HSC \textit{grizy} and $i_{\rm{fiber}}$ axes. While this can remove some selection effects to make the DESI sample more representative of an LSST-Y1 like sample with $i < 24.1$, it is only able to account for such effects that manifest in the this color-space, is limited by the resolution of the SOM, and does not properly account for objects observed with DESI for which we could not obtain a reliable redshift.


\subsection{Spectroscopic Redshift Comparison}

\begin{figure}
    \centering
    \includegraphics[width=\linewidth]{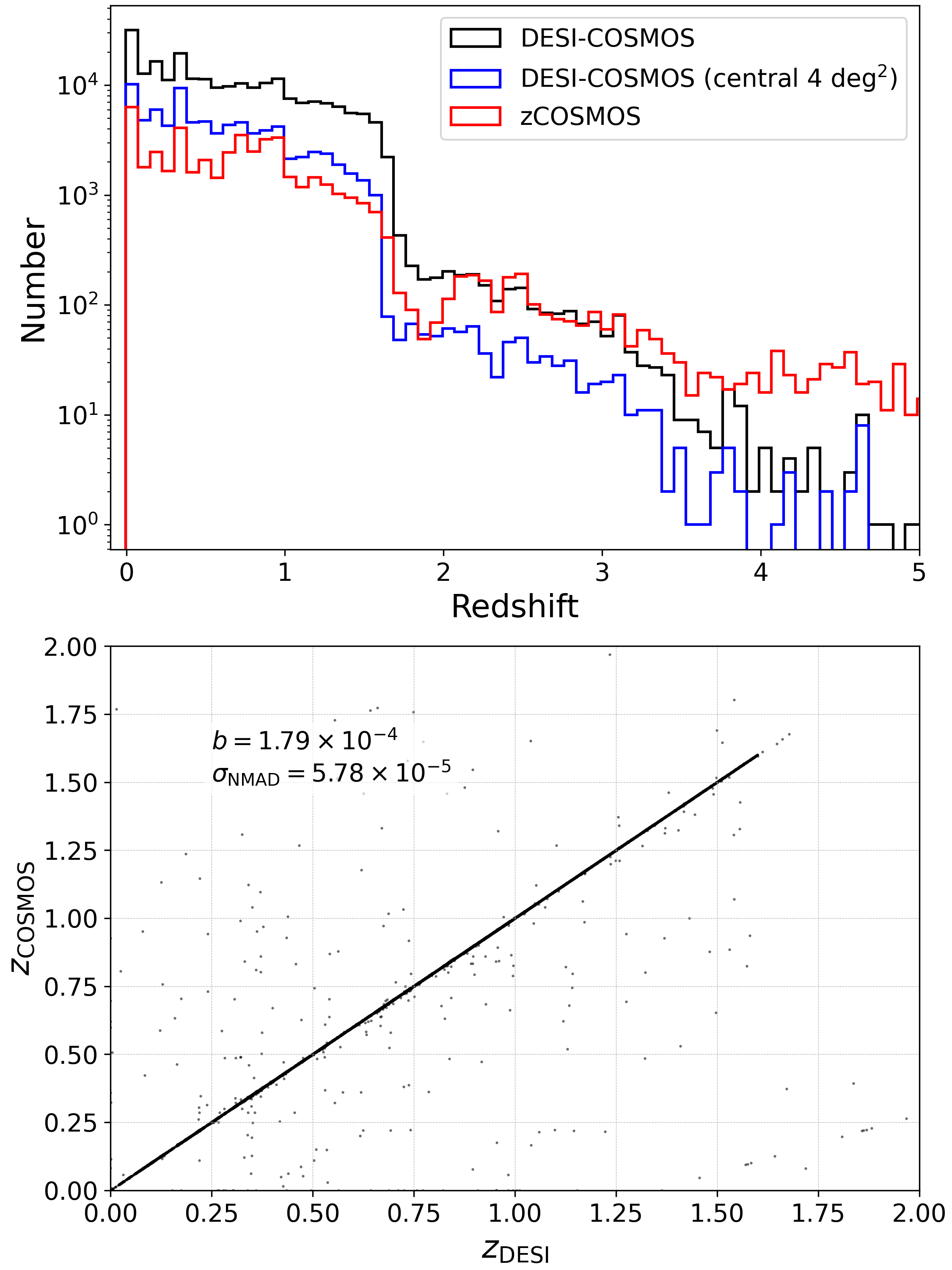}
    \caption{Spectroscopic redshift comparison between the DESI-COSMOS catalog and the zCOSMOS catalog\citep{COSMOS_spec}. \textbf{Top}: The redshift distributions for the DESI-COSMOS and zCOSMOS catalogs. The DESI-COSMOS redshift distribution (black) encompasses the full 16 deg$^2$ footprint.  The DESI-COSMOS (blue) and zCOSMOS (red) show the redshift distribution for the central 4 deg$^2$ covered by the zCOSMOS catalog. \textbf{Bottom}: The spectroscopic redshift comparison between the DESI-COSMOS and zCOSMOS catalogs. The comparison yields small bias ($b = 2.01\times10^{-4}$) and 99.3\% of redshifts agree within 1000 km s$^{-1}$. } 
    \label{fig:spec_comp}
\end{figure}

Complementing the DESI-COSMOS catalog, the spectroscopic redshift compilation from the COSMOS collaboration provides 48,575 quality redshifts ($CL ~\textrm{(confidence level)} \geq 95\%$) from spectroscopy of 97,929 unique objects in the central 4 deg$^{2}$ region \citep{COSMOS_spec}. In the same footprint, the DESI-COSMOS catalog reports 83,925 quality redshifts.  For objects classified as galaxies, the catastrophic rate ($\Delta v < 1000$ km s$^{-1}$) from the DESI catalog is expected to be less than 0.5\%. Matching across the central 4 deg$^2$ region, we find the 28,030 objects in common between the two catalogs. In Figure~\ref{fig:spec_comp}, a comparison between spectroscopic redshifts yields a 99.2\% agreement within 1000 km s$^{-1}$.  

The zCOSMOS catalog has 20,173 galaxies with high quality redshifts unique to the compilation, of which 3,333 are galaxies with $z > 1.6$ that cannot be classified by the current version of the DESI pipeline due to limitations in the galaxy template set.
The recent zCOSMOS catalog also includes physical parameter estimates such as stellar mass and star formation rates. 
The DESI-COSMOS offers 45,196 quality
 redshifts for $z < 1.6$ galaxies in the central 4 deg$^{2}$ footprint that are distinct from the COSMOS compilation. The DESI-COSMOS catalog includes an additional 128,630 unique redshifts in an extended 16 deg$^{2}$. Both the DESI-COSMOS and DESI-XMMLSS catalogs provide prominent emission line flux estimates for all objects, well-defined target selection algorithms, and row matched photometry from public DECam, HSC and COSMOS2020 catalogs.

\subsection{Photometric Redshift Comparison}

The COSMOS2020 catalog \citep{Weaver22} provides a large collection of photometry within the 2 deg$^{2}$ COSMOS field, completely within the 16 deg$^2$ footprint of our catalog. Photometric redshifts were computed following the method in \cite{laigle2016} where both galaxy and stellar templates are fitted to the observed photometry using the code LePhare \citep{lephare1,lephare2}. A comprehensive review of the photometric redshift (photo-z) measurements can be found in \cite{Weaver22}. 

\begin{figure}
    \centering
    \includegraphics[width=\linewidth]{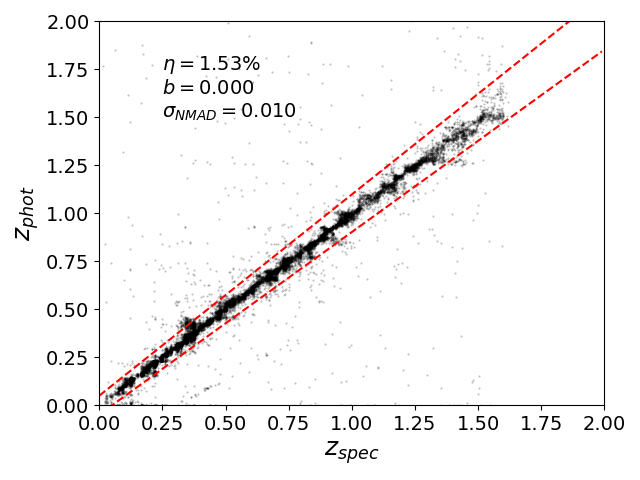}
    \caption{Comparison of DESI spectroscopic redshifts to photometric redshift estimates using COSMOS2020 LePhare photometric redshifts. Objects in the spectroscopic catalog have quality redshifts and are matched to the COSMOS2020 catalog within 1".  The red dashed lines indicate the upper and lower $5\sigma_{NMAD}$ bounds. $\sim$93\% of matched objects fall within the $5\sigma_{NMAD}$ bounds. The outlier fraction ($\eta$), bias ($b$), and $\sigma_{NMAD}$ are also given. }
    \label{fig:photz_me}
\end{figure}

Using the matched COSMOS2020 catalog, we compared all  robust spectroscopic redshifts to the measured \texttt{lp\_zBEST} estimates of the photo-z (Figure~\ref{fig:photz_me}).  We only include spectra with extended morphologies for our calibration (\texttt{SOLUTION\_MODEL} $\neq$ `PointSource'). For each object, we measure the $\frac{\Delta z}{1+z}$ between matched objects from the COSMOS2020 catalog using $z_{spec} =$ \texttt{BEST\_Z}:

\begin{equation}\label{eq:dz}
    \frac{\Delta z}{1+z} = \frac{z_{phot}-z_{spec}}{1+z_{spec}}
\end{equation}

We quantify the precision and accuracy of the photo-z measurements with our spectroscopic measurements through three metrics (dispersion, outlier fraction and bias). For dispersion, we use the normalized median absolute deviation (NMAD), defined as:
\begin{equation}
    \sigma_{\mathrm{NMAD}} = 1.48 \cdot \mathrm{median} \left( \left| \frac{\Delta z}{1+z} - \mathrm{median} \left( \frac{\Delta z}{1+z} \right) \right| \right)
\end{equation}
following \cite{brammer_2008} as it is less sensitive to outliers than the standard deviation or other similar measure of the dispersion. We report the fraction of outliers in two ways. First, denoted as $\eta$, the outlier fraction is defined as the fraction of  galaxies whose photo-z deviate from their spec-z by $|\Delta z| > 0.15 \cdot (1 + z_{spec})$ \citep{hildebrandt2012}. 
Second, using $\sigma_{NMAD}$ as the characteristic scatter of the photo-z calibration, we determine the fraction of outliers that fall within $5\sigma_{NMAD}$. Lastly, the bias, $b$, is computed as the median  of $\frac{\Delta z}{1+z}$.

Across the whole sample, as shown in Figure~\ref{fig:photz_me}, we find $\eta = 1.53\%$, $b = 0.0004$ and $\sigma_{NMAD} = 0.010$.  In addition we find that 93.4\% of photometric redshifts fall within $5\sigma_{NMAD}$ of the spectroscopic redshift. A summary of the comparison from the total sample and primary samples can be found in Table \ref{tab:photz}.

\begin{table}[h]
\centering
\begin{tabular}{lccccc}
\hline
Sample & Number & $\sigma_{NMAD}$ & $\eta$ & $b$ & $<5\sigma_{NMAD}$ \\
\hline\hline
Total  & 18294 & 0.010 & $1.53\%$ & 0.0004 & $93.7\%$ \\
LRG   & 1135 & 0.008 & $1.42\%$ & 0.012 & $76.5\%$ \\
ELG   & 2615 & 0.004 & $1.53\%$ & 0.003 & $85.9\%$ \\
BGS   & 1422 & 0.010  & $2.53\%$ & -0.004 & $78.1\%$ \\
\hline
\end{tabular}

\caption{Photo-z vs Spec-z comparison for LRG, ELG, and BGS. Provided are the number of objects in each sample, the characteristic scatter ($\sigma_{NMAD}$), the outlier fraction ($\eta$), the bias, ($b$), and the fraction of photometric redshifts that agree with the spectroscopic redshift within 5$\sigma_{NMAD}$.}
\label{tab:photz}
\end{table}

\begin{figure*}
    \centering
    \includegraphics[width=\textwidth]{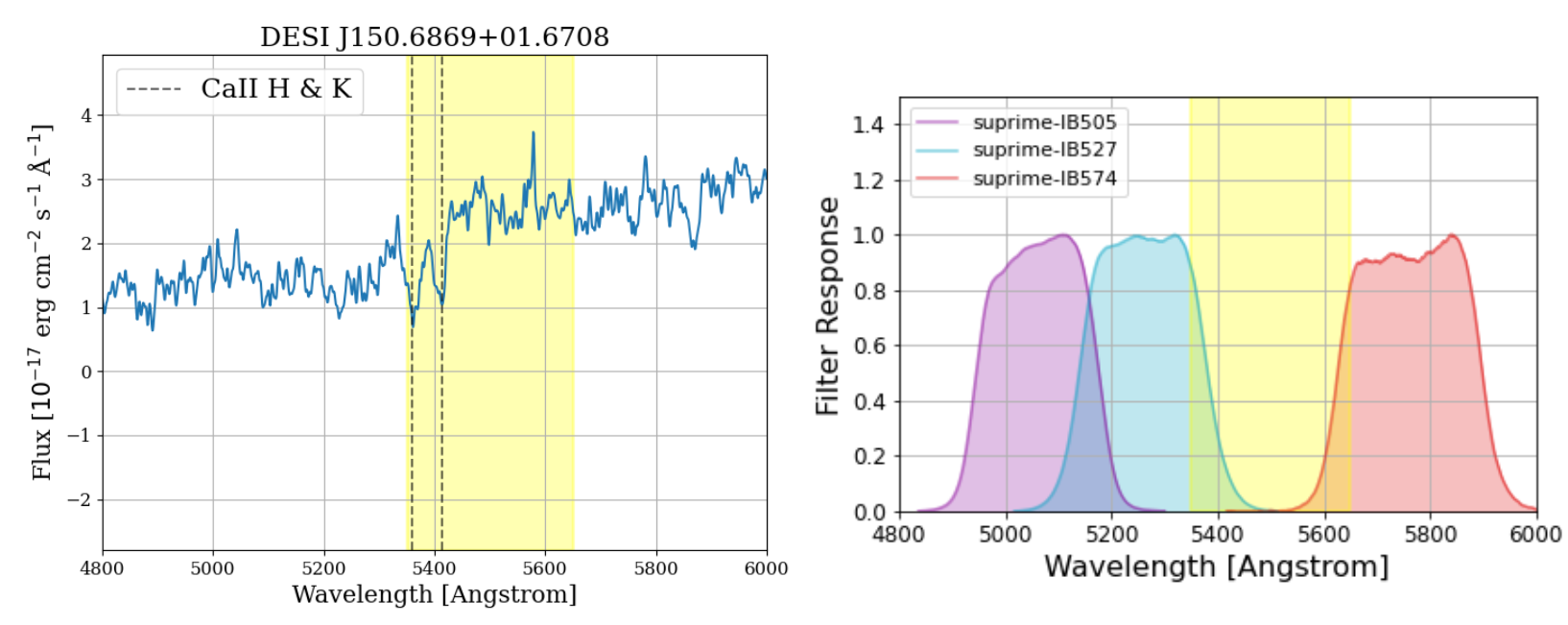}
    \hfill
    \caption{\textbf{Left}:  Spectrum of a galaxy with $z_{spec} = 0.362$ and a photometric redshift estimate at $z_{phot} = 0.4299$. The black dashed lines display strong Ca II H and K absorption is present indicating a confident spectroscopic redshift. The prominent absorption features are located around 5400 \AA. The \texttt{DESINAME} of the public spectrum is located in the title of the left panel. The spectrum was observed during SV3 during the dark program. \textbf{Right}: Intermediate band filter response curves used for photometric redshift estimations from COSMOS2020 \citep{Weaver22}. In both panels, the gap in intermediate band coverage is highlighted in yellow.}
    \label{fig:photz_filter} 
\end{figure*}

Examining the comparison of the full sample in Figure~\ref{fig:photz_me}, there is a significant feature around spectroscopic redshift of z = 0.34, where there is notable disagreement between the redshift measurements. Inspecting a random sample of spectra from this `cloud', the spectra exhibit strong Ca \rm{II} H and K absorption and a strong continuum break around 5400 \AA\ in the observer frame. 
An example spectrum is shown in Figure~\ref{fig:photz_filter}. As shown in the right hand panel of Figure~\ref{fig:photz_filter}, there is a gap in the intermediate band filters\footnote{\url{http://svo2.cab.inta-csic.es/theory/fps/index.php?mode=browse&gname=Subaru&gname2=Suprime&asttype=}} covering roughly 5400 \AA\ to 5600 \AA.  When the Ca II H and K feature is located in this gap, the photometric redshift estimates tend towards higher redshifts that place the Ca II H and K break at the longer wavelengths in the gap.

\begin{figure*}[h]
    \centering
    \includegraphics[width=\linewidth]{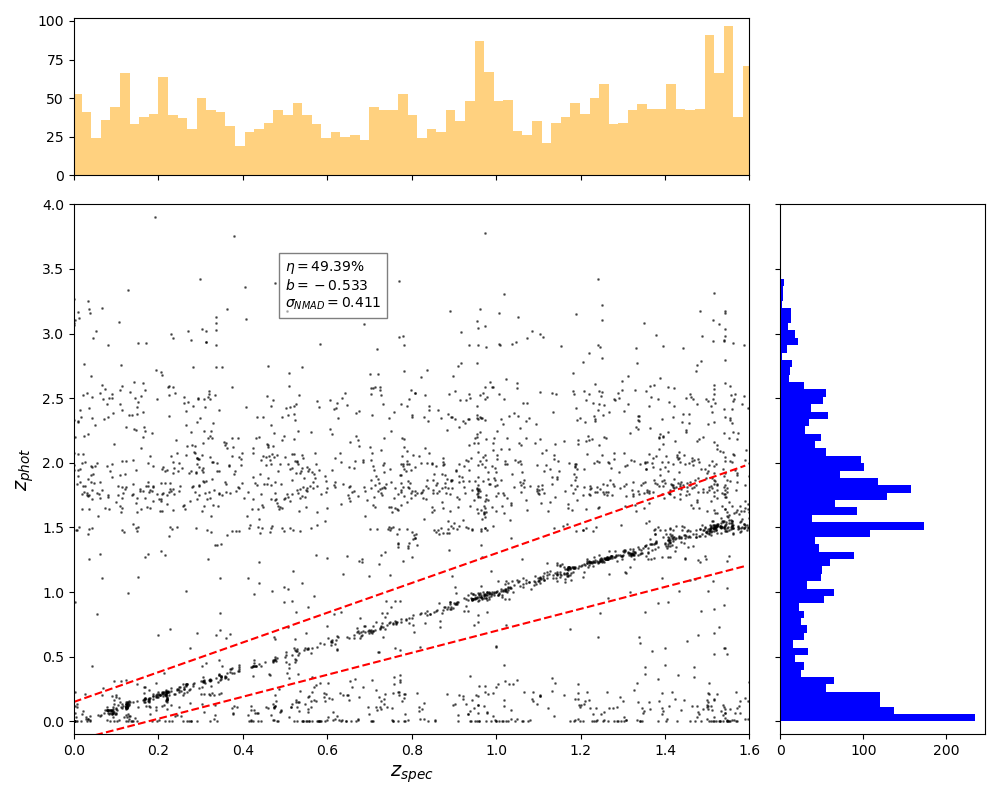}
    \hfill
    \caption{Photo-z comparison of objects with poor quality (\texttt{QUALITY\_Z = False} )spectroscopic redshifts in the DESI-COSMOS catalog. The objects outside the red dashed lines deviate from the spectroscopic redshift by more than 15\%. The top histogram shows the distribution of spectroscopic redshifts. The right side panel displays the distribution of photometric redshifts. The outlier fraction ($\eta$), bias ($b$), and $\sigma_{NMAD}$ are also given.}
    \label{fig:bad_spec_calib} 
\end{figure*}

In addition to comparing photometric redshift estimates to objects with quality spectroscopic redshifts, we also examined the photometric redshift estimates of galaxies that have poor quality spectroscopic redshifts (\texttt{QUALITY\_Z} = False). Figure~\ref{fig:bad_spec_calib} shows the comparison of the \texttt{lp\_zBEST} estimates to those DESI spectroscopic redshifts.  Not surprisingly, we find a twenty-fold increase in $\eta$, a large bias, and an intrinsic scatter that is forty-times larger than the case where DESI redshifts were reliable. Three distinct features are apparent in the comparison. First, a large fraction of photometric redshifts are above $z_{phot}>1.6$, which is beyond the range of DESI spectral templates for galaxies. It is likely that these photometric redshift estimates are accurate but DESI lacks the wavelength coverage to properly determine the redshift. Second, a significant number of galaxies have photometric redshift estimates consistent with zero or significantly below the DESI spectroscopic redshift estimate. Finally, a significant fraction of galaxies agree in the redshift estimates, implying the DESI redshifts are  accurate even though we have low confidence in classification. It would be valuable to consider these objects that could not be classified by DESI in future photometric redshift training.

\subsection{Galaxy Cluster Validation}
In combination with galaxy cluster finding algorithms, galaxy redshifts can be used to spectroscopically confirm galaxy clusters, estimate cluster mass, and characterize member galaxies. 

As an example, clusters in the SPIDERS cluster catalog \citep{SPIDERS_Cluster} were identified using CODEX \citep{codex}, an algorithm that identifies galaxy clusters from their X-Ray properties. The SPIDERS \citep{clerc2016} spectroscopic followup program was conducted within SDSS to obtain redshifts for galaxy cluster members to validate cluster membership and location, including observations in the XMM-LSS field. These SPIDERS observations were also used to estimate galaxy cluster mass \citep{gal_clus_mass}. 

\begin{figure}[h]
    \centering
    \includegraphics[width=\linewidth]{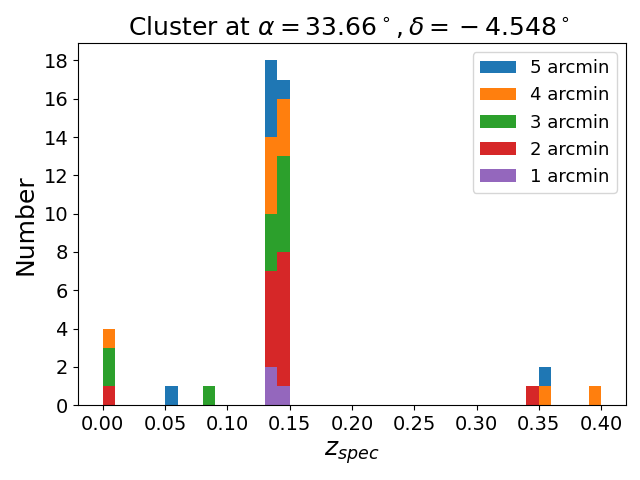}
    \caption{The redshift distribution of galaxies as a function of radial separation from the center of SPIDERS Cluster 1\_24090. We report the number of galaxies in redshift intervals of $\Delta z = 0.01$, varying the radii from 1 (purple) to 5 (blue) arcmin around the cluster center at $\alpha = 33.66^\circ, \delta = -4.548^\circ$. A clear cluster is identified with a mean redshift of $z = 0.1398 \pm 0.002$, matching the spectroscopic redshift of the cluster from external spectroscopic catalogs.}
    \label{fig:cluster}
\end{figure}

With a large spectroscopic sample in the XMM-LSS field, the DESI-XMMLSS catalog can be used to supplement the information about galaxy clusters in the SPIDERS cluster catalog and othe cluster catalogs, and potentially identify new clusters at higher redshift. To demonstrate the utility of the DESI-XMMLSS catalog, we selected one cluster with a high number of spectroscopically confirmed members from SDSS ($N_{spec}$ = 30) located in the XMM-LSS footprint. We report the number of DESI spectroscopically confirmed galaxies as function of redshift in the vicinity of the X-ray cluster center.  Using the DESI-XMMLSS catalog, we clearly identify 35 galaxies within 5 arcmin as likely members of the cluster, with a mean redshift of $z = 0.1398 \pm 0.002$, as shown in Figure~\ref{fig:cluster}. Our estimate of the cluster's redshift matches the spectroscopically validated redshift of the cluster at $z = 0.1398 \pm 0.0006$ from \cite{SPIDERS_Cluster}. Further studies into the redshifts of these cluster galaxies can inform our understanding galaxy cluster dynamics but are outside the scope of this paper.

\section{Summary}
\label{summary}

This paper presents the DESI spectroscopic redshift catalogs for the DESI-COSMOS and DESI-XMMLSS fields. We developed robust spectroscopic redshift classification based on the spectroscopic properties of each target class in these two fields. In combination, the two catalogs present spectroscopic redshifts for more than 304,000 unique objects across two 16 deg$^2$ fields with over 257,000 quality redshifts.  The spectroscopic redshifts in this catalog are highly robust, yielding less than 1\% catastrophic failure rates across various samples. These samples include, but are not limited to, blue-star-forming galaxies, passive red galaxies, active galactic nuclei, Type Ia supernovae hosts, low redshift dwarf galaxies and many other target selections. Lastly, we showed the utility of these heterogeneous samples through broad applications to completeness studies, spectroscopic and photometric redshift validation, and galaxy cluster confirmation.

As of the writing of this paper, DESI is progressing into its fourth year of observations, with regular data releases planned over the coming years. Future data from DESI releases will include spectra and redshifts for more galaxies, stars and quasars in these fields. In addition, the existing data on LAEs and LBGs will continue to be studied and characterized. The additional observations and LBG/LAE characterizations in these fields will be appended to these catalogs regularly as these data are processed and assessed for redshift robustness. The data model and redshift robustness procedures will minimally change as new data is added. In this way these catalogs will serve as a living record of DESI observations that will continue to serve as a resource for the DESI collaboration and the broader astronomical community.

\section*{ACKNOWLEDGEMENTS}
The work of Joshua Ratajczak and
Kyle Dawson was supported in part by U.S. Department of Energy, Office of Science, Office of High Energy
Physics, under Award No. DESC0009959.

This material is based upon work supported by the U.S. Department of Energy (DOE), Office of Science, Office of High-Energy Physics, under Contract No. DE–AC02–05CH11231, and by the National Energy Research Scientific Computing Center, a DOE Office of Science User Facility under the same contract. Additional support for DESI was provided by the U.S. National Science Foundation (NSF), Division of Astronomical Sciences under Contract No. AST-0950945 to the NSF’s National Optical-Infrared Astronomy Research Laboratory; the Science and Technology Facilities Council of the United Kingdom; the Gordon and Betty Moore Foundation; the Heising-Simons Foundation; the French Alternative Energies and Atomic Energy Commission (CEA); the National Council of Humanities, Science and Technology of Mexico (CONAHCYT); the Ministry of Science, Innovation and Universities of Spain (MICIU/AEI/10.13039/501100011033), and by the DESI Member Institutions: \url{https://www.desi.lbl.gov/collaborating-institutions}. Any opinions, findings, and conclusions or recommendations expressed in this material are those of the author(s) and do not necessarily reflect the views of the U. S. National Science Foundation, the U. S. Department of Energy, or any of the listed funding agencies.
The authors are honored to be permitted to conduct scientific research on I'oligam Du'ag (Kitt Peak), a mountain with particular significance to the Tohono O’odham Nation.

\section*{DATA AVAILABILITY}
All data points shown in the published graphs are available in the public VACs found here\footnote{\url{https://data.desi.lbl.gov/public/papers/c3/cosmos-xmmlss/}}. Spectra shown are available here\footnote{\url{https://data.desi.lbl.gov/public/dr1/spectro/redux/iron/healpix/}} as part of the public DESI data release. \texttt{DESINAME} and observational program for each spectrum is included in the figure caption for traceability. Supplemental data for selected plots are publicly available at
\url{https://doi.org/10.5281/zenodo.16808719}.

\newpage

\newpage

\onecolumngrid

\appendix

\section{Target Selections}
\label{appendix}
 
The algorithms for targeting various classes of galaxy and quasar targets described in \S\ref{target} and several additional classes of object for calibration are all recorded in the various DESI\_TARGET bits. A brief description of those objects, taken directly from their associated targeting papers, is found below with a full description found in the cited papers.\\

\textbf{PRIMARY TARGETS:} The following targets are DESI's primary targets with a brief description of their targeting algorithm outlined below:
\begin{itemize}
    \item \textbf{LRG} \citep{zhou_2023}: The LRG targets are selected using optical photometry in the $grz$ bands and near-infrared photometry in the WISE $W1$. The LRG selection cuts for the South are:

        \begin{itemize}[label=\textbullet]
            \item $z_{\rm{fiber}} < 21.60$ 
            \item $z - W1 > 0.8 \cdot (r - z) - 0.6$
            \item $g - W1 > 2.9$  OR  $r - W1 > 1.8$
            \item $r - W1 > 1.8 \cdot (W1 - 17.14)$ AND
            \item $r - W1 > W1 - 16.33)$ OR $r - W1 > 3.3$
        \end{itemize}
        
        where $g, r, z$, and $W1$ are magnitudes and $z$-fiber is the $z$-band fiber magnitude, i.e., the magnitude corresponding to the expected flux within a DESI fiber.
        
        The photometry in the North is slightly different from the South, and the selection cuts are tuned to match the number density and the redshift distribution in the South. The cuts for the North are:
        
        \begin{itemize}[label=\textbullet]
            \item $z_{\rm{fiber}} < 21.61$ 
            \item $(z - W1) > 0.8 \cdot (r - z) - 0.6$
            \item $(g - W1) > 2.97$  OR  $(r - W1) > 1.8$
            \item $(r - W1) > 1.83 \cdot (W1 - 17.13)$ AND
            \item $(r - W1) > W1 - 16.31)$ OR $(r - W1) > 3.4$
        \end{itemize}

    \item \textbf{ELG} \citep{Raichoor_2023}: The Main Survey ELG selection cuts are of three kinds: (1) quality cuts to ensure that the photometry is reliable; (2) a cut in the $g$-band fiber magnitude; (3) a selection box in the $g - r$ vs. $r - z$ diagram. Likewise there is an extension of cuts for the\textbf{\texttt{ ELG\_VLO}} and \textbf{\texttt{ ELG\_LOP}} classes summarized below in Table~\ref{elg_cuts}.

\begin{table}
    \setlength{\tabcolsep}{4pt}
    \begin{tabular}{|l|l|l|}
    \hline
    Sample & Cut & Comment \\
    \hline
    \textbf{ELG} & brick primary = True & Brick primary selection \\
    & $nobs_{grz} >0$ & Observed in $grz$-bands \\
    & $flux_{grz} \cdot \sqrt{flux\_ ivar_{grz}} > 0$ & Positive SNR in $grz$-bands \\
    & $(maskbits \&2^1) = 0$, $(maskbits \&2^{12}) = 0$, $(maskbits \&2^{13}) = 0$ & Not close to bright star/galaxy \\
    \hline
    \textbf{ELG\_LOP} & $(g > 20)\ and\ (g_{fib} < 24.1)$ & Magnitude constraints \\
    & $0.15 < r - z$ & Color condition 1 \\
    & $(g - r) < 0.5 \cdot (r - z) + 0.1$ & Color slope condition 1 \\
    & $(g - r) < -1.2 \cdot (r - z) + 1.3$ & Color slope condition 2 \\
    \hline
    \textbf{ELG\_VLO} & $(g > 20)\ and\ (g_{fib} < 24.1)$ & Magnitude constraints \\
    & $0.15 < (r - z)$ & Color condition 1 \\
    & $(g - r) < 0.5 \cdot (r - z) + 0.1$ & Color slope condition 1 \\
    & $(g - r > -1.2 \cdot (r - z) + 1.3)\ and\ (g - r < -1.2 \cdot (r - z) + 1.6)$ & Color slope condition 2 \\
    \hline
    
    \end{tabular}
\caption{The cuts are the same for the North and South-DECaLS/DES regions. We use the following definitions: \textbf{${grz} = 22.5 - 2.5 \cdot log10($flux\_{grz}/mw\_transmission\_{grz}), $g_{fib} = 22.5 - 2.5 \cdot log10($fiberflux\_g/mw\_transmission\_g)}. The \textbf{\texttt{brick primary, nobs\_{grz}, flux\_{grz}, fiberflux\_g, flux\_{grz}, mw\_transmission\_{grz}, maskbits}} columns are described here: https://www.legacysurvey.org/dr9/catalogs/}
\label{elg_cuts}
\end{table}
    
\textbf{ELG\_HIP},  is a ten percent random subsampling of the \textbf{\texttt{ ELG\_LOP}} and \textbf{\texttt{ELG\_VLO}} samples, with the same fiber assignment priority as one of the LRG targets
    
    \item \textbf{QSO} \citep{Chaussidon_2023}:The DESI quasar target selection is a combination of optical-only and optical+IR colors.  Two colors are used, $grz -W$ vs. $g - z$ where $grz$ is a weighted average of the $grz$ band fluxes and W a weighted average of $W1$ and $W2$ fluxes:

        \begin{equation}
            \texttt{flux}_{grz} = [\texttt{flux}_{g} + 0.8 \cdot \texttt{flux}_{r} + 0.5 \cdot \texttt{flux}_{z}] / 2.3
        \end{equation}
        \begin{equation}
            \texttt{flux}_{W} = 0.75 \cdot \texttt{flux}_{W1} + 0.25 \cdot \texttt{flux}_{W2}
        \end{equation}
     To improve the success rate for DESI, we use a machine-learning algorithm based on Random Forests (RF).we restrict the selection to objects with stellar morphology (’PSF’ in DR9), to avoid an almost 10-fold contamination by galaxies that otherwise enter our selection region, and we impose $16.5 < rAB < 23.0$. In addition, to reject stars, we apply a cut on the (WISE) magnitudes ($W1 < 22.3$ and $W2 < 22.3$). This cut is particularly efficient at getting rid of stars in the Sagittarius Stream, a region which exhibits an overdensity of QSO targets. We also require that the targets are not in the vicinity of bright stars, globular clusters, or large galaxies. Such “masked” sources have \textbf{\texttt{MASKBITS}} of 1, 12 or 13 set in Legacy Surveys catalogs.

    \item \textbf{BGS} \citep{Hahn_2023}:
        \begin{itemize}
        \item \textbf{BGS\_ANY}:
            The \textbf{\texttt{ BGS\_ANY}} sample is any object selected from the BGS criterion outlined below.
            First a spatial mask is applied. masks are compiled using 773,673 Gaia DR2 objects with $G_{\rm{Gaia}} < 13$ and 3,349 Tycho-2 objects with visual magnitude brighter than \textbf{\texttt{MAG\_VT}} $<$ 13. The masks have radii: 
            \begin{equation}
                R_{BS}(m) = 815 \cdot 1.396^{-m} arcsec
            \end{equation}

            where m is either $G_{\rm{Gaia}}$ or Tycho-2 \textbf{\texttt{MAG\_VT }}magnitude. We use $G_{\rm{Gaia}}$ when both are available.We do not apply spatial masking around stars fainter than 13th magnitude. Next, for GCs, we apply a circular mask with radius defined by the major axis of the object. We apply this masking around all of the GCs and PNe in the list compiled from the OpenNGC catalog. In total, our bright star and GC masks exclude 0.87 and 0.01\% of the initial area, respectively.

            For the BGS targets, we only want to include galaxies and exclude stars. An object is considered a BGS target if either of the following conditions are met:
            \begin{enumerate}
                \item object is not in the Gaia catalog;
                \item object is in Gaia and has $(G_{\rm{Gaia}} - r_{\rm{raw}}) > 0.6$.
            \end{enumerate}
            $G_{\rm{Gaia}}$ is the G band magnitude from Gaia and $r_{\rm{raw}}$ is the LS r band magnitude that is not corrected for galactic
            extinction. Our ($G_{\rm{Gaia}} - r_{\rm{raw}}$) criterion takes advantages of the fact that $G_{\rm{Gaia}}$ is measured assuming that the object is a point source for a (narrow) diffraction-limited point spread function (PSF) measured from space.

            Some of the objects in the LS are imaging artifacts or fragments of ‘shredded’ galaxies. In order to remove these spurious objects from our BGS target catalog, we apply the fiber magnitude cut (FMC) below:

           \begin{equation}
                r_{\rm{fiber}} = 
                \left\{
                    \begin{array}{lr}
                        22.9 + (r - 17.8) , \text{for} \hspace{.2em} r < 17.8\\
                        22.9, \text{for }  17.8 < r < 20
                    \end{array}
                \right\}
            \end{equation}

            $r_{\rm{fiber}}$ is the $r$-band fiber magnitude derived from the predicted $r$-band flux of the object within a 1.500 diameter fiber and
            $r$ is the total $r$-band magnitude

            We want BGS Bright to be complete in all three optical bands of its imaging: $g$, $r$, and $z$. We therefore require that there is at least one photometric observations in each of the bands:

                \begin{equation}
                    nobs_{i} > 0\ for\ i\ = g, r, z
                \end{equation}

            $nobs_{i}$ represents the number of observations (images) at the central pixel of the source in each band. We also exclude spurious objects (e.g. imaging artifacts or stars) with extreme colors by requiring:

                \begin{equation}
                    -1 < (g - r) < 4
                \end{equation}
                \begin{equation}
                    -1 < (r - z) < 4
                \end{equation}

            Lastly, flux from very bright objects on neighboring fibers can contaminate the traces of faint objects on the spectrograph CCD, and pollute their observed fluxes. This is particularly problematic at $ > 9000$ \AA, where 10\% of the flux of the contaminating source is scattered into the wings of its PSF. Since the faintest fiber magnitudes in BGS have $r_{\rm{fiber}} \approx 21.5$, we remove all objects that meet

            \begin{equation}
                (r > 12)\ \& \ (r_{\rm{fibertot}} < 15)
            \end{equation}

            from the BGS target catalog. Here, $r_{fibertot}$ is the total fiber magnitude derived from the predicted $r$-band flux within a 1.500diameter fiber from all sources.

        \item \textbf{BGS\_ BRIGHT}:
        For \textbf{\texttt{ BGS\_BRIGHT}} targets we impose  a simple $r < 19.5$ magnitude limit to select the BGS Bright targets.

        \item \textbf{BGS\_FAINT}:
        BGS includes galaxies with magnitudes fainter than $r > 19.5$. This fainter sample will substantially increase the overall BGS target density and, thus, enable small-scale clustering measurements with higher signal-to-noise. The selection for these is following $r_{\rm{fiber}}$-color cut:

        \begin{equation}
                r_{\rm{fiber}} <
                \left\{
                    \begin{array}{lr}
                        20.75,\ \text{if}\ (z - W1) - 1.2(g - r) + 1.2 < 0\\
                        21.5,\ \text{if}\ (z - W1) - 1.2(g - r) + 1.2 \geq 0
                    \end{array}
                \right\}
            \end{equation}

            Here, $W1$ is the magnitude in the WISE $W1$ band, 3.4µm at 6.100 angular resolution. We also include a $19.5 < r < 20.175$ magnitude limit in order to satisfy a $\sim1,400$ targets/deg$^2$ constraint on the total target density imposed by the survey fiber budget allocated to BGS.

        \item \textbf{BGS\_WISE}:
        To select targets with AGN, we exploit optical and infrared colors that trace the signatures of hot, AGN-heated dust in the spectral energy distribution. The primary AGN selection criteria are:
            \begin{itemize}
                \item $(z - W2) - (g - r) > -0.5$
                \item $(z - W1) - (g - r) > -0.7$
                \item $W1 - W2) > -0.2$
                \item $(G_{\rm{Gaia}} - r) < 0.6$
            \end{itemize}
        We also require a SNR $> 10$ detection in both $W1$ and $W2$ bands to ensure a robust constraint in the infrared regime, and apply quality and magnitude cuts.
        
        \item \textbf{BGS\_FAINT\_HIP}:
        In the DESI catalogs these \texttt{BGS\_FAINT} targets with higher priority are labeled under the \texttt{BGS\_FAINT\_HIP} bitname.
        \end{itemize}
    \item \textbf{MWS} \citep{Cooper_2023}:

        \begin{itemize}
            \item \textbf{MWS\_ ANY}:
            The MWS main sample is selected from the LS DR9 source catalog combined with astrometric measurements from Gaia EDR3. The following definition encompases any target with the \texttt{MWS\_ANY} bit:

            \begin{itemize}
                \item $16 < r < 19$
                \item $r_{\rm{obs}} < 20$
                \item \texttt{type = PSF}
                \item \texttt{gaia\_astrometric\_excess\_noise} $< 3$
                \item \texttt{gaia\_duplicated\_source} = False
                \item \texttt{brick\_primary} = True
                \item \texttt{nobs\_\{g,r\}} $> 0$
                \item \texttt{\{g,r\}\_flux} $> 0$
                \item \texttt{fracmasked\_\{g,r\}} $< 0.5$
            \end{itemize}

            The observed $r$-band magnitude is obtained from the LS flux in nanomaggies as $r_{\rm{obs}} = 22.5 - 2.5\log_{10}(r_{flux})$. The extinction-corrected magnitude is computed as $r = r_{\rm{obs}} + 2.5log10 $(\texttt{mws\_transmission\_r})

            \item \textbf{MWS\_MAIN\_BLUE}
            The \texttt{MAIN\_BLUE} subsample is approximately magnitude-limited selection, comprised mainly of main-sequence turnoff stars and bluer subgiants and defined by the color criterion:

                \begin{equation}
                     g - r < 0.7
                \end{equation}
            
            with $g$ defined in the same way as $r$ above.

            \item \textbf{MWS\_MAIN\_RED}
            The \texttt{MWS\_MAIN\_RED} subsample is defined by the following additional criteria to the  \texttt{MAIN\_ANY} criteria :

            \begin{itemize}
                \item $g - r \geq 0.7$
                \item \texttt{astrometric\_params\_solved} $\geq 31$
                \item Gaia parallax $\pi < 3\sigma_{\pi} + 0.3$ mas
                \item Gaia proper-motion: $|\mu| < 5\sqrt{f_{r}/f_{19}}$ mas/yr \ $(f_{r}=10^((22.5-r)/2.5)$
            \end{itemize}

            \item \textbf{MWS\_MAIN\_BROAD}

            The \texttt{MWS\_MAIN\_BROAD} sample comprises stars with color $g - r \geq 0.7$ that satisfy the same magnitude and data quality requirements as the other two categories, but do not satisfy one or more of the astrometric criteria that define main-red (stars that do not have well-measured astrometric parameters in the Gaia catalog are therefore included in this selection).

            \item \textbf{MWS\_WD}

            The \texttt{MWS\_WD} has an extensive selection summarized here. White dwarfs are selected using a set of BP - RP color and $G$-band absolute magnitude criteria based on \cite{fusillo_19} that identify the white dwarf cooling sequence in the Gaia photometric catalog alone. This selection is applied to the LS catalog using the properties of cross-matched Gaia EDR3 sources; the LS photometry is not used in this selection. Photometric measurements are taken from Gaia DR2 and astrometric measurements from EDR3.
            
             \begin{itemize}
                 \item $G_{abs} > 5$
                 \item $BP - RP < 1.7$
                 \item $G_{abs} > 5.93 + 5.047(BP-RP)$
                 \item $G_{abs} > 6(BP - RP)^{3} - 21.77(BP - RP)^{2} + 27.91(BP - RP) + 0.897$
             \end{itemize}

             These criteria are applied to a sample defined by extinction-corrected Gaia G flux (\texttt{gaia\_phot\_g\_mean\_mag} in the LS catalog) at high latitude, with mild parallax and proper-motion cuts to select against nearby luminous blue stars (early-type main-sequence stars, horizontal branch stars, and subdwarfs) as well as QSOs:

             \begin{itemize}
                 \item $G < 20.0$
                 \item $|b| > 20$ degrees
                 \item $\pi/\sigma_{\pi} > 1$ mas
                 \item \texttt{astrometric\_params\_solved} $\geq 31$
                 \item $|\mu| > 2$ mas/yr
             \end{itemize}

             We impose the following photometric quality cut, because failing this criterion results in poor astrometry:

            \begin{itemize}
                \item  \texttt{phot\_bp\_rp\_excess\_factor} $< 1.7 + 0.06(BP - RP)^{2}$
            \end{itemize}
            However, we retain objects that have reliable parallaxes and significant proper-motions that meet either of the following criteria:

            \begin{itemize}
                \item \texttt{astrometric\_sigma5d\_max} $< 1.5$
                \item \texttt{(astrometric\_excess\_noise} $< 1)\ \& \ (\pi/\sigma_{\pi} > 4)\ \& \ (|\mu| > 10$ mas/yr)
            \end{itemize}

            \item\textbf{MWS\_RR\_LYRAE}
    
                We target Gaia DR2 sources with magnitudes $14 < G < 19$ that are classified as RR Lyrae by the general variability classifier and the Special Object Studies classifier \citep{holl_2018,clementini_19}, comprising all objects from the table \texttt{vari\_rrylrae} and any sources from the table \texttt{vari\_classifier\_result} for which \texttt{best\_class\_name} includes ”RR”.

            \item \textbf{MWS\_NEARBY}
            For \texttt{MWS\_NEARBY} targets, we select dwarf stars within 100 pc of the Sun based only on apparent Gaia magnitude and parallax, allowing for moderate parallax uncertainties:

            \begin{itemize}
                \item $16 < G < 20$
                \item $\pi + \sigma_{\pi} > 10 mas$
                \item \texttt{astrometric\_params\_solved} $\geq 31$
            \end{itemize}

            \texttt{MWS\_NEARBY} targets are prioritized over all other MWS targets except \texttt{MWS\_WD} and \texttt{MWS\_RR\_LYRAE}.

            \item \textbf{MWS\_BHB}
            BHBs are selected starting from the basic definition of main sample stars given at the start of this section (common to main-blue and main-red), with the following additional criteria:

            \begin{itemize}
                \item $G > 10$
                \item $\pi \leq 0.1 + 3\sigma_{\pi} mas$
                \item $-0.35 \leq g - r \leq -0.02$
                \item $-0.05 \leq X_{BHB} \leq 0.05$
                \item $r - 2.3(g - r) -W1_{faint} < -1.5$
            \end{itemize}
            
            These criteria exclude nearby stars and select around the BHB locus in a combined LS and WISE color space, defined by:
            
            \begin{equation}
            X_{BHB} = (g - z) - [1.07163(g - r)^{5} - 1.42272(g - r)^{4} \\
                      +0.69476(g - r)^{3} - 0.12911(g - r)^{2}+0.66993(g - r) - 0.11368]
            \end{equation}
            
            and
            
            \begin{equation}
                W1_{faint} = 22.5 - 2.5\log_{10}(W1 - 3\sigma_{W1})
            \end{equation}
            
             where W1 and $\sigma_{W1}$ are the WISE 3.4µm flux and its error, respectively.
        \end{itemize}
\end{itemize}

\textbf{SECONDARY TARGETS:} The following targets are DESI's secondary targets with a brief description of their targeting algorithm outlined below. These targets are flagged as \texttt{SCND\_ANY} bit in the main desi mask and can be found in the scnd mask:

\begin{itemize}
    \item \textbf{UDG}:
    Targets are selected from the SMUDGes catalog of the DECaLS data. SMUDGes analyzes the DECaLS data images to identify low surface, diffuse galaxies (central surface brightness $>$ 24.0 mag/sq. arcsec and half light radii $>$ 5.3 arcsec). The procedure involves various steps described in Zaritsky et al. 2019 and in another paper in progress. The specific subsample presented here are those ultradiffuse galaxies in the SMUDGes catalog that have $g-r < 0.3$, making it more likely that they have emission lines from which measuring the redshift would be simplified. Some of these sources are already included in the BGS Bright and Faint samples, but many are not (we set override = 0 so as to not affect targeting for those galaxies previously identified).

    \item \textbf{QSO\_RED}: A selection of dust-reddened quasars with selection using DR8 quasars as a base sample:
    \begin{itemize}
        \item $r$-band magnitude cut ($r < 23$): ensures high SN sources observed
        \item require \texttt{MORPHTYPE==‘PSF’}: to select QSO-like sources
        \item remove sources within the main DESI survey color selection ($- 0.4 < r - z < 1.1 \ \& \ g - r < 1.3$): this reduces the chance of re-observing a blue QSO that will already be targeted in the main survey
        \item require W1, W2 and W3 detections and a $SNR > 3$ in all three bands
        \item remove sources that do not satisfy the \cite{mateos_2012} WISE AGN wedge using the W1~-~W2 and W2~-~W3 colours: this important step significantly reduces the number of non-QSO contaminants
    \end{itemize}

    \item \textbf{FIRST\_MALS}:These targets are faint radio sources that have 21-cm and OH absorption spectra from MALS generated by cross-matching the radio source positions with DR8 source locations with a positional accuracy of 1 arcsecond. Initially we are focusing on MALS points with RA in the range $120 - 210$ deg and DEC in the range $-10$ to $+20$ deg. There are 85 MALS pointing overlapping with DESI foot prints in the above mentioned RA range. Within the field of view of MALS observations we have identified 19107 radio sources with peak flux density of $>10$ mJy using FIRST survey. The flux limit is set by the SNR requirements from the 21-cm absorption line searches. While cross-matching with DR8 catalog of DECALS we have identified 8300 sources that are listed in the above file. We just want to mention that about 10\% of our MALS FOV have poor coverage in the FIRST survey.

    \item \textbf{MWS\_CLUS\_GAL\_DEEP}:  We have 3 different classes of objects here, dwarf galaxies, globular clusters and open clusters. Since they have different properties, we select the targets with different criteria, detailed below
    \begin{itemize}
        \item  Dwarf Galaxies (DC):

        Two dwarf galaxies are selected for SV, Draco and Ursa Major II (Uma II).
        
        We select field centers to be $(ra+0, dec+0.4)$ where ra/dec are the center of the galaxies. This offset is chosen so that the dwarf center fall on one of DESI petal instead in the center of the field (to avoid possible fiber gaps). We first select targets from DECaLS DR8.Then targets are selected with 1.7 deg from the field center, with brick\_primary = True and type=`PSF'. The later is to ensure the targets are stellar like object. All photometry used for selection are dereddened using SFD98. We first select targets have $r > 16$ and $g-r < 1.2$.We then select targets with two passes:
        
        First pass, for stars that have a proper motion (pm) and parallax (w) measurements, we select targets with:
        \begin{itemize}
            \item Proper motion cut: $|\mu-\mu_{0}| < 2$ where $\mu_{0}$ is the proper motion of the cluster
            \item Parallax cut: $\pi - 3 \cdot \sigma_{\pi} < 1/distance $ 
            \item Stellar quality cut: (\texttt{gaia\_astrometric\_excess\_noise} $< 1)$ \ \& (\texttt{gaia\_phot\_bp\_rp\_excess\_factor}$<1.3 + 0.06 \cdot$(\texttt{gaia\_phot\_bp\_mean\_mag-gaia\_phot\_rp\_mean\_mag}$)^{2})$
        \end{itemize}

        Second pass, for stars that have no PM or parallax measurements, we select targets with DECaLS photometry. Specifically we select targets using a metal poor isochrone to select stars are consistent to isochrone within certain width. Finally we combine the two passes and put stars with $G > 18$ as high priority targets and $G < 18$ as low priority targets. (Because $G < 18$ stars should mostly observed by regular MWS SV). We have a total of 4079 targets for Draco and 2306 targets for Uma II.

        \item Globular Clusters(GC):
        
        Two GC are selected, Pal 13 and M 53. M 53 field has two GC, M 53 and NGC 5053, which can be observed together with one DESI field.

        Selection for GC is very similar to DG, with the only exception is:

        \begin{enumerate}
            \item In the photometry cut, we used $r > 16$ and $g-r < 1.1$
            \item In first pass, proper motion cut is $|\mu-\mu_{0}| < 3$
        \end{enumerate}
        
        We have a total of  2741 targets in Pal 13 and 11994 targets in M53. Note that although there are a lot of targets in this field, if we remove the crowded cluster core, the target number drops to 4027 in total. Therefore, the outskirt of the GCs will still be able to observed by DESI. 

        \item Open clusters (OC):
        Fields centered around the central cluster coordinates using DECALS (and the Gaia astrometry/photometry in the DECALS files).

        For each cluster, two catalogs are created. One with high priority containing astrometric cluster members, and a second one with targets that match the g-r and r-z colors as a function of g magnitude.

    \end{itemize}

    \item \textbf{LOW\_MASS\_AGN}:
    We propose to target AGN candidates selected from the DESI tractor photometry in the z and WISE bands, in order to search for AGN in low-mass galaxies.
    The target selection steps are as follows:
    
    Consider sources with the following conditions:
    \begin{itemize}
        \item \texttt{brick\_primary = 1, maskbits = 0,  wisemask\_w1 = 0 and wisemask\_w2 = 0,  anymask\_g = 0 and anymask\_r = 0 and anymask\_z = 0}
        \item  0.02 $\leq$ \texttt{z\_phot\_median} $\leq$ 0.3

        \item \texttt{flux\_X} and \texttt{flux\_ivar\_X} columns selected where X = g,r,z,w1,w2,w3,w4

        \item  \texttt{snr\_X} = \texttt{flux\_X}$\cdot \sqrt{\texttt{flux\_ivar\_X}}$ for all the filters is calculated.

        \item Considering sources that are detected in $g,r,z,w1,w2,$ and $w3$ bands: applying snr $\geq$ 3 in these filters. Flux is corrected for reddening using \texttt{mw\_transmission\_X } column values: \texttt{flux\_dereddened\_X = flux\_X/mw\_transmission\_X}. Dereddened Magnitudes in the AB system computed as \texttt{m\_X} = -2.5$\cdot$log10(\texttt{flux\_dereddened\_X}) + 22.5. Removing sources with confirmed spectroscopic redshift i.e., where `survey' column from photometric redshift catalog = `BOSS' or `SDSS' or `eBOSS'. Removing all the sources with non-zero proper motion in ra and dec. Keeping the sources with:
        pmra = 0 and pmdec = 0

        \item Removing extreme color sources. Keeping only sources with:
        \begin{itemize}
            \item $-2.0 \leq z - W1 \leq 2.0$
            \item $ -2.0 \leq W1 - W2 \leq 2.0$
            \item  $-2.0 \leq W2 - W3 \leq 4.0$
        \end{itemize}

        \item   Computing the absolute magnitude in $z$-band, $Mz$ from $z$mag and \texttt{z\_phot\_median} or selecting low-mass faint galaxies, we make a cut in $Mz$ and $r$:

        \begin{itemize}
            \item $Mz \geq -21 $
            \item $ r \geq 19 $
        \end{itemize}

        \item Using the new WISE color-color AGN diagnostics, we make two agn candidate selections:
        \begin{itemize}
            \item AGN1 \ selection: $(W1-W2 < (0.5667\cdot(z-W1)) + 0.5)$  and  $(W1-W2 > (-1.5714\cdot(z-W1))+0.5)$  and  $(W1-W2 > (0.4375\cdot(z-W1))-0.90625)$
            \item AGN2 \ selection: $(W1-W2 > -0.4)$  and $(W1-W2 > (0.2\cdot(W2-W3)) - 0.6)$ and  $(W1-W2 > (4\cdot(W2-W3)) - 8.2)$
        \end{itemize}

        \item We define three priority classes:
        \begin{itemize}
            \item High: Candidates that satisfy both agn1 and agn2 selection and have $W3 \leq 17$.
            \item Medium: Candidates that satisfy both agn1 and agn2 selection and have $W3 > 17$
            \item Low: Candidates that satisfy only of the selections, i.e., agn1 or agn2, but not both.
        \end{itemize}

    \end{itemize}

    \item \textbf{FAINT\_HPM}:
    \begin{enumerate}
        \item HPM-G : All Gaia DR2 stars with $G \geq 19$ and Gaia proper motion $>$ 100mas/yr
        \item  HPM-N : Sources from the NOIRLab Science Catalog DR2 with NSC proper motion $ >$ 100mas/yr and $(15<g<23)$ or $(15<r<23)$ or $(15<z<22))$ where g,r,z are NSC mags.
        \item HPM-BD All brown dwarf candidates from WISE, DES, Ultracool, etc.  (no proper motion cut)
    \end{enumerate}

    \item \textbf{LOW\_Z\_TIER\_1/2/3}: This program uses spare fibers in dark time to pursue a complete survey of the low-z Universe. After generating the list of objects passing the SAGA II cuts \citep{Mao_2021}, the target lists are then sorted by the following three tiers:

    Tier 1: CNN predicted objects from SAGA II cuts + SAGA II extended cuts (10\% extra in color and magnitude)
    \begin{itemize}
        \item Sorting: pCNN value (CNN probability)
        \item Density: 16 per square degree
    \end{itemize}
    
    Tier 2: SAGA II cuts outside of BGS sample ($r < 19.5 \ and \ 19.5 < r < 20.3$ + color cuts)
    \begin{itemize}
        \item Sorting: pSAT value (satellite probability from \cite{Mao_2021})
        \item Density: 123 per square degree
        \item ($r_{fib} > 23$): 3 per square degree
    \end{itemize}
    
    Tier 3: BGS sample + SAGA II extended cuts
    \begin{itemize}
        \item Sorting: random
        \item Density: 200 per square degree
        \item BGS galaxies: 72 per square degree
        \item SAGA II extended cuts: 128 per square degre
    \end{itemize}

    \item \textbf{BHB}: A complement to the DESI Milky Way (MW) Survey by observing faint Blue Horizontal Branch (BHB) and RR Lyrae stars in the dark observing time of DESI \citep{bhb}. The target selection is as follows:
    \begin{enumerate}
        \item Selection of blue sources:$ -0.3 < g - r < 0$
        \item Magnitude limit: $19 < g < 21.5$
        \item Color selection to separate BHB from BS/WD and QSO $Poly1(g - r) < r - z < Poly2(g - r)$ where Poly1 and Poly2 are quadratic polynomials
        \item WISE selection: $f_{W1}/f_{G} < 0.3(g - r) + 0.15 + \sigma[_{W1}/f_{G}]$ where $f_{W1}$ and $f_{G}$ are fluxes in Gaia G and WISE W1 and $\sigma[_{W1}/f_{G}]$ is the uncertainty on the flux ratio. This is necessary to remove residual contamination from QSOs.
        \item Quality cuts: ANYMASK G/R/Z and GALAXY maskbits are not set.
    \end{enumerate}

    As RR Lyrae are essentially pulsating counterparts to blue horizontal branch stars we also include a small number of distant RR Lyrae stars that have been identified by Gaia \citep{gaia_2019} and PS1 \citep{sesar_2017} within the same magnitude limits.

    \item \textbf{PSF\_OUT\_BRIGHT/DARK}: Program to obtain spectroscopy of all point sources of unusual color in order to find missing quasars and to be a discovery engine for new, interesting objects. We simply target any sources of type PSF that are inconsistent with the colors of main sequence stars and that are missed by other selection criteria. This includes the 18 deg$^{2}$ at $15 < r < 19$ as bright time spare fibers, and the 130 deg$^{2}$ at $16 < r < 22$ as dark time spare fibers.

    \item \textbf{HPM\_SOUM}:A selection of high proper motion stars. To make our target list, we used the list of HPM objects from \cite{segev_soum}. We calculated the current locations of all the objects in the list (using the proper motion provided). We then cross-matched the current locations with the DR8 bricks corners to make a list of targets that are currently in the DR8 footprint.

    \item \textbf{HSC\_HIZ\_SNE}:
    Our candidates are selected by the presence of a SN in the host
galaxy within the HSC COSMOS deep field \citep{2019PASJ...71...74Y}. There
is no magnitude or color cut on the targeted host but we cut at
$r=25$ for the host galaxies of less certain SNe. Internally the
targets are prioritized according to their likely scientific value
for constructing a Hubble diagram and measuring the dark energy
equation of state.  Highest priority went to host galaxies of SNe\'Ia
observed with HST as part of programs GO14808 and GO15363 (PI
Suzuki); priority then decreased in the order of: other likely
SNe\,Ia; including the HST SNe\,Ia), possible SNe\,Ia, and less
certain SNe\,Ia. We then removed the targets whose redshifts we had
previously obtained with AAOmega \citep{2006SPIE.6269E..0GS}, and
targets with redshifts from the literature. To make a more complete
Hubble diagram, we also include host galaxies for confirmed orsuspected SNe Ia lacking redshifts with the SNLS D2 deep \citep{sn_hosts3}.

    \item \textbf{SN\_HOSTS}:
    A set of targets for SNe host to understand SN/host correlations, SNe Ia cosmology, fundamental plane astrophysics and other scientific pruposes. The selection is summarized below:
    \begin{enumerate}
        \item We computed a list of transients based on spectroscopically classified transients from,ZTF, PTF/iPTF, SDSS II, Greg Aldering's sample including historical SNe, ZTF QSOs and made sure we have no duplicates
        \item For the non-QSOs sources, we ran a host identifier code which: (a)looked at possible hosts from the local universe and (b) looked at possible hosts from dr8\_north and dr8\_south.
        "Possible hosts" are selected using two criteria: (1)light radius proximity (2)redshift proximity. When the two criteria do not agree, the host is ambiguous and we request to observe both possible  hosts.
        \item For the QSOs sources, we simply give their RA and DEC
        \item We add 259 hosts from Greg Aldering's sample which disagree with our host identifications, for reasons we believe are related to DR8 and should disappear in DR9.
    \end{enumerate}

    \item \textbf{STRONG\_LENS}:Our targets are selected from strong lensing candidate systems discovered in DESI Legacy Imaging Surveys. They target the $\sim 70$ lenses not currently in the DESI target list and the brightest lensed image for every source (typically one per lensing system) to the same depth as other DESI targets (4 passes, $NUMOBS=1$) in dark time, to be revisited based upon SV observations. We will provide coordinates for all our targets.

    \item \textbf{GAL\_CLUS\_SAT}
     This sample is a volume limited complete spectroscopic sample of BCGs and the bright cluster members. The BCG is the SDSS-redMaPPer most probable central. To determine the fraction of SDSS-redMaPPer BCGs and members fibered by DESI, we do the following. We note that galaxies not fibered likely result from fiber collisions.
     \begin{itemize}
         \item We identify SDSS-redMaPPer BCGs and galaxies with $P_{mem} > 0.90$ in the DESI northern footprint.
         \item We use fiber assignments provided to BGS and C3 by Jaime Forero-Romero todetermine which galaxies will be fibered. In that simulation all tiles for the entire surveyare fed into fiberassign in a single run (i.e., we do not simulate the survey progress nor any type of cadence to update the Merged Target List (MTL)). We use DR9 targets and incorporate multiple passes. Since some galaxies may be observed during dark time, we also determine which DESI LRG targets correspond to our galaxies.
         \item For the BCGs we extend our analysis out to $z < 0.35$. For the satellite galaxies, we only go out to $z < 0.30$.
         \item In total, there are 202 BCGs, 260 second brightest galaxies, and 10787 satellite galaxies with $P_{mem} > 0.90$.
     \end{itemize}

     This sample is all the satellite galaxies (excluding the 2nd brightest) in the galaxy cluster.

    \item \textbf{MWS\_FAINT\_BLUE/RED}
    
    The faint-blue and faint-red samples extend the main-blue and main-red classes, respectively, to the fainter magnitude range $19 < r < 20$. The selection criteria are otherwise identical to those for the corresponding main classes, except for a slightly fainter limit on the observed $r$-band magnitude, $r_{\rm{obs}} < 20.1$

    \item \textbf{WISE\_VAR\_QSO}:
    QSO selection based on variability in WISE light curves. Used the combine band W for variable selection:
    \begin{itemize}
        \item $N_{epoch}>=9$
        \item $A>0.001$
        \item $(\gamma >-28.0*A+1.5)$
        \item $\chi^{2} >1.0$
    \end{itemize}

    where A is the amplitude of the light curves, $\gamma$ is the time duration exponent, and $\chi^{2}$ is the $\chi^{2}$ of the lightcurves compared to a flat distribution.

    \item \textbf{Z5\_QSO}: Proposed targets for DESI observations to carry out a systematic survey of $z \sim 5 - 6.5$ quasars to probe both the evolution of early super-massive black holes (SMBHs) and the cosmic reionization history. The selection using photometric data from only the legacy imaging survey has already discovered 7 quasars at $4.4 < z < 5.2$ during SV. 
    By adding i band photometry from Pan-STARR1 (PS1), the updated selection will improve the successful rate without significant lost of completeness. PS1 y data and J band data from public NIR surveys are also used to further reject contaminants. The updated selection is based on the DELS DR9, PS1,  DR1 \& DR2, and public
    NIR J (e.g., UHS, ULAS, and VHS) data. We build up selection criteria for quasars at redshift ranges of $z \sim 4.8 - 5.3, 5.7 - 6.4$, and $6.4 - 6.8$, respectively. The main color-color selections are the $(r - i)/(i - z)$ for $z \sim 5$ and 6, the $(i - z)/(z - y)$ for $z \sim 6$ and 6.5, and $z/y - W1/W1 - W2$ for all redshifts. Within the DESI footprint, we obtain
    $\sim 3,500$ targets for $z \sim 5$ quasar, 3,000 targets from $z \sim 6$ selection, and 450 targets for $z > 6.4$ quasars, down to a depth of z band magnitude 21.4 for $z \sim 5$ quasars and 21.5 for $z \sim 6$ and 6.5 quasars. We test our selections with both known quasar sample and simulation and estimate the completeness as $60-90\%$ for different redshifts.

    \item \textbf{MWS\_RR\_LYRAE}:
    We target Gaia DR2 sources with magnitudes $14 < G < 19$ that are classified as RR Lyrae by the general variability classifier and the Special Object Studies classifier.

    \item \textbf{UNWISE\_BLUE/GREEN}:
    
     unWISE catalogs are matched against HSC catalogs with a 2.75" match radius (1 WISE pixel). The unWISE samples are defined in 1901.03337 and 1909.07412.  They consist of sources passing a color cut in the WISE bands, and not matched to Gaia stars.  Additionally we remove ghosts, latents, and diffraction spikes, and mask around bright infrared stars.  For the secondary call, we consider only the green and blue samples. We require deep optical photometry to contain all of the unWISE sources. 
     
     By matching unWISE sources to deep imaging in COSMOS, we find that nearly all of the green (blue) sample has $y < 24$ (22.5).  Therefore, the y-band depth of the HSC imaging must reach 23.7 for point sources (accounting for the fact that extended source depths are typically 0.3 magnitudes fainter).  We make a healpix map of the y magnitude PSF depth from HSC's ``patch\_qa" files. However, we find that there is still significant incompleteness towards the edge of each HSC region, where the y-band limiting magnitude drops. Therefore, we also manually define an allowed region for each of the 7 HSC regions that contains HSC imaging to sufficient depth.

     We also use the HSC randoms to remove all regions near a bright star in HSC and require that the randoms are "primary" sources. We also provide an optional flag indicating that a source lies within an nside=8192 healpix pixel with less than 50\% of the median number of randoms (i.e. suggesting that it is near a moderately bright star that decreases the number density of randoms).  We do not remove sources lying in these regions, as this would reduce the catalog size by $\sim20\%$, but do flag them for caution in later analyses.Within the green sample, we find 98.3-98.5\% of unWISE sources are matched to an HSC source after these cuts. Completeness in blue is 99.3-99.5\%. The non-matches are distributed randomly throughout the footprint, and visual inspection reveals that they all either lack a nearby HSC source of sufficient brightness or are weird masking edge cases (i.e. where the unWISE source is unmasked but the nearby HSC source falls in a different healpix pixel and thus is masked).

     Finally, because of the large match radius required by the broad WISE pixels, WISE sources often have multiple HSC matches.  To help reduce the number of spurious matches, we remove all HSC sources with $y > 24$ in green (22.5 in blue) (this cut is determined by COSMOS matches of unWISE sources, which would suffer from the same spurious-match problem as the HSC sources, except that we can additionally use deep Spitzer imaging to exclude sources where the Spitzer flux is too faint, i.e. they are unlikely to contribute much to the unWISE flux). After this cut, for unWISE sources with multiple matches, we randomly select with probability $exp(-0.5 * (\delta/2.72)^2)$, where Delta is the positional offset between unWISE and HSC, and 2.72" is the 1-sigma width of the WISE PSF. That is, we randomly select the HSC objects with a probability corresponding to the probability of the position scattering by a specified amount just due to the unWISE PSF.

    \item \textbf{WD\_BINARIES\_BRIGHT}:

    These targets are for the White Dwarf binary survey have been selected as follows by cross-matching Gaia and GALEX GR6 and applying the following three cuts: (1) require a 5-sigma parallax detection: parallax/parallax\_error $>$ 5 and (2) select targets in a FUV vs FUV-G color magnitude diagram that lie below the main sequence:  $FUV_{mag} + 5\cdot log10(parallax/1000) + 5 > -0.3 + 1.5 \cdot (FUV_{mag} - phot\_ g\_ mean\_ mag )$ and (3) visible for DESI: dec $>$ -20

    This results in 164,503 targets. Of these, $\sim$ 34,000 overlap with the single white dwarfs that DESI will target for flux calibration purposes, and as science targets in the bright time MWS observations, so that there are $\sim$130,000 net targets for this project.

    \item \textbf{PV\_BRIGHT/PV\_DARK}:
    
    DESI aims to produce the largest catalog of peculiar velocities ever assembled. This sample will augment constraints on the growth rate of structure from the DESI BGS, reducing the uncertainty by a factor of $\sim 2.5$ at $z < 0.1$ compared to the BGS selection and numbers is based on a combination of DR8 and the Siena Galaxy Atlas (SGA), an angular diameter-limited catalog of objects within the DESI footprint that have diameters $D(25) > 20”$ (the diameter of the surface brightness isophote at 25 mag/arcsec2 in the $r$-band. Bright and dark delineate if the object was observed during bright or dark time.

    \item \textbf{GC\_BRIGHT/DARK}

    A target selection of Milky Way globular clusters (GCs). Here, we selected all the high probable member stars (P$>0.3$) \citep{gc_scnd}. Since the selection in this work is based on Gaia EDR3, some GC members might be outside the DESI footprint and may not in LS DR9.

    Then we break the target list into bright time and dark time survey
    \begin{itemize}
        \item  Bright: $16 < r_0 < 20$ for bright time survey
        \item  Dark:   $19 < r_0 < 21$ for dark time survey
    \end{itemize}

    \item \textbf{TOO\_HIP}

    These targets are attributed to spare fibers when spare fibers are available.

\end{itemize}

\textbf{Special Program Targets}: The following targets are part of various spare fiber programs with a brief description of their target selection outlined below. These targets are flagged as TOO\_HIP bit in the scnd  mask and can be
found in the \texttt{TERTIARY\_TARGET} column.

\begin{itemize}
    \item \textbf{TERTIARY 1: High Density Low-Z}

The \texttt{PRINCIPAL} sample is a simple magnitude-limited sample of $z_{\rm{fiber}} < 21.6$ with a stellar rejection cut in $r-z$ and $z-w1$ (similar to the DESI LRGs, although slightly relaxed to accommodate lower redshift galaxies):
\begin{itemize}
    \item $(z-w1) > 0.8 \cdot (r-z) -1.1 $
\end{itemize}
Unlike the DESI LRG selection, we also remove objects with extended morphology (i.e., objects that are not ``PSF'' type) from the \texttt{PRINCIPAL} sample.

The filler (\texttt{FILLER\_HIP},\texttt{FILLER\_LOP}) sample extends the magnitude limit to $z_{\rm{fiber}} < 22.4$, and we use color cuts in $g-r$ and $r-z$ to assign a higher priority to ($z >\sim 0.7$) galaxies (which are roughly half of the $21.6 \leq z_{\rm{fiber}} < 22.4$ target sample; few lower-priority filler targets were assigned fibers):
\begin{itemize}
    \item $(g-r) < 1.2 \cdot (r-z)$
    \item $(r-z) >1.3$
\end{itemize}

For both the principal and filler samples, we exclude from them objects that have already been observed by DESI with DELTACHI2 $>40$. We also excluded SV3 BGS targets with $r_{\rm{fiber}}<21.0$ (fainter BGS galaxies are retained here because they have lower redshift success rates in SV3) and any SV3 LRG and QSO targets as we already observed them with high completeness and high success rates in SV3.

The combined sample is observed with 11 DESI tiles in a ``rosette'' pattern in the COSMOS field centered at $R.A.=150.100^{\circ}, Dec.=2.182^{\circ}$. One 1200s (effective time) exposure is requested for each \texttt{PRINCIPAL} target. Four 1200s exposures are requested for each filler target (although only a small fraction of them were observed to full depth with the 11 tiles).

    \item \textbf{TERTIARY 5/6/7/8: Calibration}
    These fields are ordinary dark and bright tiles that get repeatedly re-observed over the course of the survey. We observe one dark and bright calibration field every lunation. On each field location, we design a series of tiles with slight offsets from one another so that the same targets can be imaged on different fibers.

    \item \textbf{TERTIARY 15: SN-HOSTS/DESI\_DEEP\_WL/ELG}

    \texttt{DESI\_DEEP}: We selected objects with extinction-corrected $i$-band magnitudes between 22 and 24.5 and $i$-fiber magnitude between 22 and 25 for the COSMOS field and 22 and 24.75 for the XMM-LSS field from the HSC-PDR3 wide catalog located. The faint limit was set to be similar to the depth of LSST Year-1 data, and the HSC weak lensing sample. The number of objects was then subsampled to have a uniform distribution of $i$-band magnitudes. The targets were then divided into two bins of priority, more objects with $i$-band magnitude between 22 and 23 were observed but for shorter exposure times whereas a smaller number of objects with $i$-band magnitude between 23 and 24.5 were observed but with larger exposure time. Further details on the target selection can be found in Dey et al. 2025 (in prep).

     \texttt{SN\_HOSTS}: Our candidates are selected by the presence of a SN in the host galaxy within the HSC COSMOS deep field \citep{2019PASJ...71...74Y}. There is no magnitude or color cut on the targeted host but we cut at $r=25$ for the host galaxies of less certain SNe. Internally the targets are prioritized according to their likely scientific value for constructing a Hubble diagram and measuring the dark energy equation of state.  Highest priority went to host galaxies of SNeIa observed with HST as part of programs GO14808 and GO15363 (PI Suzuki); priority then decreased in the order of: other likely SNe\,Ia; including the HST SNe\,Ia), possible SNe\,Ia, and less certain SNe\,Ia. We then removed the targets whose redshifts we had previously obtained with AAOmega \cite{2006SPIE.6269E..0GS}, and targets with redshifts from the literature. To make a more complete Hubble diagram, we also include host galaxies for confirmed or suspected SNe\,Ia lacking redshifts with the SNLS D2 deep field \cite{sn_hosts3}.

    \item \textbf{TERTIARY 19: DESI II Bright}

    We propose to observe $\sim$4000 galaxies in the XMM-LSS field in $i \in [19, 22]$ during typical or favorable bright time conditions, fairly drawn from a parent target list with density 3125 per square degree to fill the available number of fibers (resulting in a final density of $\sim$480 per square degree). We propose 4000 sec of effective time, with dithering. Our selection observes objects at $i > 22$. We estimate the necessary observing time based on the “filler hip” high density sample from the COSMOS field, which peaks at $i \sim 22$ and achieved $99\%$ redshift success rate with 4800s of effective time. Similarly to that sample, we will apply a cut in $i$-band fiber magnitude to remove objects with very low surface brightness that are very unlikely to have successful redshifts after our full integration time. We also subsample galaxies with an analytic exponential weight such that they are roughly uniformly distributed in magnitude.

    \item \textbf{TERTIARY 23: 4-in-1}
    
    LOWZ: See LOWZ

    MERIAN: The Merian Ancillary Program consists of 3 selections

    \begin{itemize}
        \item MS1: dwarf galaxies with $7.8<logM_{star}/M_{sun}<9.5$ at $z<0.25$ and $19.5 < i < 23$. The mass and redshift for this sample is based on the COSMOS2015 catalog 30-band SED fitting.
        \item MS2: all galaxies with $i < 25$ at $z<0.12$ based on Merian 7-band photo-z’s.
    \end{itemize}

    DESI\_DEEP: See TERTIARY 15

    ELG: Two extended selection of ELGs are a simple grizy cut based on HSC/WIDE
    
    \begin{enumerate}
        \item Selection 1:
        \begin{itemize}[label=\textbullet]
            \item $g_{cmodel\_mag} - a_g > 22.0$
            \item $g_{cmodel\_mag} - a_g < 24.5$
            \item $r_{cmodel\_mag} - a_r - i_{cmodel\_mag} + a_i > 0.1$
            \item $r_{cmodel\_mag} - a_r - i_{cmodel\_mag} + a_i < 0.4$
            \item $g_{cmodel\_mag} - a_g - r_{cmodel\_mag} + a_r > -0.1 + 0.7 \cdot (r_{cmodel\_mag} - a_r - i_{cmodel\_mag} + a_i)$
            \item $g_{cmodel\_mag} - a_g - r_{cmodel\_mag} + a_r < 0.1 + 0.7 \cdot (r_{cmodel\_mag} - a_r - i_{cmodel\_mag} + a_i)$
            \item $z_{cmodel\_mag} - a_z - y_{cmodel\_mag} + a_y > 0.35$ 
        \end{itemize}

        \item Selection 2:
        \begin{itemize}[label=\textbullet]
            \item ($i-y > 0.35$) \& $(r-i < i-y -0.19)$
            \item ($g < 24$ \& $g_{\rm{fiber}} < 24.3$) OR $(r < 24$ \& $r_{\rm{fiber}} < 24.3)$
        \end{itemize}
    
    \end{enumerate}

    \item \textbf{TERTIARY 26: ELG (Special)}

    Selection of Special Program ELGs(see TERTIARY 23 ELGs) that received less than 4 observations.

    \item \textbf{TERTIARY 38: ELG Systematics}

    This sample is an expanded ELG selection for the use of forward-modeling studies of the ELG imaging systematics \citep{zhou_2025}. The imaging data comes from DECam deep field imaging and HSC PDR3 deep/ultra-deep imaging in COSMOS. 
    
    DECam selection:
    \begin{itemize}[label=\textbullet]
        \item $19.5 <\texttt{gfibermag} < 24.5$
        \item  $g - r < 0.9$
        \item $g - r < 1.32 - 0.7 \cdot (r-z)$
    \end{itemize}

    HSC selection:
    \begin{itemize}[label=\textbullet]
        \item $19.5 < \texttt{gfibermag} < 24.55$
        \item  $g - r < 0.8$
        \item $g - r < 1.22 - 0.7 \cdot (r-z)$
    \end{itemize}
    
    where $\texttt{gfibermag}$ is based on $\texttt{g\_apertureflux\_15\_flux}$.
    The combined target sample (excluding duplicates from the two selections) has a density of roughly 24000/sq. deg. It is observed by 25 tiles in a ``rosette" pattern in the COSMOS field centered at $R.A.=150.11^{\circ}, Dec.=2.3^{\circ}$, each tile with 1000s effective exposure time. A total of 97800 unique objects in the target sample were observed.

\end{itemize}

\section{Targeting Information}
\label{appendix2}
The tables below summarize all the target classes within each catalog, the number of quality redshifts, and the fraction of quality redshifts for that class. Note that objects in these tables are represented by their target selection, not by spectroscopic classification. As there is overlap between galaxy, star and quasar spectroscopic classification within a target class, numbers for total objects and quality redshift may appear higher than previously mentioned. Targets may share multiple target classes as their photometric properties satisfy multiple selection algorithms.

\begin{longtable}{lllrrl}
\caption{COSMOS Target Classification and Redshift Robustness} \label{tab:cosmos_analysis_updated} \\
\hline
 Target Class & Total Objects & Quality Redshifts & Quality Redshift (\%) \\ 
\hline
\endfirsthead

\hline
Target Class & Total Objects & Quality\_Z True & Quality\_Z (\%) \\ 
\hline
\endhead

\hline
\endfoot

\hline
LRG & 11720 & 11565 & 98.7\% \\
ELG & 36128 & 30264 & 83.8\% \\
QSO & 4873 & 4471 & 91.8\% \\ 
BGS\_ANY & 21865 & 21684 & 99.2\% \\ 
MWS\_ANY & 19424 & 19401 & 99.9\% \\ 
STD\_FAINT & 1191 & 1189 & 99.8\% \\
STD\_BRIGHT & 618 & 618 & 100.0\% \\ 
STD\_WD & 88 & 87 & 98.9\% \\
SCND\_ANY & 146452 & 113971 & 77.8\% \\ 

\hline
 \textbf{Secondary Targets} \\
 \hline
DARK\_TOO\_HIP & 139840 & 108331 & 77.5\% \\ 
LOW\_Z\_TIER3 & 1909 & 1875 & 98.2\% \\ 
WISE\_VAR\_QSO & 1581 & 1494 & 94.5\% \\ 
PSF\_OUT\_DARK & 1414 & 1342 & 94.9\% \\ 
LOW\_Z & 856 & 842 & 98.4\% \\ 
PV\_DARK\_HIGH & 816 & 528 & 64.7\% \\ 
PV\_BRIGHT\_HIGH & 809 & 519 & 64.2\% \\ 
HSC\_HIZ\_SNE & 753 & 625 & 83.0\% \\ 
LOW\_Z\_TIER2 & 686 & 655 & 95.5\% \\ 
PSF\_OUT\_BRIGHT & 651 & 640 & 98.3\% \\ 
UNWISE\_BLUE & 298 & 296 & 99.3\% \\ 
PV\_DARK\_MEDIUM & 208 & 201 & 96.6\% \\ 
PV\_BRIGHT\_MEDIUM & 201 & 200 & 99.5\% \\ 
LOW\_Z\_TIER1 & 162 & 141 & 87.0\% \\ 
PV\_DARK\_LOW & 130 & 126 & 96.9\% \\ 
PV\_BRIGHT\_LOW & 129 & 124 & 96.1\% \\ 
BRIGHT\_TOO\_HIP & 140 & 35 & 25.0\% \\
MWS\_FAINT\_BLUE & 61 & 59 & 96.7\% \\ 
ISM\_CGM\_QGP & 60 & 55 & 91.7\% \\ 
UNWISE\_GREEN & 54 & 53 & 98.1\% \\  
QSO\_RED & 26 & 22 & 84.6\% \\ 
MWS\_CALIB & 24 & 24 & 100.0\% \\ 
BACKUP\_CALIB & 24 & 24 & 100.0\% \\ 
FAINT\_HPM & 23 & 18 & 78.3\% \\ 
SN\_HOSTS & 22 & 22 & 100.0\% \\ 
STRONG\_LENS & 22 & 17 & 77.3\% \\ 
MWS\_CLUS\_GAL\_DEEP & 18 & 15 & 83.3\% \\ 
MWS\_FAINT\_RED & 18 & 18 & 100.0\% \\ 
HPM\_SOUM & 5 & 4 & 80.0\% \\ 
WD\_BINARIES\_BRIGHT & 4 & 4 & 100.0\% \\ 
BHB & 3 & 3 & 100.0\% \\ 
LOW\_MASS\_AGN & 3 & 3 & 100.0\% \\ 
Z5\_QSO & 2 & 2 & 100.0\% \\ 
MWS\_RR\_LYRAE & 3 & 3 & 100.0\% \\ 
MWS\_RRLYR & 1 & 1 & 100.0\% \\ 

\hline
 \textbf{Special Targets} \\
\hline
ELG\_PRINCIPAL & 78072 & 58057 & 74.4\% \\ 
PRINCIPAL & 32881 & 32124 & 97.7\% \\ 
ELG & 18603 & 13671 & 73.5\% \\ 
FILLER\_HIP & 5240 & 4051 & 77.3\% \\ 
LOWZ\_FAINT & 2760 & 1819 & 65.9\% \\ 
BGS\_BGS\_BRIGHT & 1754 & 1752 & 99.9\% \\ 
FILLER\_LOP & 1752 & 1679 & 95.8\% \\ 
DESI\_LRG & 1114 & 1112 & 99.8\% \\ 
MERIAN\_MS2 & 984 & 530 & 53.9\% \\ 
MERIAN\_MS1 & 938 & 805 & 85.8\% \\ 
DESI\_DEEP\_HIP & 932 & 657 & 70.5\% \\ 
BGS\_BGS\_FAINT & 680 & 679 & 99.9\% \\ 
DESI\_ELG\_LOP & 645 & 547 & 84.8\% \\ 
DESI\_QSO & 627 & 596 & 95.1\% \\ 
LOWZ\_BRIGHT & 589 & 569 & 96.6\% \\ 
DESI\_DEEP\_LOP & 576 & 484 & 84.0\% \\ 
DESI\_ELG\_HIP & 470 & 420 & 89.4\% \\ 
MWS\_MWS\_MAIN\_BLUE & 265 & 265 & 100.0\% \\ 
BGS\_BGS\_FAINT\_HIP & 263 & 263 & 100.0\% \\ 
FILLER\_LOP & 1752 & 1679 & 95.8\% \\
MWS\_MWS\_BROAD & 188 & 188 & 100.0\% \\ 
DESI\_DEEP\_HIP2 & 136 & 132 & 97.1\% \\ 
DESI\_ELG & 85 & 85 & 100.0\% \\ 
MWS\_MWS\_MAIN\_RED & 65 & 65 & 100.0\% \\ 
MWS\_MWS\_WD & 15 & 15 & 100.0\% \\ 
MWS\_MWS\_BHB & 7 & 7 & 100.0\% \\ 
BGS\_BGS\_WISE & 4 & 4 & 100.0\% \\ 
MWS\_MWS\_NEARBY & 1 & 1 & 100.0\% \\ 
\end{longtable}

\begin{longtable}{lllrrl}
\caption{XMM Target Classifications and Redshift Robustness} 
\label{tab:xmm_analysis} \\
\hline
 Target Class & Total Objects & Quality Redshifts & Quality Redshift (\%) \\ 
\hline
\endfirsthead

\hline
Target Class & Total Objects & Quality\_Z True & Quality\_Z (\%) \\ 
\hline
\endhead

\hline
\endfoot

 LRG & 9985 & 9736 & 97.5\% \\
 ELG & 28056 & 22427 & 79.9\% \\ 
 QSO & 5251 & 4711 & 89.7\% \\
 BGS\_ANY & 18357 & 18021 & 98.2\% \\ 
 MWS\_ANY & 10092 & 10057 & 99.7\% \\  
 STD\_FAINT & 770 & 768 & 99.7\% \\ 
 STD\_BRIGHT & 425 & 423 & 99.5\% \\ 
 STD\_WD & 66 & 64 & 97.0\% \\
 SCND\_ANY & 10926 & 9232 & 84.5\% \\ 

\hline
\textbf{Secondary Targets} \\
\hline
DARK\_TOO\_HIP & 3345 & 2404 & 71.9\% \\
PSF\_OUT\_DARK & 2468 & 2335 & 94.6\% \\
WISE\_VAR\_QSO & 1856 & 1741 & 93.8\% \\
LOW\_Z\_TIER3 & 1488 & 1425 & 95.8\% \\
BRIGHT\_TOO\_HIP & 1320 & 1156 & 87.6\% \\
PSF\_OUT\_BRIGHT & 719 & 683 & 95.0\% \\
PV\_DARK\_HIGH & 353 & 267 & 75.6\% \\
PV\_BRIGHT\_HIGH & 319 & 261 & 81.8\% \\
LOW\_Z\_TIER2 & 258 & 202 & 78.3\% \\
PV\_DARK\_MEDIUM & 145 & 140 & 96.6\% \\
LOW\_Z\_TIER1 & 140 & 97 & 69.3\% \\
PV\_BRIGHT\_MEDIUM & 140 & 140 & 100.0\% \\
STRONG\_LENS & 78 & 62 & 79.5\% \\
FAINT\_HPM & 76 & 19 & 25.0\% \\
MWS\_FAINT\_BLUE & 66 & 64 & 97.0\% \\
PV\_DARK\_LOW & 55 & 51 & 92.7\% \\
PV\_BRIGHT\_LOW & 51 & 49 & 96.1\% \\
QSO\_RED & 37 & 29 & 78.4\% \\
MWS\_FAINT\_RED & 35 & 35 & 100.0\% \\
SN\_HOSTS & 16 & 16 & 100.0\% \\
LOW\_MASS\_AGN & 12 & 12 & 100.0\% \\
BHB & 4 & 4 & 100.0\% \\
WD\_BINARIES\_BRIGHT & 2 & 2 & 100.0\% \\
Z5\_QSO & 2 & 2 & 100.0\% \\
MWS\_RR\_LYRAE & 2 & 2 & 100.0\% \\
WD\_BINARIES\_DARK & 1 & 1 & 100.0\% \\
BRIGHT\_HPM & 1 & 1 & 100.0\% \\

\hline
\textbf{Special Targets} \\
\hline
BGS\_BGS\_BRIGHT & 2662 & 2652 & 99.6\% \\
DESI\_DEEP\_LOP & 1979 & 1603 & 81.0\% \\
DESI\_LRG & 1890 & 1875 & 99.2\% \\
DESI\_ELG\_LOP & 1733 & 1444 & 83.3\% \\
DESI\_QSO & 1584 & 1436 & 90.7\% \\
PRINCIPAL & 1140 & 1101 & 96.6\% \\
FILLER & 1136 & 1126 & 99.1\% \\
DESI\_DEEP\_HIP & 1025 & 624 & 60.9\% \\
DESI\_ELG\_HIP & 839 & 748 & 89.2\% \\
BGS\_BGS\_FAINT & 689 & 687 & 99.7\% \\
MWS\_MWS\_BROAD & 434 & 434 & 100.0\% \\
BGS\_BGS\_FAINT\_HIP & 414 & 411 & 99.3\% \\
MWS\_MWS\_MAIN\_BLUE & 362 & 362 & 100.0\% \\
FILLER\_LOP & 322 & 306 & 95.0\% \\
SNHOST & 285 & 187 & 65.6\% \\
DESI\_ELG & 149 & 145 & 97.3\% \\
SNHOST\_HIP & 124 & 92 & 74.2\% \\
MWS\_MWS\_MAIN\_RED & 103 & 103 & 100.0\% \\
MWS\_MWS\_WD & 15 & 14 & 93.3\% \\
SNHOST\_VERY\_HIP & 10 & 9 & 90.0\% \\
BGS\_BGS\_WISE & 8 & 8 & 100.0\% \\
MWS\_MWS\_BHB & 4 & 4 & 100.0\% \\
MWS\_MWS\_NEARBY & 1 & 1 & 100.0\% \\
\hline
\end{longtable}

\onecolumngrid
\section{Data Model}
\label{appendix3}

\begin{longtable}[h!]{| l | l | l | >{\raggedright\arraybackslash}p{8cm} |}
\hline
\textbf{Column Name} & \textbf{Unit} & \textbf{Dtype} & \textbf{Description} \\
\hline
\endfirsthead
\hline
\textbf{Column Name} & \textbf{Unit} & \textbf{Dtype} & \textbf{Description} \\
\hline
\endhead
\hline
\endfoot
\hline
DESINAME & -- & |S22 & Human readable identifier of a sky location DESI $JXXX.XXXX[+/-]YY.YYYY$, where X,Y=truncated decimal TARGET\_RA, TARGET\_DEC, precise to 0.36 arcsec. Multiple objects can map to a single DESINAME if very close on the sky.
 \\
TARGET\_RA & deg & $>f8$ & Barycentric right ascension in ICRS \\
TARGET\_DEC & deg & $>f8$ & Barycentric declination in ICRS \\
OBJTYPE & -- & |S3 & Object type: TGT, SKY, NON, BAD \\
PMRA & $\rm{mas}\,yr^{-1}$ & $>f4$ & Proper motion in the +RA direction (already including cos(dec)) \\
PMDEC & $\rm{mas}\,yr^{-1}$ & $>f4$ & Proper motion in the +Dec direction \\
SURVEY & -- & |S7 & Survey when the object's best spectrum was observed (Main, SV1, SV3, Special) \\
PROGRAM & -- & |S6 & Program when the object's best spectrum was observed (bright, dark, backup, other) \\
HEALPIX & -- & $>i4$ & HealPIX pixel of the object's best spectrum \\
Z & -- & $>f8$ & Redshift estimation from \textit{Redrock} or \textit{Redrock}+afterburners for quasar targets for the object's best spectrum \\
ZERR & -- & $>f8$ & Redshift uncertainty estimation from \textit{Redrock} or \textit{Redrock}+afterburners for quasar targets for the object's best spectrum \\
ZWARN & -- & $>i8$ & Warning flag bit for \textit{Redrock} for the object's best spectrum; 0 indicates no issue; 4 indicates low DELTACHI2 \\
DELTACHI2 & -- & $>f8$ & Difference in $\chi^2$ from \textit{Redrock}'s best PCA template fit and second-best fit for the object's best spectrum (used for quality assessments) \\
CHI2 & -- & $>f8$ & Non-reduced $\chi^2$ from \textit{Redrock}'s best PCA template fit for the object's best spectrum \\
NPIXELS & -- & $>i8$ & Number of unmasked pixels contributing to the \textit{Redrock} fit \\
SPECTYPE & -- & |S6 & Spectral classification from \textit{Redrock} or \textit{Redrock}+afterburners for quasar targets for the object's best spectrum \\
SUBTYPE & -- & |S3 & Spectral subtype of the object for the object's best spectrum (may be blank) \\
BEST\_Z & -- & $>f8$ & Best reported redshift for this object \\
DZ & -- & $>f8$ & Largest difference in redshifts between objects that were merged \\
QUALITY\_Z & -- & bool & Flag (0 or 1) indicating if the object has a quality redshift \\
TSNR2\_BGS & -- & $>f4$ & Target signal-to-noise ratio squared for Bright Galaxy Survey targets for the best spectrum \\
TSNR2\_LRG & -- & $>f4$ & Target signal-to-noise ratio squared for Luminous Red Galaxy targets for the best spectrum \\
TSNR2\_ELG & -- & $>f4$ & Target signal-to-noise ratio squared for Emission Line Galaxy targets for the best spectrum \\
TSNR2\_QSO & -- & $>f4$ & Target signal-to-noise ratio squared for Quasar targets for the best spectrum \\
TSNR2\_LYA & -- & $>f4$ & Target signal-to-noise ratio squared for Lyman-alpha targets for the best spectrum \\
IS\_QSO\_MGII & -- & bool & Indicates if a quasar target was flagged as a quasar by the MgII afterburner for the object's best spectrum \\
A & -- & $>f4$ & Fitted parameter by MgII \\
B & -- & $>f4$ & Fitted parameter by MgII \\
Z\_NEW & -- & $>f8$ & New redshift computed with \textit{Redrock} with QN prior and only QSO templates \\
ZERR\_NEW & -- & $>f4$ & Redshift error from the new run of \textit{Redrock} \\
Z\_QN & -- & $>f4$ & Redshift computed with Quasarnp \\
IS\_QSO\_QN\_NEW\_RR & -- & bool & Indicates if the object is detected as a QSO with Quasarnp and a new redshift fit with prior is performed for the object's best spectrum \\
DESI\_TARGET & -- & $>i8$ & DESI (dark time program) target selection bitmask \\
BGS\_TARGET & -- & $>i8$ & BGS (Bright Galaxy Survey) target selection bitmask \\
MWS\_TARGET & -- & $>i8$ & Milky Way Survey targeting bits \\
SCND\_TARGET & -- & $>i8$ & Target selection bitmask for secondary programs\\
SV1\_DESI\_TARGET & -- & $>i8$ & DESI (dark time program) target selection bitmask for SV1 \\
SV1\_BGS\_TARGET & -- & $>i8$ & BGS (Bright Galaxy Survey) target selection bitmask for SV1 \\
SV1\_MWS\_TARGET & -- & $>i8$ & Milky Way Survey targeting bits for SV1 \\
SV1\_SCND\_TARGET & -- & $>i8$ & Target selection bitmask for secondary programs for SV1 \\
SV3\_DESI\_TARGET & -- & $>i8$ & DESI (dark time program) target selection bitmask for SV3 \\
SV3\_BGS\_TARGET & -- & $>i8$ & BGS (Bright Galaxy Survey) target selection bitmask for SV3 \\
SV3\_MWS\_TARGET & -- & $>i8$ & Milky Way Survey targeting bits for SV3 \\
SV3\_SCND\_TARGET & -- & $>i8$ & Target selection bitmask for secondary programs for SV3 \\
SPECIAL\_TARGET & -- & |S100 & Target selection description for special programs \\
LINENAME\_FLUX & $ 10^{-17} \mathrm{erg} / (\mathrm{s} \,\mathrm{cm}^2)$ & $>f4$ & Gaussian-integrated emission-line flux measurement from \textit{FastSpecFit} of the best spectrum \\
LINENAME\_FLUX\_IVAR & $10^{34} \mathrm{cm}^4 \mathrm{s}^2 / \mathrm{erg}^2$ & $>f4$ & Inverse variance of integrated flux measurement from \textit{FastSpecFit} of the best spectrum \\
TOTAL\_NUM\_COADD & -- & $>f8$ & Number of exposures used in the coadded spectrum for the object's best spectrum \\
TOTAL\_COADD\_EXPTIME & s & $>f8$ & Total exposure time across all coadded exposures \\
LRG\_MASK & -- & $uint8$ & Bright star mask used for LRG targets applied to the catalog \\
ELG\_MASK & -- & $uint8$ & Bright star mask used for ELG targets applied to the catalog \\
HAS\_RVS & -- & bool & Boolean flag indicating RVS measurements for object \\
VRAD & $\rm{km}\,s^{-1}$ & $>f8$ & Radial velocity \\
VRAD\_ERR & $\rm{km}\,s^{-1}$  & $>f8$ & Radial velocity error \\
VRAD\_SKEW & -- & $>f8$ & Radial velocity skewness \\
VRAD\_KURT & -- & $>f8$ & Radial velocity kurtosis \\
LOGG & -- & $>f8$ & Log of surface gravity \\
TEFF & K & $>f8$ & Effective temperature \\
FE\_H & -- & $>f8$ & [Fe/H] from template fitting \\
LOGG\_ERR & -- & $>f8$ & Log of surface gravity uncertainty \\
TEFF\_ERR & K & $>f8$ & Effective temperature uncertainty \\
FE\_H\_ERR & -- & $>f8$ & [Fe/H] from template fitting uncertainty \\
VSINI & $\rm{km}\,s^{-1}$ & $>f8$ & Stellar rotation velocity \\
RVS\_WARN & -- & $>i8$ & \textit{RVSpecfit} warning flag \\
HAS\_DECALS\_DR9 & -- & bool & Flag indicating if the object has corresponding photometry from DECaLS \\
HAS\_DECAM\_DR10 & -- & bool & Flag indicating if the object has corresponding photometry from DECam \\
HAS\_HSC\_UD\_PDR3 & -- & bool & Flag indicating if the object has corresponding photometry from HSC Ultra Deep \\
HAS\_HSC\_WIDE\_PDR3 & -- & bool & Flag indicating if the object has corresponding photometry from HSC WIDE \\
HAS\_COSMOS2020 & -- & bool & Flag indicating if the object has corresponding photometry from COSMOS2020 \\
HAS\_MERIAN & -- & bool & Flag indicating if the object has corresponding photometry from Merian \\
HAS\_VI\_Z & -- & bool & Boolean flag indicating if object has visually inspected redshift \\
VI\_Z & -- & $>f8$ & Visually inspected redshift \\
VI\_SPECTYPE & -- & |S6 & Visually inspected spectroscopic classification \\
VI\_QUALITY & -- & $>f8$ & Visually inspected redshift quality (3 and 4 indicate quality visually inspected redshift) \\
\hline
\end{longtable}

\end{document}